\documentclass{article}

\usepackage[nonatbib,preprint]{neurips_2022}

\usepackage[utf8]{inputenc} 
\usepackage[T1]{fontenc}    
\usepackage{hyperref}       
\usepackage{url}            
\usepackage{booktabs}       
\usepackage{amsfonts}       
\usepackage{nicefrac}       
\usepackage{microtype}      
\usepackage{xcolor}         

\usepackage{ifthen}
\usepackage{cite}
\usepackage[cmex10]{amsmath} 
\usepackage{physics}

\usepackage[shortlabels]{enumitem}
\usepackage{array}
\usepackage{makecell}
\usepackage{verbatim}
\usepackage{graphicx}
\usepackage{caption}
\usepackage{float}
\usepackage[colorinlistoftodos]{todonotes}
\usepackage{amssymb,bm}
\usepackage{amsthm}
\usepackage{tcolorbox}
\usepackage[framemethod=tikz]{mdframed}
\usepackage{mathtools,xparse}
\usepackage{tikz} 
\usepackage{multicol} 
\usetikzlibrary{arrows}
\usetikzlibrary{shapes}
\usepackage{pgfplots}
\usetikzlibrary{plotmarks}
\pgfplotsset{compat=newest}

\newtheorem{theorem}{Theorem}
\newtheorem{lemma}[theorem]{Lemma}

\def\one{\mathbf{1}}
\def\zero{\mathbf{0}}

\def\cD{\mathcal{D}}
\def\cX{\mathcal{X}}
\def\cY{\mathcal{Y}}

\def\cP{\mathcal{P}}
\def\cQ{\mathcal{Q}}

\def\bpi{\bm{\pi}}

\def\pr{\text{Pr}}

\def\GT{\widehat{M}_0^{\text{GT}}}

\def\tIID{\text{i.i.d.}}
\def\tMarkov{\text{Markov}}

\def\wh{\widehat}

\DeclarePairedDelimiter\floor{\lfloor}{\rfloor}

\title{How good is Good-Turing for Markov samples?} 

\author{%
  Prafulla Chandra and Andrew Thangaraj\\
  Department of Electrical Engineering\\
  Indian Institute of Technology Madras\\
  Chennai, India 600036 \\
  \texttt{ee16d402,andrew@ee.iitm.ac.in} \\
   \And
    Nived Rajaraman \\
  Department of Electrical Engineering and Computer Science \\
  University of California, Berkeley \\ 
Berkeley, CA 94720, USA \\
  \texttt{nived.rajaraman@gmail.com} \\
}

\begin{document}

\maketitle

\begin{abstract}

   The Good-Turing (GT) estimator for the missing mass (i.e., total probability of missing symbols) in $n$ samples is the number of symbols that appeared exactly once divided by $n$. For i.i.d. samples, the bias and squared-error risk of the GT estimator can be shown to fall as $1/n$ by bounding the expected error uniformly over all symbols. In this work, we study convergence of the GT estimator for missing stationary mass (i.e., total stationary probability of missing symbols) of Markov samples on an alphabet $\cX$ with stationary distribution $[\pi_x:x\in\cX]$ and transition probability matrix (t.p.m.) $P$. This is an important and interesting problem because GT is widely used in applications with temporal dependencies such as language models assigning probabilities to word sequences, which are modelled as Markov. We show that convergence of GT depends on convergence of $(P^{\sim x})^n$, where $P^{\sim x}$ is $P$ with the $x$-th column zeroed out. This, in turn, depends on the Perron eigenvalue $\lambda^{\sim x}$ of $P^{\sim x}$ and its relationship with $\pi_x$ uniformly over $x$. For randomly generated t.p.ms and t.p.ms derived from New York Times and Charles Dickens corpora, we numerically exhibit such uniform-over-$x$ relationships between $\lambda^{\sim x}$ and $\pi_x$. This supports the observed success of GT in language models and practical text data scenarios. For Markov chains with rank-2, diagonalizable t.p.ms having spectral gap $\beta$, we show minimax rate upper and lower bounds of $1/(n\beta^5)$ and $1/(n\beta)$, respectively, for the estimation of stationary missing mass. This theoretical result extends the $1/n$ minimax rate for i.i.d. or rank-1 t.p.ms to rank-2 Markov, and is a first such minimax rate result for missing mass of Markov samples.

\end{abstract}

\section{Introduction} 
When observing a sequence of symbols from an unknown alphabet and distribution, we are often interested in the probability that the next sampled symbol is going to be new, i.e. a symbol that has not been seen so far. This probability is the sum of the probabilities of all the symbols missing in the sequence of samples observed so far, and is called the \emph{missing mass}. Good and Turing \cite{Good1953} had originally studied this problem in the context of solving the enigma code. The popular Good-Turing (GT) estimator estimates missing mass as the ratio of the number of symbols seen exactly once in the samples to the sample size. Today, estimation of missing mass finds applications in language modelling \cite{Gale1991,Gale1995,Chen1996}, ecology \cite{Chao1992,Shen_03} and in entropy estimation \cite{vu_07}, and it has been studied in the i.i.d. samples setting by multiple authors \cite{McAllester2000, berend2013, Chandra19, Ohannessian12, Ohannessian19, Orlitsky2015, Rajaraman17, Acharya17, Kontorovich_20, Painsky22}. The mean-squared error for estimating missing mass using the GT estimator in the i.i.d. setting \cite{Rajaraman17} falls as $(\text{sample-size})^{-1}$ with no further assumptions on the alphabet size or restrictions to the distribution. So, whenever the missing mass is expected to be non-vanishing, it can be reliably estimated in the i.i.d. case. 

While missing mass and the GT estimator are well-studied in the i.i.d. samples regime, their definitions and properties in cases where the samples have memory have not been extensively considered in the literature. 
Many applications like natural language text involve data with temporal dependencies, which are often modelled as Markov chains \cite{Chen1996}. 
In this work, we study missing mass and its estimation through GT estimator in cases when the samples form a Markov chain. Estimation from Markov samples has been considered in \cite{wolfer-kontorovich19, Hao2018learning,Weissman_18,Kontorovich_19,han2021optimal_pred} and forays towards missing mass estimation from  Markov chains were made in \cite{skorski_Markov_M0,Chandra20_ISIT,Chandra22}. Though GT missing mass estimates are used as part of distribution estimation in cases where the samples have memory, theoretical properties of the estimation of missing mass in such scenarios is, to the best of our knowledge, considered for the first time here.

\section{Missing stationary mass of a Markov chain}
A sequence $X^n = (X_1,X_2,\ldots,X_n)$, $X_i\in\cX$, is said to be a Markov chain if 
$$\pr(X_{i}{=}x_i|X_{i-1}{=}x_{i-1},\ldots,X_1{=}x_1)=\pr(X_2{=}x_i|X_1{=}x_{i-1})$$ 
for $i=2,\ldots,n$ and all $x_i\in\cX$. The $X_i$'s are called states and $\cX$ is called the state space. $K \triangleq |\cX|$ is the size of the alphabet $\cX$. The transition probability matrix (t.p.m.) of the Markov chain, denoted $P$, is the $K \times K$ matrix with $(i,j)$-th element $P_{ij} \triangleq \pr(X_2 = j | X_1 = i)$. 
A distribution $\bpi=[\pi_1,\ldots,\pi_{K}]$ on $\cX$ is said to be a stationary or invariant distribution of the Markov chain if $\bpi P = \bpi$ \cite{markov_book}.  A Markov chain $X^n$ is said to be stationary if $X_1\sim\bpi$, which implies that $X_i\sim\bpi$ for all $i$. We denote by $X^n\sim\tMarkov(P,\bpi)$, a stationary Markov chain with t.p.m. $P$ and state distribution $\bpi$.

 Let $I(\cdot)$ and $E[\cdot]$ denote the indicator random variable and expectation, respectively and $[K]$ denote the set $\{1,2,\ldots,K\}$. 
 For $x \in \cX$, $N_x(X^n) \triangleq \sum_{i=1}^n I(X_i = x)$ is the number of occurrences of $x$ in $X^n$, also called the frequency of $x$. 
    For $l =0,1,2,\ldots$, $\phi_l(X^n) \triangleq \sum_{x \in \cX} I(N_x(X^n) = l)$ is the number of letters that have occurred $l$ times in $X^n$. 
 
 The \emph{missing stationary mass} of a Markov chain $X^n\sim\tMarkov(P,\bpi)$, which is the missing mass of  $\bpi$ in $X^n$, is defined as
\begin{equation}
    M_{0}(\bpi,X^n) \triangleq \sum_{x \in \cX} \pi_x \ I(N_x(X^n) = 0).
    \label{eq:M0_stat_def}
\end{equation}

Estimation of the missing mass $M_0(\bpi,X^n)$ when $X^n\sim\tMarkov(P,\bpi)$ and the quantities $\cX$, $K$, $\bpi$ and $P$ are unknown is important for many applications and in theory (see the role of missing mass in excess risk of competitive distribution estimation in \cite{Orlitsky2015}). Note that $M_0(\bpi,X^n)$ is a random variable that is a function of both the samples $X^n$ and the distribution $\bpi$. This makes estimation of missing mass and its analysis non-trivial even in the classical i.i.d. regime where $X^n\sim$ i.i.d. $\bpi$ or a Markov chain with $P = \one \bpi$. For $X^n \sim \tMarkov(P,\bpi)$ with a general $P$, the samples are not drawn exactly as per $\bpi$, which is the weight for measuring missing mass. This makes the study of missing stationary mass of a Markov chain more challenging, when compared to an i.i.d. sequence.

The Good-Turing (GT) estimator \cite{Good1953} for the missing mass $M_0(\bpi,X^n)$ is defined as
\begin{equation}
    \GT(X^n) \triangleq \frac{\phi_1(X^n)}{n}, \label{eq:GT_def}
\end{equation}
which is the fraction of symbols that have appeared exactly once in the $n$ samples. 

The minimax squared-error risk\footnote{In cases where missing mass is expected to vanish with $n$, relative error is more meaningful as shown in \cite{Ohannessian19}. However, in many interesting large alphabet scenarios, missing mass is non-vanishing for large $n$, and estimation is critical in the non-vanishing case. So, we consider additive error in this work.} of estimating missing mass over a class of distributions $\cP$, denoted $R^{*}_{n}(\cP)$, is defined as
\begin{align}
    R^{*}_{n}(\cP) 
 = \min_{\text{Estimator }\wh{M}_0}\ \max_{ (P,\bpi) \in \cP } \ E_{X^n\sim \tMarkov(P,\bpi)}[(\wh{M}_0(X^n)-M_0(\bpi,X^n))^2].
    \label{eq:M0_minimax}
\end{align}
 

In this work, we make two main contributions. Firstly, we study the Good-Turing (GT) estimator and attempt to characterise the classes of Markov chains or t.p.ms for which it converges to missing stationary mass. 

Secondly, on the theoretical side, we characterise the minimax squared-error risk of estimating missing stationary mass over a class of rank-2 Markov t.p.ms with a spectral gap. In the i.i.d. case, when $X^n \sim$ i.i.d. $\bpi$, the minimax rate is $1/n$ as shown in \cite{Rajaraman17,Acharya17}. To the best of our knowledge, this work presents the first minimax rate result for a Markov case.

\section{Main Results}

\subsection{Convergence of GT estimator}
We first provide simplified expressions for the bias of the GT estimator for missing stationary mass, using which convergence analysis becomes possible. We require the following notation.

For $x\in\cX$, let $P^{\sim x}$ be a modified transition matrix equal to $P$ in all positions except the $x$-th column, which is set as the all-$0$ vector. So, under $P^{\sim x}$, the symbol $x$ is never observed. Let $P_x\triangleq P-P^{\sim x}$ be all-zero except for the $x$-th column, which is set as the $x$-th column of $P$. Let $\bpi^{\sim x}$ be equal to the vector $\bpi$ in all positions except the $x$-th entry, which is set as $0$. Let $\one$ be the $|\cX| \times 1$ vector with all entries as 1.  
\begin{lemma}
    For a stationary Markov chain $X^n \sim \tMarkov(P,\bpi)$, the bias of Good-Turing estimator can be expressed as follows:
    \begin{align}
         &E[ \GT(X^n) - M_0(\bpi,X^n) ] = \sum_{x \in \cX} \frac{1}{n} \sum_{m=1}^n \bigg[ \bpi^{\sim x} (P^{\sim x} )^{(m-2)} (P_x - \pi_x P^{\sim x}) (P^{\sim x} )^{(n-m)}  \one   \bigg] \label{eq:Bias_P_no_x}\\
         & = \sum_{x \in \cX} \ \Bigg[ \sum_{i \in [K] : \lambda_{i,x} \neq 0} ( \lambda_{i,x} )^{n-1} \  (  u^{\sim x}_{i} \one )\ ( u^{\sim x}_{i} P_x v^{\sim x}_{i} - \pi_x  \lambda_{i,x}  ) \nonumber \\
        & \qquad \qquad \qquad  + \sum_{\substack{ i,j \in [K]: \  i \neq j, \\ \lambda_{i,x} \neq 0,  \lambda_{j,x} \neq 0 }}  \Big( \frac{1}{n} \ \sum_{m=1}^n \ ( \lambda_{i,x} )^{m-1}  \ ( \lambda_{j,x} )^{n-m} \Big) \ (  u^{\sim x}_{j} \one )\ ( u^{\sim x}_{i} P_x v^{\sim x}_{j}) \Bigg], \label{eq:Bias_lmda_x}
    \end{align}
where, for the second expression, we suppose that $P^{\sim x}=\sum_{i\in[K]}\lambda_{i,x}v^{\sim x}_{i} u^{\sim x}_{i}$ is diagonalizable with eigenvalues $\{ \lambda_{i,x} : i \in [K] \}$, corresponding left eigenvectors $\{u^{\sim x}_{i}\}$ and right eigenvectors $\{v^{\sim x}_{i}\}$ satisfying $u^{\sim x}_{i} \ v^{\sim x}_{j} = I(i=j)$ and $\bpi \ v^{\sim x}_{i} = 1$\footnote{Any eigenvector $v^{\sim x}_{i}$ of $P^{\sim x}$ can be made to satisfy  $\bpi \ v^{\sim x}_{i} = 1$ by scaling $v^{\sim x}_{i}$ with $(\bpi \ v^{\sim x}_{i})^{-1}$.}. 
\label{lem:bias_exp}
\end{lemma}
Since $P^{\sim x}$ is non-negative, we will suppose that $\lambda_{1,x}$ is its unique positive Perron eigenvalue \cite{pillai2005perron} i.e. $\lambda_{1,x} = \max_{i \in [K]} |\lambda_{i,x}|$. If these Perron eigenvalues $\lambda_{1,x}$ for all $x\in\cX$ satisfy a linear bound relationship with the stationary probability $\pi_x$ (along with some further technical conditions), the absolute bias can be shown to converge to zero. This is presented in the next theorem.



\begin{theorem}
Under the notation of Lemma \ref{lem:bias_exp}, the absolute bias of the Good-Turing estimator $\GT(X^n)$ converges to $0$ if the following conditions are satisfied:
    \begin{enumerate}
        \item For all $x\in\cX$ there exists $c_0\in(0,1)$ such that  $\lambda_{1,x} \leq 1-c_0 \pi_x$.
        \item For $i$ in $[K]$ such that $\lambda_{i,x} \neq 0$,
        $$|(u^{\sim x}_{i}\one)(u^{\sim x}_{i} P_x v^{\sim x}_{i} - \pi_x  \lambda_{i,x}  )| \leq q_x \pi_x +  q'_x (1-\lambda_{1,x}),$$  
        where $\sum_{x \in \cX} q_x$ and $\sum_{x \in \cX} q'_x$ are small constants.
        \item For $i,j$ in $[K]$ such that $i \neq j$ and both $\lambda_{i,x}, \lambda_{j,x}$ are non-zero, 
        $$ |\lambda_{i,x} - \lambda_{j,x}|^{-1} \  |(  u^{\sim x}_{j} \one ) \ (u^{\sim x}_{i} P_x v^{\sim x}_{j})| \leq q''_x,$$ 
        where $\sum_{x \in \cX} q''_x $ is a small constant.
        \item The number of non-zero eigenvalues of $P^{\sim x}$ is $O(n^{c})$ with $c < 1/4$.
    \end{enumerate}
    \label{thm:Bias_cnvg}
\end{theorem}
\begin{proof}
\begin{align*}
   & \big \vert E[ \GT(X^n) - M_0(\bpi,X^n) ] \big \vert \\ 
   & \ \overset{(a)}{\leq} \ \sum_{x \in \cX} \ \Bigg[ \sum_{i \in [K] : \lambda_{i,x} \neq 0} \big \vert( \lambda_{i,x} )^{n-1}\big \vert \  (q_x \pi_x +  q'_x (1-\lambda_{1,x}) ) \nonumber \\
        & \qquad \qquad \qquad  + \sum_{\substack{ i,j \in [K]: \  i \neq j, \\ \lambda_{i,x} \neq 0,  \lambda_{j,x} \neq 0 }}  \Big \vert \frac{1}{n} \ \sum_{m=1}^n \ ( \lambda_{i,x} )^{m-1}  \ ( \lambda_{j,x} )^{n-m} \Big \vert \
         \Big \vert (  u^{\sim x}_{j} \one ) \ ( u^{\sim x}_{i} P_x v^{\sim x}_{j}) \Big \vert \Bigg] \\ 
   & \ \overset{(b)}{\leq} \  \sum_{x \in \cX} \Bigg[ \sum_{i \in [K] : \lambda_{i,x} \neq 0} (  \lambda_{1,x} )^{n-1}  \ ( q_x \ \pi_x + q'_x (1-\lambda_{1,x}) ) \nonumber \\
    & \qquad \qquad \qquad + \sum_{\substack{ i,j \in [K]: \  i \neq j, \\ \lambda_{i,x} \neq 0,  \lambda_{j,x} \neq 0 }} \ \frac{1}{n} \ \Big \vert  (\lambda_{i,x} )^{n}  - (\lambda_{j,x} )^{n}  \Big \vert \ \Big \vert \lambda_{i,x} - \lambda_{j,x} \Big \vert^{-1} \  \Big \vert(  u^{\sim x}_{j} \one ) \ (u^{\sim x}_{i} P_x v^{\sim x}_{j}) \Big \vert  \Bigg] \\ 
    & \ \overset{(c)}{\leq}  \sum_{i \in [K] : \lambda_{i,x} \neq 0} \Bigg[ \sum_{x \in \cX} (q_x   \pi_x  (1-c_0 \pi_x)^{n-1} + q'_x  (1-\lambda_{1,x})  (  \lambda_{1,x} )^{n-1} ) \Bigg]  +   \sum_{\substack{ i,j \in [K]: \  i \neq j, \\ \lambda_{i,x} \neq 0,  \lambda_{j,x} \neq 0 }}  \Bigg[ \sum_{x \in \cX} \frac{2}{n} \ q''_x \Bigg] \\
     & \ \overset{(d)}{\leq} \  \sum_{i \in [K] : \lambda_{i,x} \neq 0} \  O(1/n) \ \Big[ \sum_{x \in \cX} (q_x + q'_x) \Big] \ + \ \sum_{\substack{ i,j \in [K]: \  i \neq j, \\ \lambda_{i,x} \neq 0,  \lambda_{j,x} \neq 0 }}  \  O(1/n) \ \Big[ \sum_{x \in \cX}   q''_x \Big] \\  
    & \ \overset{(e)}{\leq} \ O(1/n^{1-c}) \  \Big[ \sum_{x \in \cX} (q_x + q'_x) \Big] \ + \  O(1/n^{1-2c}) \ \Big[ \sum_{x \in \cX}   q''_x \Big],
\end{align*}
where we get $(a)$ by using the condition $(2)$ from the above theorem, $(b)$ by using $ \big \vert \lambda_{i,x} \big \vert  \leq \lambda_{1,x}$ and
$\sum_{m=1}^n \ ( \lambda_{i,x} )^{m-1}  \ ( \lambda_{j,x} )^{n-m} $ $ =   (\lambda_{i,x} - \lambda_{j,x} )^{-1} \ ((\lambda_{i,x} )^{n-1} - (\lambda_{j,x} )^{n-1})$, $(c)$ by using the conditions $(1)$ and $(3)$ from the above theorem, $(d)$ by using $t (1-mt)^{n-1} = O(1/n)$  for $m,t$ in $[0,1]$, and $(e)$ by using the condition $(4)$ from the above theorem. 
\end{proof}
In the sufficient conditions above, 
the most crucial one is the Perron eigenvalue of $P^{\sim x}$ i.e. $\lambda_{1,x}$ being bounded above by $1-c_0 \pi_x$. The other conditions involve lower eigenvalues and associated eigenvectors and could play a role in some cases. We illustrate such convergence using several synthetic and corpora-based t.p.ms in a later section.

 It is easy to retrieve the classical result that the bias of the GT estimator for missing mass in $n$ i.i.d. samples is $O(1/n)$ by checking that the conditions in Theorem \ref{thm:Bias_cnvg} are satisfied in the i.i.d. case with $P = \one \ \bpi$. Note that the i.i.d. t.p.m. has rank 1. As a natural next step, we consider t.p.ms with rank equal to 2. 

\subsection{Rank-2 Markov chains}
Consider a Markov chain with a rank-2 t.p.m $P$, which we will call, loosely, as a rank-2 Markov chain. Since $P$ has all entries in $[0,1]$ with each row adding to 1 and it has rank 2, the eigenvalues of $P$ will be 1, $\lambda_2$, 0, $\ldots$, 0, and $-1\le\lambda_2\le 1$ (by Perron-Frobenius theorem) \cite{pillai2005perron}. The value of $\lambda_2$ determines several important properties of the chain. If $\lambda_2=1$, the chain is reducible. If $\lambda_2=-1$, the chain is periodic with period 2. If $\lambda_2=0$, the chain is i.i.d.
For $-1<\lambda_2<1$, the chain is irreducible and aperiodic. We define the spectral gap of a t.p.m $P$ as
    \begin{equation} 
     \beta(P) \triangleq 1-\lambda_2\in[0,2].
     \label{eq:beta}
    \end{equation}
In this section, we let $\cP_{2,\beta}$ denote the family of rank-2 diagonalizable t.p.ms with spectral gap $\beta$.


\subsubsection{Bias of GT estimator}
The absolute bias of the GT estimator converges as $1/(n\beta^2)$ for rank-2 Markov chains with spectral gap $\beta$. This is shown in the next theorem.
\begin{theorem}\label{thm:Bias_r2}
        For $P\in \cP_{2,\beta}$ with stationary distribution $\pi$ and $\beta \geq \Big[ 30 \ \ln n / (n-3)  \Big]^{1/2}$,  there exists universal constants $c_1, c_2 > 0$, such that 
    \begin{align}
         \left \vert E[\GT(X^n) - M_0(\bpi,X^n)] \right \vert  \ & \leq \ \frac{c_1}{n \beta^2} +  \frac{c_2}{n \beta^3}  + \ O(1/n). \label{eq:R2_Bias}
    \end{align}
\end{theorem}
\begin{proof}
For a rank-2 diagonalizable t.p.m. $P$, $P^{\sim x}$ is diagonalizable for all $x \in \cX$ and has two non-zero eigenvalues $\lambda_{1,x}$ and $ \lambda_{2,x}$ with left eigenvectors $ u^{\sim x}_{1} $ and $  u^{\sim x}_{2}$, and right eigenvectors $ v^{\sim x}_{1}$ and $ v^{\sim x}_{2}$. Let $\Delta_x \triangleq \lambda_{1,x} - \lambda_{2,x}$. For $i,j$ in $\{1,2\}$, we can show that (proof given in supplementary material) 
\begin{enumerate}
\item  $\lambda_{i,x} \ \leq \ 1 - c_0 \pi_x$, where $c_0 = \beta/2$ for $\beta \in [0,1]$ and $c_0 = 1/2$ for $\beta \geq 1$, 
\item $ \Big \vert (  u^{\sim x}_{j} \one ) \ (u^{\sim x}_{i} P_x v^{\sim x}_{j}) \Big \vert \  \lesssim \ \pi_x \ \Delta_x^{-2}$, and
\item $ \Big \vert (  u^{\sim x}_{i} \one ) \ ( u^{\sim x}_{i} P_x v^{\sim x}_{i} - \pi_x  \lambda_{i,x}  ) \Big \vert \ \lesssim \  \Delta_x^{-2} \  ( t_x \ \pi_x + \pi_x (1-\lambda_{1,x}) ) $ with $\sum_{x \in \cX} t_x$ being a finite constant.
\end{enumerate}
In the above, $a \lesssim b$ denotes that $a \leq c b$ for some $c > 0$. We see that the above bounds have the same form as the conditions in Theorem \ref{thm:Bias_cnvg} with $q_x = t_x/\Delta_x^2$, $ q'_x = t_x/\Delta_x^2$ and $q''_x = \pi_x/\Delta_x^3$. Using these bounds in \eqref{eq:Bias_lmda_x}, we get  
$$\big \vert E[ \GT(X^n) - M_0(\bpi,X^n) ] \big \vert \ {\lesssim} \  \ (1/n) \ \sum_{x \in \cX} \Big[ \Delta_x^{-2} \ [ t_x   + \pi_x ] \  + 2 \  \Delta_x^{-3} \  \pi_x \Big].$$  
Using $\beta$ to bound $\Delta_x$ for $x$ with low stationary probabilities i.e. the letters that contribute significantly to the missing stationary mass (this is skipped), we get the claimed bound on the bias of GT.     
\end{proof}

\subsubsection{Minimax rate}
For rank-2 chains, a much stronger result than convergence of bias can be shown. We next characterise the minimax rate $R_n^*(\cP_{2,\beta})$ of the squared error risk of estimating $M_{0}(\bpi,X^n)$ of $X^n \sim \tMarkov(P,\bpi)$ for the class $\cP_{2,\beta}$.
\begin{theorem}
The minimax squared error risk $R^*_{n}(\cP_{2,\beta})$ is bounded as follows:
\begin{enumerate}
    \item For $\beta \geq  \left( (160/(n-5)) \ {\ln n} \right)^{1/3}$, 
    \begin{align}
        R^*_{n}(\cP_{2,\beta}) \ & \leq \ O(1/{n \beta^5})  \label{eq:stat_ub}
    \end{align}
    \item For $n$ sufficiently large, there is a constant $c$ such that
    \begin{equation}
        R^*_{n}(\cP_{\beta})  \ \geq \ R^*_{n}(\cP_{2,\beta}) \ \geq \ \frac{c}{n\beta}.
        \label{eq:stat_lb}
    \end{equation}
\end{enumerate}
\label{thm:minimax_stat}
\end{theorem}
We get the upper bound in \eqref{eq:stat_ub} by analyzing the worst case MSE (over $\cP_{2,\beta}$) of the GT estimator $\GT(X^n) = \phi_1(X^n)/n$  in estimating $M_0(\bpi,X^n)$ from $X^n \sim \tMarkov(P,\bpi)$ with $P \in \cP_{2,\beta}$. The simplifications in the upper bound are fairly involved and an outline of the proof is presented in Section \ref{sec:Thm1_Prf}. The complete proof is provided in the supplementary material.

To get the lower bound in \eqref{eq:stat_lb}, we modify the Le Cam two point method. The usual Le Cam method for lower bounds on minimax risk \cite{yu1997lecam} directly applies when the estimand does not depend on the samples, and is a parameter of the distribution alone. Using concentration properties, we extend the Le Cam two-point method to the case of estimating missing stationary mass $M_{0}(\bpi,X^n)$ that  clearly depends on the samples $X^n$. By constructing two Markov chains (with t.p.ms in $\cP_{2,\beta}$) close in distribution and separated in  $M_{0}(\bpi,X^n)$, we get the lower bound in \eqref{eq:stat_lb} on $R^*_{n}(\cP_{2,\beta})$. Since $\cP_{2,\beta}$  is contained in $\cP_{\beta}$, the class of all Markov chains with spectral gap $\beta$, the same lower bound extends to $R^*_{n}(\cP_{\beta})$ as well. Specific details are provided in the supplementary material.

Overall, we see that the minimax rate behaves as $1/n$ for rank-2 chains. This extends the previously known $1/n$ rate for i.i.d. samples. The behaviour with respect to $\beta$ differs in the upper and lower bounds ($1/(n\beta^5)$ vs $1/(n\beta)$), and this gap could be closed in future work.

\subsection{Synthetic and corpora-based illustrations} \label{sec:Simulations}
In this section, we present the results of our simulations studying the performance of the GT estimator in estimating the missing stationary mass of Markov sequences drawn using rank 2 t.p.ms, randomly generated t.p.ms, and empirical t.p.ms built over natural language text. 
\subsubsection{Rank-2 synthetic t.p.m.}
We consider a $K\times K$ rank-2 t.p.m. with spectral gap $\beta$ formed by 4 $K/2\times K/2$ blocks as follows:
$$\left[\begin{array}{cccc|cccc}
c_{K,\beta} &\cdots & c_{K,\beta} & c_{K,\beta}&\beta/2 & 0 &\cdots & 0\\
    \vdots  &\ddots & \vdots      & \vdots  &\vdots &\vdots&\ddots&\vdots\\
c_{K,\beta} &\cdots & c_{K,\beta} & c_{K,\beta}&\beta/2& 0&\cdots & 0\\[2pt] \hline
    0 &\cdots& 0 & \beta/2&c_{K,\beta}& c_{K,\beta} &\cdots & c_{K,\beta}\\
\vdots&\ddots&\vdots&\vdots&\vdots    & \vdots      &\ddots & \vdots     \\
    0 &\cdots& 0 & \beta/2&c_{K,\beta}& c_{K,\beta} &\cdots & c_{K,\beta}
\end{array}\right],$$
where $c_{K,\beta}=(1-\beta/2)(2/K)$. 
We generated stationary Markov sequences of lengths $n  = 30, 60, 120, 240, 500, 1000$ from the rank$-2$ t.p.m. specified above with $K=1.2n$ and with $\beta=1/n^{0.2}$, $\beta=1/n$ and computed the missing mass $M_0$ and the GT estimate over 5000 trials. The mean values with standard deviation bars are shown in Fig. \ref{fig:exp_value_sd}.
\begin{figure}[htb]
    \centering
    \input{expected_value_R2.tex}
    \caption{Missing mass and GT estimates for rank-2 chain and i.i.d. sequence.}    
    \label{fig:exp_value_sd}
\end{figure}
For comparison, similar results are shown for i.i.d. sequences from the same stationary distribution. As predicted by the theoretical results, the GT estimate converges in the rank-2 case for $\beta=1/n^{0.2}$, while it is away by a constant value for $\beta=1/n$. In the i.i.d. case, there is convergence in all cases. We see that the variances in the rank-2 case are noticeably higher when compared to the i.i.d. case.  



A plot of MSE of GT versus $n$ is shown in Fig. \ref{fig:R2_MSE} for a larger range of values for $n$ for the rank-2 chain considered above.
\begin{figure}[htb]
    \centering
    \definecolor{mycolor1}{rgb}{1.00000,0.00000,1.00000}%
\begin{tikzpicture}

\begin{axis}[%
width=2.0in,
height=1.25in,
at={(0.75in,0.516in)},
scale only axis,
xmode=log,
xmin=100,
xmax=10000,
xtick={100,400,1600,6400},
xticklabels={{100},{400},{1600},{6400}},
xminorticks=true,
xlabel={Sample size $n$},
ymode=log,
ymin=1e-05,
ymax=1,
yminorticks=true,
ylabel style={font=\color{white!15!black}},
ylabel={MSE},
axis background/.style={fill=white},
xmajorgrids,
xminorgrids,
ymajorgrids,
yminorgrids
]
\addplot [color=black, mark=asterisk, mark options={solid, black}, forget plot]
  table[row sep=crcr]{%
100	0.00557907415653476\\
200	0.00278405679802805\\
400	0.00140876596181349\\
800	0.000697719457991264\\
1600	0.000354384366989862\\
3200	0.000176515654600605\\
6400	9.07211183944709e-05\\
};
\addplot [color=blue, dashed, mark=x, mark options={solid, blue}, forget plot]
  table[row sep=crcr]{%
100	0.0151159561848958\\
200	0.00844807424983113\\
400	0.00470614974365243\\
800	0.00247854709243393\\
1600	0.00126530381742282\\
3200	0.000641839464105597\\
6400	0.000326538016922323\\
};
\addplot [color=mycolor1, mark=diamond, mark options={solid, mycolor1}, forget plot]
  table[row sep=crcr]{%
100	0.0694445422576367\\
200	0.0595946899428748\\
400	0.0525104429522162\\
800	0.044130214723101\\
1600	0.039897850503719\\
3200	0.0314753053821247\\
6400	0.026859008676111\\
};
\addplot [color=red, dashdotted, mark=star, mark options={solid, red}, forget plot]
  table[row sep=crcr]{%
100	0.113836024607314\\
200	0.112554098978516\\
400	0.111352333513318\\
800	0.112068802439604\\
1600	0.111644458896155\\
3200	0.110819579787333\\
6400	0.1114534132392\\
};
\end{axis}

\begin{axis}[%
width=2.0in,
height=1.25in,
at={(0.75in,0.516in)},
scale only axis,
xmin=0,
xmax=1,
ymin=0,
ymax=1,
axis line style={draw=none},
ticks=none,
axis x line*=bottom,
axis y line*=left
]
\node[below right, align=left, draw=white]
at (rel axis cs:0.3,0.29) {$\beta = 1/n^{0.2}$};
\node[below right, align=left, draw=white]
at (rel axis cs:0.65,0.52) {$\beta= 1/n^{0.5}$};
\node[below right, align=left, draw=white]
at (rel axis cs:0.25,0.95) {$\beta=1/n$};
\node[below right, align=left, draw=white]
at (rel axis cs:0.65,0.7) {$\beta = 1/n^{0.8}$};
\end{axis}


\end{tikzpicture}%
    \caption{MSE of GT for a rank-2 chain with spectral gap $\beta$.}
    \label{fig:R2_MSE}
\end{figure}
The parameters were chosen as $K = 1.2n, \beta = 1/n^{0.2}, 1/n^{0.5}, 1/n^{0.8},1/n$, and the MSE was averaged over 16000 trials for each $n$. We observe that the MSE falls with $n$ for $\beta$ higher than $1/n$, while it stays flat for $\beta=1/n$.  
 


\subsubsection{T.p.ms generated at random and from corpora} 
Analytical characterisation of t.p.ms that satisfy the conditions in Theorem \ref{thm:Bias_cnvg} beyond the cases of rank-1 and rank-2 is a challenging problem. One interesting example is the sticky chain with a full-rank t.p.m. considered in \cite{Chandra22}, where Condition 2 of Theorem \ref{thm:Bias_cnvg} is not satisfied. The authors of \cite{Chandra22} suggest a scaling of the GT estimator, which converges to missing mass in that case. While there are such examples, the GT estimator appears to work in practice for several text-based corpora, which behave like Markov chains. To study this phenomenon under the setting of Theorem \ref{thm:Bias_cnvg}, we construct some random t.p.ms and some t.p.ms from text corpora and attempt to verify the main condition $\lambda_{1,x} \leq 1-c_0 \pi_x$ on the Perron eigenvalue and stationary probability numerically. 
 
We consider two classes of randomly generated t.p.ms, and two classes of t.p.ms from text corpora.
\begin{itemize}
    \item $P_{\text{unif-gen}}$: Each entry is first drawn i.i.d. from the uniform distribution over $[0,1]$ and each row is then scaled to make the row sum equal to 1.
    \item $P_{\text{pwr-gen}}$: Each row is a power law distribution with the $(i,j)$-th entry being proportional to $1/j^{a_i}$, where $a_i$ are drawn i.i.d. from the uniform distribution over $[0,1]$.
    \item NYT: The New York Times (NYT) corpus \footnote{available under CC0:Pubilc domain license at https://www.kaggle.com/datasets/mathurinache/10700-articles-from-new-york-times} consists of randomly collected articles from the front pages of New York Times from the years 2017 and 2018. To build an empirical t.p.m, we consider an article in the NYT corpus and set the empirical transition probability (from word $w_1$ to $w_2$) $P_{w_1,w_2}$ as $N_{w_1, w_2}/N_{w_1}$,  where $N_{w_1, w_2}$ is the number of times the word $w_2$ follows $w_1$ in the article and $N_{w}$ is the number of occurences of the word ${w}$ in the article \footnote{We change all the words in the article to lower case, lemmatize the words and ignore punctuations.}. The empirical probability $\pi_{w}$ is set to $N_w/$the total wordcount of the article. 
    \item GE: The novel \textit{Great Expectations} (GE) by Charles Dickens is available under the project Gutenberg (https://www.gutenberg.org/). To construct a t.p.m, we consider a chapter from the novel and repeat the same process as NYT above.
\end{itemize}

 Figure \ref{fig:scatter} shows scatter plots of $1-\lambda_{1,x}$ against the stationary probability $\pi_x$ for the randomly generated t.p.ms $P_{\text{unif-gen}}$ and $P_{\text{pwr-gen}}$ with $K=1250$, and against the empirical probability $\pi_x$ for three empirical t.p.ms, each built using a chapter of the novel \textit{Great Expectations} and five empirical t.p.ms, each built using an article with more than 1600 words from the NYT corpus. 
 \begin{figure}[htb]
    \begin{center}
        \input{scatter}
   \end{center}
\caption{$1-\lambda_{1,x}$ vs $\pi_x$ for randomly generated t.p.ms and t.p.ms from corpora.}
\label{fig:scatter} 
\end{figure}
All the plots show a linear relation between $1-\lambda_{1,x}$ and $\pi_x$.

Figure \ref{fig:rand_corp_MSE} plots the MSE of the GT estimator for $M_0(\bpi,X^n)$ of a stationary Markov chain $X^n$ generated using the t.p.ms $P_{\text{unif-gen}}$, $P_{\text{pwr-gen}}$ with $K=1250$, the three empirical t.p.ms from GE and the five empirical t.p.ms from NYT corpora, for $n = 100,200,400,800,1600,3200,6400$ and averaged over 16000 trials.    
\begin{figure}[htb]
\begin{center}
        \definecolor{mycolor1}{rgb}{0.64000,0.24700,0.74100}%
\definecolor{mycolor2}{rgb}{0.85000,0.32500,0.09800}%
\definecolor{mycolor3}{rgb}{0.92900,0.69400,0.12500}%
\begin{tikzpicture}

\begin{axis}[%
width= 2.25in,
height=1.5in,
at={(0.751in,0.481in)},
scale only axis,
xmode=log,
xmin=100,
xmax=7500,
xtick={100,200,400,800,1600,6400},
xticklabels={{100},{200},{400},{800},{1600},{6400}},
xminorticks=true,
xlabel style={font=\color{white!15!black}},
xlabel={Sample size $n$},
ymode=log,
ymin=2e-06,
ymax=5e-03,
yminorticks=true,
ylabel style={font=\color{white!15!black}},
ylabel={MSE},
axis background/.style={fill=white},
xmajorgrids,
xminorgrids,
ymajorgrids,
yminorgrids,
legend style={at={(0.45,0.5)},anchor=north east,draw=black,fill=white,legend cell align=left,font=\footnotesize}
]

\addplot [color = blue, dashed, mark=asterisk, mark options={solid, blue}]
  table[row sep=crcr]{%
100	0.00145293018197539\\
200	0.0013039899144563\\
400	0.00107139137106069\\
800	0.000730815392414544\\
1600	0.000359214174875308\\
3200	8.38710771752689e-05\\
6400	5.61958420184922e-06\\
};
\addlegendentry{$P_{\text{unif-gen}}$}

\addplot [color = black, dashed, mark=x, mark options={solid, black}]
  table[row sep=crcr]{%
100	0.00269533119267298\\
200	0.00183601832484105\\
400	0.00115828183465621\\
800	0.000640485671517793\\
1600	0.000288008192582664\\
3200	8.64505481837129e-05\\
6400	1.29672961764692e-05\\
};
\addlegendentry{$P_{\text{pwr-gen}}$}

\addplot [color = mycolor1, dashed, mark=asterisk, mark options={solid, mycolor1}]
  table[row sep=crcr]{%
100	0.0047\\
200	0.0023\\
400	0.0011\\
800	0.0005\\
1600	0.0002\\
3200	0.0001\\
6400	5e-05\\
};
\addlegendentry{GE}

\addplot [color = red, dashed, mark=x, mark options={solid, red}]
  table[row sep=crcr]{%
100	0.0050\\
200	0.0029\\
400	0.0016\\
800	0.0008\\
1600	0.0003\\
3200	0.0001625\\
6400	8.83e-05\\
};
\addlegendentry{NYT}

\end{axis}


\end{tikzpicture}%
    \end{center}
    \captionof{figure}{MSE of GT for randomly generated t.p.ms and t.p.ms from corpora.}
    \label{fig:rand_corp_MSE}     
\end{figure}
The curves for GE and NYT correspond to MSE averaged over the three t.p.ms from \textit{Great Expectations} and the five t.p.ms from the NYT corpus. We observe that MSE falls for all of these t.p.ms.


\section{Conclusion and Future Directions} \label{sec:conclusion}
In conclusion, our study of the Good-Turing (GT) estimator for missing stationary mass of a Markov chain with t.p.m $P$ and stationary probability $\bpi$ (with support size assumed to be unknown) indicates that convergence of GT depends on relationships between the spectrum of $P^{\sim x}$ ($P$ with $x$-th column zeroed out) and $\pi_x$ to be satisfied uniformly for all $x$. We derive specific sufficient conditions for convergence of absolute bias in the case where $P^{\sim x}$ is diagonalizable. These conditions are verified numerically for t.p.ms derived from text corpora, which supports the success of GT in practice. Analytical understanding of relationships between the spectrum of $P^{\sim x}$ and $\pi_x$ for arbitrary t.p.ms is a topic for future study. In the case when $P$ has rank 2 with a spectral gap of $\beta$, we derive minimax squared-error risk lower ($c/(n\beta)$) and upper ($c'/(n\beta^5)$) bounds. The bounds extend the $1/n$ minimax rate result from the i.i.d. case to rank-2 Markov. Characterising the exact dependence on $\beta$ is a topic for future work.

\section{Proof Outline of Theorem \ref{thm:minimax_stat}, Upper bound}
\label{sec:Thm1_Prf} 
$R^{*}_{n}(P_{2,\beta})$ is upper bounded by the worst case risk (over $\cP_{2,\beta}$) of the Good-Turing estimator. 
Our next lemma gives a closed form expression for the MSE of the GT estimator.
\begin{lemma}
Consider a stationary Markov chain $X^n$ with state distribution $\bpi.$ Let $Q_x^n(a) \triangleq \pr(N_x(X^n) = a),$ and $ Q_{x,y}^n(a,b) \triangleq \pr(N_x(X^n) = a, N_y(X^n) = b) $ for $x, y \in \cX$ and $a,b \in \{ 0 ,1 , \ldots \}.$
\begin{align}
 & E[ ( M_{0}(\bpi,X^n) - \GT(X^n)  )^2 ] \nonumber \\ 
 & \quad  = \ \sum_{x \in \cX} \bigg( \pi_x^2 \ Q_x^n(0) + (1/n)^2 \  Q_x^n(1) \bigg) + \sum_{x \in \cX} \sum_{y \in \cX, y \neq x} T_{xy}^n - \frac{1}{n^2 (n-1)} \ Q_{x,y}^n(1,1)   \label{eq:MSE}
\end{align}
where 
$ T_{xy}^n \triangleq  \pi_x \pi_y \ Q_{x,y}^n(0,0) - \frac{1}{n} ( \pi_y \ Q_{x,y}^n(1,0) + \pi_x \ Q_{x,y}^n(0,1)  )  + \frac{1}{n (n-1) } \ Q_{x,y}^n(1,1). $ \\
\label{lem:MSE}
\end{lemma}
\begin{proof}
Substituting $M_{0}(X^n, \bpi) =  \sum_{x \in \cX} \pi_x I(N_x = 0)$ and $\GT(X^n) = (1/n)  \sum_{x \in \cX} I(N_x = 1)$ into $E[(M_0(X^n,\bpi) - \GT(X^n))^2] $ and taking expectation of each term in the square of the summation, 
\begin{align}
    E[(M_0 - \GT)^2] \ & = \ E\left[ \left( \sum_{x \in \cX} \pi_x I(N_x = 0) - (1/n) I(N_x = 1)  \right) ^2 \right] \nonumber \\
     & = \ \sum_{x \in \cX} \bigg( \pi_x^2 \ \pr(N_x = 0) + (1/n)^2 \ \pr(N_x = 1) \bigg) \nonumber \\ 
     & \qquad + \sum_{x \in \cX} \sum_{y \in \cX, y \neq x} \pi_x \pi_y \pr(N_x = N_y = 0) - \frac{1}{n} \pi_y \pr(N_x = 1, N_y = 0) \nonumber \\ 
     & \qquad \qquad \qquad \qquad \qquad - \frac{1}{n} \pi_x \pr(N_x = 0, N_y = 1) + \frac{1}{n^2}  \pr(N_x = N_y = 1) \nonumber \\
     & \overset{(a)}{=} \  \sum_{x \in \cX} \bigg( \pi_x^2 \ Q_x^n(0)  + (1/n)^2 \  Q_x^n(1)  \bigg) - \sum_{x \in \cX} \sum_{y \in \cX, y \neq x} \frac{1}{n^2 (n-1)} \ Q_{x,y}^n(1,1) \nonumber \\ 
     & \qquad + \sum_{x \in \cX} \sum_{y \in \cX, y \neq x} \pi_x \pi_y \ Q_{x,y}^n(0,0) - \frac{1}{n} \left( \pi_y \ Q_{x,y}^n(1,0) + \pi_x \ Q_{x,y}^n(0,1)   \right) \nonumber \\ 
     & \qquad \qquad \qquad \qquad \qquad  + \frac{1}{n (n-1) } \  Q_{x,y}^n(1,1), \label{eq:MSE_prf}
\end{align}
where we get $(a)$ by expanding $1/n^2$ as $1/n^2 = \frac{1}{n (n-1)} - \frac{1}{n^2 (n-1)}$ and using the definitions of $Q^n_{x}(a)$, $Q^n_{x,y}(a,b).$  Using the definition of $T^n_{x,y}$ in \eqref{eq:MSE_prf} completes the proof of \eqref{eq:MSE}.    
\end{proof}
In the next lemma, we bound each of the terms in \eqref{eq:MSE}. 
\begin{lemma}
    For a stationary Markov chain $X^n$ with state distribution $\bpi$ and a rank-2 diagonalizable t.p.m $P$ with spectral gap $\beta \geq  \left( (160/(n-5)) \ {\ln n} \right)^{1/3}$,
    \begin{align}
    \frac{1}{n^2} \sum_{x \in \cX} Q_x^n(1) & \leq \frac{1}{n}, \label{eq:MSE_t2}\\
    \sum_{x \in \cX} \sum_{y \in \cX, y \neq x} \frac{1}{n^2 (n-1)} Q_{x,y}^n(1,1) & \leq 1/n, \label{eq:MSE_t4}\\
    \sum_{x \in\cX} \pi_x^2 \ Q_x^n(0) & \leq 2 \frac{(\beta+2)}{ \beta^2 n} \left( 1 + \frac{9}{\beta}  \right) + \ O(1/n),   \label{eq:MSE_t1}\\
    \sum_{x \in \cX} \sum_{y \in \cX, y \neq x} T^n_{xy} & \leq O(1/{n \beta^5}).   \label{eq:MSE_t3}
\end{align}
\label{lem:MSE_term_bd}
\end{lemma}
\begin{proof}
Section \ref{subsec:MSE_term_bd_Prf} in the supplementary material.
\end{proof}
Combining  \eqref{eq:MSE_t2}, \eqref{eq:MSE_t4},  \eqref{eq:MSE_t1},  and  \eqref{eq:MSE_t3}  to upper bound \eqref{eq:MSE} completes the proof of the upper bound on $R^*_{n}(P_{2,\beta})$ in Theorem \ref{thm:minimax_stat}.

\bibliographystyle{ieeetr}
\bibliography{main}

\newpage 

\section{Proof of Theorem \ref{thm:Bias_r2}}
\label{sec:Bias_r2_Prf}
To prove \eqref{eq:R2_Bias}, we divide the alphabet $\cX$ into three sets, 
\begin{align}
    A_0 \ & \triangleq \ \{x \in \cX : \pi_x = P_{xx}  \}, \nonumber \\
    A(\delta) \ & \triangleq \ \{ x \in \cX \setminus A_0 : \pi_x < \delta \}, \ \text{ and } \nonumber \\ 
    \overline{A}(\delta) \ & \triangleq \ \{ x \in \cX \setminus A_0 : \pi_x \geq \delta \}, \nonumber 
\end{align}
with $0 \leq \delta < \beta/5$.

Let $\Gamma_x \triangleq  \frac{1}{n} \  \pr(N_x(X^n) =1) - \pi_x \ \pr(N_x(X^n) = 0)$ be the letter-wise bias of the GT estimator i.e. $E[\GT(X^n) - M_0(\bpi,X^n)] = \sum_{x \in \cX} \Gamma_x $.
The following lemma bounds the sum of $| \Gamma_x |$ over each of the three sets defined above. 

\begin{lemma} 
For a rank-2 diagonalizable t.p.m $P$ with spectral gap $\beta,$
    \begin{enumerate}
        \item For $x \in A(\delta)$,
        \begin{align}
        \sum_{x \in A(\delta)} \  \vert \Gamma_x  \vert \ {\leq} \  36 \ (1/n\beta^2) \ ( 1 + 16/c_{\beta} +  34/\beta),  \label{eq:x_Bias_A_del_fin} 
        \end{align}
        \label{lem_part:x_Bias_A_del_fin}

        \item For $x \in \overline{A}(\delta)$,

        \begin{align}
             \sum_{x \in \overline{A}(\delta)} \  \vert \Gamma_x  \vert \ {\leq} \ \Big[ g(n) \ \exp{ - (n-3) \ c_{\beta} \ \delta/2 } + (1 + (3/2)  \  n ) \ \exp{ - (n-1) \ c_{\beta} \  \delta/2 } \Big],  \label{eq:x_Bias_A_til_fin}
        \end{align}
         where $c_{\beta} =
             \beta$ for $ \beta \in (0,1]$, $
             c_{\beta} = 1$ for $ \beta \in [1,2]$,  and $g(n) = [1 + (5/2) \  (n-1) \ (1  + (5/12)  \ ( n-2)) ]$.
             \label{lem_part:x_Bias_A_til_fin}
             
        \item For $x \in A_0$,
           
        \begin{equation}
              \sum_{x \in A_0} \ \vert \Gamma_x  \vert \ \leq \  O(1/n) \label{eq:x_Bias_A0_fin}
        \end{equation}

        \label{lem_part:x_Bias_A0_fin}
    \end{enumerate}
    \label{lem:x_Bias_fin}
\end{lemma}
\begin{proof}
    Section \ref{sec:x_Bias_fin_Prf}.
\end{proof}

Taking the sum of \eqref{eq:x_Bias_A_del_fin}, \eqref{eq:x_Bias_A_til_fin} and \eqref{eq:x_Bias_A0_fin}, and choosing $\delta = (6/c_{\beta}) \ (\ln n)/(n-3)$, we get 

\begin{equation*}
 \Big \vert E[\GT(X^n) - M_0(\bpi,X^n)] \Big \vert \leq   36 \ (1/n\beta^2) \ (1 + 16/c_{\beta} +  18/\beta) + O(1/n), 
\end{equation*}
for $\beta \geq \Big[ 30 \ln n/(n-3) \Big]^{1/2}$ (since $\delta \leq \beta/5$). 

This completes the proof of \eqref{eq:R2_Bias}. 

\section{Proof of Theorem \ref{thm:minimax_stat}, Lower bound }\label{subsec:stat_lb_Prf}

To prove the lower bound on $R^*_{n}(\cP_{2,\beta})$ in \eqref{eq:stat_lb}, we use the Le Cam method. 
The standard Le Cam method \cite{yu1997lecam} is for estimating constant parameters of a distribution whereas $M_0(\bpi,X^n)$ is a function of both the distribution and the samples. To get \eqref{eq:stat_lb}, we use the following extension of Le Cam method for estimands that depend on both the distribution and its samples.

\textit{Le Cam lower bound for estimating random variables}:
Let $ \cQ $ be a family of distributions over an alphabet $\cY$ and $Y$ be a random variable distributed according to $Q \in \cQ$. Let $\theta(Y,Q)$, taking values in a pseudometric space $\cD$ with a pseudometric $d$, be a function of both $Y$ and the distribution $Q$. We assume that the set $\cD$ is bounded i.e. the distance $d(u,v)$ between any two points $u , v \in \cD$ is at most $\Delta$.  
Let $d(\cD_1, \cD_2) \triangleq \min_{u \in \cD_1, v \in \cD_2} d(u,v)$ be the distance between the subsets $\cD_1, \cD_2$ of $\cD$. Let $\widehat{\theta}(Y)$ be an estimator for $\theta(Y,Q)$ and $co(\cQ)$ denote the convex hull of $\cQ$. 

The following lemma provides a lower bound on the worst case risk (over $\cQ$) of any estimator $\widehat{\theta}(Y)$ for $\theta(Y,Q)$. 
\begin{lemma}\cite[Lemma 5]{Chandra22}
Let $ \cD_1, \cD_2$ be two subsets of $\cD$, and  $\cQ_1, \cQ_2$ be two subsets of $\cQ$ such that for any $Q_i \in \cQ_i,$ $\theta(Y,Q_i) \in \cD_i$ with probability at least $1 - \epsilon_i$, $i = 1,2$. Let  $\delta \triangleq d(\cD_1, \cD_2)/2,$ and $||Q_1 \wedge Q_2 || \triangleq 1 - ||Q_1 - Q_2||_{TV}$ denote the affinity of the two distributions $Q_1$ and $Q_2$. Then 
\begin{align}
    \sup_{Q \in \cQ} E [ d( \widehat{\theta}(Y) , \theta(Y,Q) ) ] \ 
   &  \geq \ \delta \ \left( \sup_{Q_i \in co(\cQ_i)} ||Q_1 \wedge Q_2 || \right) - (\epsilon_1 + \epsilon_2)\Delta.  \label{eq:Lecam_lb} 
\end{align}
\label{lem:Lecam_lb}
\end{lemma}

To prove the lower bound \eqref{eq:stat_lb} in theorem \ref{thm:minimax_stat}, we apply Lemma \ref{lem:Lecam_lb} on $\cP_{2,\beta},$ the family of rank-2 diagonalizable t.p.ms with spectral gap $\beta,$ using two specifically constructed t.p.ms $\in \cP_{2,\beta}$ that are hard to distinguish. 

We first prove \eqref{eq:stat_lb} for $\beta \in (0,1.6)$. \\ Let $P[\delta_1, \delta_2] \triangleq \begin{bmatrix}
1-\delta_1 & (\delta_1/R) \ \one_{1 \times K-1} \\ \delta_2 \  \one_{K-1 \times 1} & (1-\delta_2)/R \ \one_{K-1 \times K-1} \end{bmatrix}$ be a $K \times K$ t.p.m parameterized by $\delta_1, \delta_2,$  with $ \one_{1 \times K-1}$ and $\one_{K-1 \times 1}$ denoting the row and column vectors with all entries as 1, and $ \one_{K-1 \times K-1} $ denoting the $K-1 \times K-1$ matrix with all entries as 1. 

Let $\cP_1 = \{ P_1 \}$ and $\cP_2 = \{ P_2 \}$ be two subsets of $\cP_{2,\beta},$ where $P_1 = P[0.5\beta, 0.5\beta],$ $P_2 = P[0.5\beta -\alpha, 0.5\beta + \alpha]$ are two t.p.ms on the alphabet $\{ 1,2,\ldots, L+1 \}$ with  $ \bpi_1 = 0.5\begin{bmatrix} 1 & (1/L)  \one_{1 \times L} \end{bmatrix}$ and $ \bpi_2 = \begin{bmatrix}
0.5 + (\alpha/\beta) & ( 0.5 - (\alpha/\beta))/L \ \one_{1 \times L}
\end{bmatrix} $ as their respective stationary distributions. For each element of $P_2$ to lie in $[0,1]$ and each row of $P_2$ to sum up to 1, we require that $\alpha \leq \min \{0.5\beta, 1-0.5\beta \}.$
Both $P_1$ and $P_2$ have the same spectral gap $\beta$ (and hence $\in \cP_{2,\beta}$) and $\alpha, L$ are choosen appropriately to get the best lower bound on $R^*_{n}(\cP_{2,\beta})$.

For $i = 1,2,$ if $X^n$ is a stationary Markov chain with t.p.m $P_i,$ then $M_0(X^n, \bpi_i)$ satisfies  
\begin{align}
    1 - M_0(X^n, \bpi_i) \ & \overset{(a)}{\geq} \ \left( 0.5 + (i-1)\frac{\alpha}{\beta} \right) \ I(N_1 \neq 0) \label{eq:seen_mass_lb} \\
    1 - M_0(X^n, \bpi_i) \ & \overset{(b)}{\leq} \ \left(0.5 + (i-1)\frac{\alpha}{\beta}\right) \ I(N_1 \neq 0) \ + \ \frac{n}{L} \ \left( 0.5 - (i-1)\frac{\alpha}{\beta} \right) \label{eq:seen_mass_ub} \\ 
    \text{ i.e. } M_0(X^n, \bpi_i) \ & \in \ \left[ \left( 0.5 - (i-1)\frac{\alpha}{\beta} \right) \ \left( 1 - \frac{n}{L} \right), \left( 0.5 - (i-1)\frac{\alpha}{\beta} \right) \right], \nonumber \\ 
      & \ \text{ with probability }  1 -  \left( 0.5 - (i-1)\frac{\alpha}{\beta} \right)\left(1 - \frac{\beta}{2} - (i-1)\alpha \right)^{n-1}, \label{eq:M0_conf_int} 
\end{align}
where we get the bounds in $(a)$ and $(b)$ by considering the cases of : (i) the letter $1$ occuring in all the samples $X^n$ and (ii) all the $n$ samples, $X^n,$ being distinct with the letter $1$ occurring only once, respectively. We get the confidence interval in \eqref{eq:M0_conf_int} by using the bounds in \eqref{eq:seen_mass_lb} and \eqref{eq:seen_mass_ub} in the event of the letter $1$ occurring atleast once. 

Let ${\alpha} > 0.5\beta n/L,$ so that the above confidence intervals in \eqref{eq:M0_conf_int} for $M_0(X^n,\bpi_i), i =1,2,$ are non-overlapping. 
Using lemma \ref{lem:Lecam_lb} with $\cQ = \cP_{2,\beta}, Y = X^n, \theta(Y,Q) = M_{0}(\bpi,X^n), \cD = [0,1], d(u,v) = (u-v)^2, \Delta = 1, \cQ_i = \cP_i, $  and $D_i$ as the confidence interval in \eqref{eq:M0_conf_int} for $i = 1,2,$ we get
\begin{equation}
     \sup_{P \in \cP} E [ ( \widehat{M_0}(X^n) - M_{0}(\bpi,X^n) )^2 ] \  \geq \ 0.5 \ \left( \frac{\alpha}{\beta} - \frac{n}{2L} \right)^2 \ ||P_1(X^n) \wedge P_2(X^n) || - \left( 1 - \frac{\beta}{2} \right)^{n-1} \label{eq:M0_Lecam_int}
\end{equation}

Our next lemma gives an upper bound on the total variation distance between $P_1(X^n)$ and $P_2(X^n).$

\begin{lemma}
For the t.p.ms $P_1, P_2 \in \cP_{2,\beta},$ constructed above and $X^n$ being a stationary Markov chain, 
\begin{equation}
    ||P_1(X^n) - P_2(X^n) ||_{TV} \ \leq \ (\sqrt{2} {\alpha}/{\beta}) \ \left( \frac{ 1 + 0.5  (n-2)  \beta  }{ 1 - 0.5\beta } \right)^{0.5} \label{eq:TV_bd} 
\end{equation}
\label{lem:TV_bd}
\end{lemma}
\begin{proof}
Section \ref{subsec:TV_bd_Prf}.
\end{proof}
Since $ ||P_1(X^n) \wedge P_2(X^n) || = 1 - ||P_1(X^n) - P_2(X^n) ||_{TV},$ 
using the above Lemma to lower bound the R.H.S in \eqref{eq:M0_Lecam_int}, we get 
\begin{align}
    \sup_{P \in \cP} E [ ( \widehat{M_0}(X^n) - M_{0}(\bpi,X^n) )^2 ] \ & \geq \ 0.5 \ \left( \frac{\alpha}{\beta} - \frac{n}{2L} \right)^2 \ \left( 1 - (\sqrt{2} {\alpha}/{\beta}) \ \left( \frac{ 1 + 0.5  (n-2)  \beta  }{ 1 - 0.5\beta } \right)^{0.5}  \right) \nonumber \\ 
    & \qquad - \left( 1 - \frac{\beta}{2} \right)^{n-1} \label{eq:M0_Lecam_int_aperiodic}
\end{align}
Substituting $\alpha = \frac{\beta}{2 \sqrt{2}} \ \left( \frac{ 1 - 0.5\beta }{  1 + 0.5  (n-2)  \beta   } \right)^{0.5},$ and  $L =e^n$ in \eqref{eq:M0_Lecam_int_aperiodic} completes the proof of \eqref{eq:stat_lb} for $\beta \in (0,1.6)$.

We now prove \eqref{eq:stat_lb} for $\beta \in [1.6,2].$
Let $\cP'_1 = \{ P'_1 \}$ and $\cP'_2 = \{ P'_2 \}$ be two subsets of $\cP_{2,\beta},$  with $\beta \in [1.6,2],$ where \\  $P'_1 = \begin{bmatrix}
2-\beta & 0.5(\beta-1) & (0.5(\beta-1)/L) \ \one_{1 \times L} \\ \one_{1 \times L+1} &  \qquad \quad \zero_{L+1 \times L+1 }
\end{bmatrix}, \\  P'_2 = \begin{bmatrix}
2-\beta & 0.5(\beta-1)+\alpha' & ((0.5(\beta-1) - \alpha')/L) \ \one_{1 \times L} \\ \one_{1 \times L+1} & \qquad \quad \zero_{L+1 \times L+1 }
\end{bmatrix}$ are two t.p.ms on the alphabet \\ $\{ 1,2,\ldots, L+2 \}$ with $ \bpi'_1 = \begin{bmatrix}
 1/\beta & 0.5(\beta-1)/\beta & 0.5(\beta-1)/\beta L \ \one_{1 \times L}
\end{bmatrix} $ and \\ $ \bpi'_2 = \begin{bmatrix}
 1/\beta & (0.5(\beta-1)+\alpha')/\beta & (0.5(\beta-1)-\alpha')/\beta L \ \one_{1 \times L}
\end{bmatrix} $ as their respective stationary distributions and $\zero_{L+1 \times L+1 }$ being the $ L+1 \times L+1$ all zero matrix.  For each element of $P'_2$ to lie in $[0,1]$ and each row of $P'_2$ to sum up to 1, we require that $\alpha' \leq \min \{0.5(\beta-1), 1-0.5(\beta-1) \}.$ Both $P'_1$ and $P'_2$ have the same spectral gap $\beta$ (and hence $\in \cP$) and $\alpha', L$ are choosen appropriately to get the best lower bound on $R^*_{n}(\cP_{2,\beta})$.  

Following a method similar to the proof of \eqref{eq:stat_lb} for $\beta \in (0,1.6)$, we get 
\begin{equation}
     \sup_{P' \in \cP'} E [ ( \widehat{M_0}(X^n) - M_{0}(\bpi,X^n) )^2 ] \  \geq \ \frac{1}{2} \ \left( \frac{\alpha'}{\beta} - \frac{n}{2L\beta}(\beta-1) \right)^2 \ ||P'_1(X^n) \wedge P'_2(X^n) || - 9\left( 0.5 \right)^{\floor{\frac{n-3}{2}}} \label{eq:M0_Lecam_int_periodic}
\end{equation}

Our next lemma gives an upper bound on the total variation distance between $P'_1(X^n)$ and $P'_2(X^n).$

\begin{lemma}
For the t.p.ms $P'_1, P'_2 \in \cP_{2,\beta},$ constructed above and $X^n$ being a stationary Markov chain, 
\begin{equation}
    ||P'_1(X^n) - P'_2(X^n) ||_{TV} \ \leq \ \sqrt{2n}\alpha'/\sqrt{\beta(\beta-1)} \label{eq:TV_bd_periodic} 
\end{equation}
\label{lem:TV_bd_periodic}
\end{lemma}
\begin{proof}
Section \ref{subsec:TV_bd_Prf}.
\end{proof}

Using the above lemma to lower bound the R.H.S in \eqref{eq:M0_Lecam_int_periodic} and choosing $\alpha' = 0.5 \sqrt{\beta(\beta-1)}/\sqrt{2n},$ $L =e^n$ completes the proof of  \eqref{eq:stat_lb} for $\beta \in [1.6,2]$. \newline
This completes the proof of the lower bound on $R^*_{n}(\cP_{2,\beta})$.

\subsection{Proof of Lemmas \ref{lem:TV_bd}, \ref{lem:TV_bd_periodic} }
\label{subsec:TV_bd_Prf}
To show the bound in \eqref{eq:TV_bd}, we first bound the total variation distance between $P_1(X^n)$ and $P_2(X^n)$ by the KL divergence $D_{KL}(P_1(X^n) || P_2(X^n))$ using Pinsker's inequality. 
\begin{lemma}
\textbf{Pinkser's inequality} \cite[Theorem 4.19]{boucheron2013}
\begin{equation}
      ||P_1(X^n) - P_2(X^n)||_{TV} \ \leq \ \frac{1}{\sqrt{2}} \ \sqrt{ D_{KL}(P_1(X^n) || P_2(X^n)) } \label{eq:Pinsker's_ineq}
\end{equation}
where $ D_{KL}(P_1(X^n) || P_2(X^n)) $ is the KL divergence between $P_1(X^n)$ and $P_2(X^n).$
\end{lemma}
Our next lemma expresses $ D_{KL}(P_1(X^n) || P_2(X^n)) $ in terms of the KL divergence between the stationary distributions $\bpi_1, \bpi_2$ and the KL divergence between the corresponding rows of $P_1$ and $P_2.$
\begin{lemma}
For any two t.p.ms $P_1, P_2,$ on an $\cX,$ with $\bpi_1, \bpi_2$ as their respective stationary distributions, 
\begin{equation}
     D_{KL}(P_1(X^n) || P_2(X^n)) \ = \ D_{KL}(\bpi_1 || \bpi_2) + (n-1) \ \sum_{x \in \cX} \pi_{1,x} \ D_{KL}( P_1(\cdot | x) \ || \ P_2(\cdot | x) ) \label{eq:KL_div_Markov}
\end{equation}
where $P_i( \cdot |x)$ denotes the row of the t.p.m $P_i, i = 1,2,$  with transition probabilities from the state $x.$
\label{lem:KL_div_Markov}
\end{lemma}
\begin{proof}
Let $x^n \triangleq (x_1, x_2,\ldots,x_n) \in \cX^n.$
\begin{align}
    & D_{KL}(P_1(X^n) || P_2(X^n)) \nonumber \\
    & = \ \sum_{x^n \in \cX^n} P_1(x^n) \ \ln (P_1(x^n)/P_2(x^n)) \nonumber \\ 
    & \overset{(a)}{=} \ \sum_{x^n \in \cX^n}  \ P_1( x^n)  \left[ \ln \ (\pi_{1,x_1}/\pi_{2,x_1}) + \sum_{l = 2}^n \ \ln  \frac{P_1(X_l = x_l | X_{l-1} = x_{l-1})}{ P_2(X_l = x_l | X_{l-1} = x_{l-1}) }  \right] \nonumber \\
    & \overset{(b)}{=} \ D_{KL}(\bpi_1||\bpi_2) \ + \ \sum_{l=2}^n \ \sum_{x^l \in \cX^l} P_1(x^l) \ \ln \frac{P_1(X_l = x_l | X_{l-1} = x_{l-1})}{ P_2(X_l = x_l | X_{l-1} = x_{l-1}) } \nonumber \\
    & \overset{(c)}{=} \ D_{KL}(\bpi_1||\bpi_2) \nonumber \\ 
    & \  + \ \sum_{l=2}^n  \ \sum_{x_{l-1}, x_l \in \cX}  \pi_{1,x_{l-1}}   P_1(X_l = x_l | X_{l-1} = x_{l-1})  \ln \frac{P_1(X_l = x_l | X_{l-1} = x_{l-1})}{ P_2(X_l = x_l | X_{l-1} = x_{l-1}) } \nonumber \\ 
    & = \ D_{KL}(\bpi_1||\bpi_2) + (n-1) \ \ \sum_{ x \in \cX} \pi_{1,x} \ D_{KL}( P_1(\cdot | x) \ || \ P_2(\cdot | x) ) \nonumber 
\end{align}
where we get $(a)$ by using the Markov property $ P_i(x^n) = \pi_{i,x_1} \ \prod_{l-2}^n P_i(X_l = x_l| X_{l-1} = x_{l-1}),$ $i = 1,2,$ $(b)$ and $(c)$ by appropriately marginalizing $P_1(x^n).$
\end{proof} 
\subsubsection{Proof of Lemma \ref{lem:TV_bd}} \label{subsec: Lemma TV_bd_prf}
Using the values specified for $\pi_{i,x},$  $P_i(X_2 = y|X_1 = x)$ for $x,y \in \{1,\ldots, L+1 \},$ $i = 1,2,$ in the section \ref{subsec:stat_lb_Prf}, we get
\begin{align}
    D_{KL}(\bpi_1||\bpi_2) \ & = \ -0.5 \ln \left( { 1 - {4 \alpha^2}/{\beta^2} } \right) \nonumber \\
    D_{KL}(P_1(\cdot|1) \ || \ P_2(\cdot|1) ) \ & = \ -(1 - 0.5\beta) \ \ln  \left( {1+\alpha/(1-0.5\beta)}\right) - 0.5 \ \beta \ \ln \left( {1 - 2\alpha/\beta} \right) \nonumber \\
    D_{KL}(P_1(\cdot|x) \ || \ P_2(\cdot|x) ) \ & = \ -(1 - 0.5\beta) \ \ln  \left( {1-\alpha/(1-0.5\beta)}\right) - 0.5 \ \beta \ \ln \left( {1 + 2\alpha/\beta} \right),  \nonumber \\
    \text{ for }  x \in \{ 2,\ldots, L+1 \}.\nonumber 
\end{align}
Using the above three equations in \eqref{eq:KL_div_Markov}, we get 
\begin{align}
     & D_{KL}(P_1(X^n) || P_2(X^n)) \nonumber \\ 
     & = \ - 0.5 \ (n-1) \left[ (1 - 0.5\beta) \ \ln  \left( {1-\alpha^2/(1-0.5\beta)^2}\right)  + 0.5 \ \beta \ \ln \left( {1 - 4\alpha^2/\beta^2} \right) \right] \nonumber \\  & \qquad -0.5 \ln \left( { 1 - {4 \alpha^2}/{\beta^2} } \right) \nonumber \\
     & \overset{(a)}{\leq} \  \ (n-1) \left[ \frac{\alpha^2}{1 - 0.5\beta} + 2\frac{\alpha^2}{ \beta}  \right] + 4 \frac{\alpha^2}{ \beta^2} \ = \  4 \frac{\alpha^2}{\beta^2 ( 1 - 0.5\beta ) } \ \left[ 1 + 0.5 (n-2) \beta \right] \nonumber 
\end{align}
where we get $(a)$ by using $-\ln(1-x) \leq 2x,$ for $x \in (0,0.5).$ Plugging the above bound into \eqref{eq:Pinsker's_ineq} completes the proof of Lemma \ref{lem:TV_bd}. 

\subsubsection{Proof of Lemma \ref{lem:TV_bd_periodic}} 
Using the values specified for $\pi'_{i,x},$  $P'_i(X_2 = y|X_1 = x)$ for $x,y \in \{1,\ldots, L+2 \},$ $i = 1,2,$ in the section \ref{subsec:stat_lb_Prf}, we get
\begin{align}
    D_{KL}(\bpi_1||\bpi_2) \ & = \ -\frac{(\beta-1)}{2\beta} \ \ln\left[1-\left(\frac{2\alpha'}{\beta-1}\right)^2\right] \nonumber \\
    D_{KL}(P'_1(\cdot|1) \ || \ P'_2(\cdot|1) ) \ & = \ -0.5(\beta-1) \ \ln\left[1-\left(\frac{2\alpha'}{\beta-1}\right)^2\right] \nonumber \\
    D_{KL}(P'_1(\cdot|x) \ || \ P'_2(\cdot|x) ) \ & = \ 0, \ 
    \text{ for }  x \in \{ 2,\ldots, L+2 \}.\nonumber 
\end{align}
Using the above three equations in \eqref{eq:KL_div_Markov} and following a method similar to \ref{subsec: Lemma TV_bd_prf}, we get  $||P'_1(X^n) - P'_2(X^n)||_{TV} \leq \sqrt{2n} \alpha'/\sqrt{\beta(\beta-1)}.$ This completes the proof of Lemma \ref{lem:TV_bd_periodic}.  

\appendix

\section{Appendix A}
In this appendix, we provide proofs for the technical lemmas involved in the proof of theorem \ref{thm:minimax_stat}. 

\subsection{Proof of Lemma \ref{lem:MSE_term_bd} }
\label{subsec:MSE_term_bd_Prf}
The first two parts of Lemma \ref{lem:MSE_term_bd} are relatively easy to prove. 

\subsubsection{Lemma \ref{lem:MSE_term_bd}, part 1} 
To prove \eqref{eq:MSE_t2}, we begin with the following expression for $Q^n_x(1).$  
\begin{align}
   Q^n_x(1)  \  = \ \pr(N_x(X^n) = 1) \ & = \  \sum_{ l = 1 }^{n} \pr(X_l = x; X_m \neq x, m \neq l) \ \leq \ \sum_{ l = 1 }^{n} \pr(X_l = x) \ = \ n \pi_x. \label{eq:P(Fx = 1)_bd} 
\end{align}

This implies $ (1/n^2) \ \sum_{x \in \cX} Q^n_x(1) \leq 1/n.$ 

\subsubsection{Lemma \ref{lem:MSE_term_bd}, part 2}

To prove \eqref{eq:MSE_t3}, we begin with the following expression for $Q^n_{x,y}(1,1).$ 
\begin{align}
  Q^n_{x,y}(1,1) =   \pr(N_x = N_y = 1) \ & = \  \sum_{ l_1 = 1 }^{n-1} \sum_{ l_2 = l_1 + 1 }^{n}  \pr(X_{l_1} = x, X_{l_2} = y, X_m \neq x,y \ ; m \neq l_1, l_2) \nonumber \\ 
    & \qquad + \sum_{ l_1 = 1 }^{n-1} \sum_{ l_2 = l_1 + 1 }^{n}  \pr(X_{l_1} = y, X_{l_2} = x, X_m \neq x,y \ ; m \neq l_1, l_2)   \label{eq:P(Fx = N_y = 1)} 
\end{align}
For  $l_2 > l_1 \geq 1,$
\begin{align}
    \pr(X_{l_1} = x, X_{l_2} = y, X_m \neq x,y \ ; m \neq l_1, l_2) \ & \leq \ \pr( X_{l_1} = x, X_{l_2} = y) \nonumber \\ 
    & = \ \pi_x \ \pr( X_{l_2} = y | X_{l_1} = x) \label{eq:cond_prob_bd_x}  \\ 
    \text{ Similarly, } \pr(X_{l_1} = y, X_{l_2} = x, X_m \neq x,y \ ; m \neq l_1, l_2) \ & \leq \ \pi_y \ \pr( X_{l_2} = x | X_{l_1} = y) \label{eq:cond_prob_bd_y}
\end{align}
Plugging \eqref{eq:cond_prob_bd_x} and \eqref{eq:cond_prob_bd_y} into \eqref{eq:P(Fx = N_y = 1)}, 
\begin{align}
   Q^n_{x,y}(1,1)  \ & \leq \  \sum_{ l_1 = 1 }^{n-1} \sum_{ l_2 = l_1 + 1 }^{n} \ \pi_x \ \pr( X_{l_2} = y | X_{l_1} = x) \nonumber \\ 
    & \qquad + \sum_{ l_1 = 1 }^{n-1} \sum_{ l_2 = l_1 + 1 }^{n} \ \pi_y \ \pr( X_{l_2} = x | X_{l_1} = y) \nonumber 
\end{align}
Since $ \sum_{x \in \cX} \sum_{y \in \cX, y \neq x} \ \pi_x \ \pr( X_{l_2} = y | X_{l_1} = x) \ \leq \ \sum_{x \in \cX} \ \pi_x \ \leq \ 1,$ we get \\ 
$ \sum_{x \in \cX} \sum_{y \in \cX, y \neq x} \ Q^n_{x,y}(1,1) \ \leq \ n(n-1).$ Dividing by $ {n^2 (n-1)}$ completes the proof. 

\subsubsection{ Lemma \ref{lem:MSE_term_bd}, part 3}
\label{subsec:Qx_0_Prf}
To bound $\sum_{x \in \cX} \pi_x^2 \ Q_x^n(0)$, we consider the sets $A_0, A(\delta)$ and $ \overline{A}(\delta)$ (with $0 \leq \delta < \beta/5$) defined in section \ref{sec:Bias_r2_Prf}, that make up $\cX$. The following lemma provides bounds on $Q_x^n(0)$ for $x$ in each of these sets. 

\begin{lemma}
For a rank-2 diagonalizable t.p.m $P$ with spectral gap $\beta,$ $\delta \in [0, \beta/5),$

\begin{enumerate}
    \item For $x \in  A(\delta),$ 
    \begin{equation}
        Q_x^n(0) \ \leq \ \left( 1 + \frac{9}{\beta} \right) \ \left( 1 - \frac{\beta^2}{2(\beta + 2)} \pi_x   \right)^n, \label{eq:P(N_x=0)_bd1}
    \end{equation}
    
    \item For $x \in \overline{A}(\delta),$ 
    \begin{equation}
          Q_x^n(0) \ \leq \ \left( 1 + \frac{3n}{2} \right) \  e^{-0.5 (n-1) \frac{\beta^2}{\beta + 2} \delta } \label{eq:P(N_x=0)_bd2} 
    \end{equation}
    
    \item For $x \in A_0,$ 
    \begin{equation}
        Q_x^n(0) \ = \ (1-\pi_x)^n \label{eq:P(N_x=0)_bd3} 
    \end{equation}
\end{enumerate}

\label{lem: P(N_x=0)_bd}

\end{lemma}

\begin{proof}
The proof given in section \ref{subsec:P(N_x=0)_bd_Prf} uses a closed form expression for $Q_x^n(0)$ obtained using the eigenvalues $\lambda_{i}^{\sim x},i = 1,2,$ (and their corresponding left and right eigenvectors) of the matrix $P^{\sim x}$. \eqref{eq:P(N_x=0)_bd1} and \eqref{eq:P(N_x=0)_bd2} are obtained by using the bounds $|\lambda_i^{\sim x}| \leq 1 - \frac{\beta^2}{\beta+2}\pi_x, i= 1,2,$ $x \in \cX,$ and $\Delta_x \triangleq \lambda_1^{\sim x} - \lambda_2^{\sim x} > \beta/3,$ for $x \in A(\delta),$ in the closed form expression for $Q_x^n(0).$
\end{proof}

We now use the above bounds on $ Q_x^n(0) $ to bound $ \sum_{x \in \cX} \pi_x^2 \ Q_x^n(0) $ as 
\begin{enumerate}
    \item \begin{align}
        \sum_{x \in A(\delta)} \pi_x^2 \ Q_x^n(0) \ & \overset{(a_1)}{\leq} \  \left( 1 + \frac{9}{\beta} \right) \sum_{x \in A(\delta)} \pi_x^2 \  \left( 1 - \frac{\beta^2}{2(\beta + 2)} \pi_x  \right)^n \nonumber \\
       &  \overset{(b_1)}{\leq}  \  2\left( 1 + \frac{9}{\beta} \right) \ \frac{(\beta  +2)}{\beta^2 n}  \ \sum_{x \in A(\delta)} \pi_x \  \overset{(c_1)}{\leq}  \ 2\left( 1 + \frac{9}{\beta} \right) \ \frac{(\beta  +2)}{\beta^2 n}  \label{eq:t1_A}
    \end{align}
    where we get $(a_1)$ by using \eqref{eq:P(N_x=0)_bd1}, $(b_1)$ by using $\max_{p \in (0,1)} p \ (1-cp)^n = \frac{1}{c(n+1)} \ \left(1 - \frac{1}{n+1} \right)^n \leq \frac{1}{cn},$ and $(c_1)$ by using $ \sum_{x \in  A(\delta)}  \pi_x \leq 1.$
    \item \begin{align}
        \sum_{x \in \overline{A}(\delta)} \pi_x^2 \ Q_x^n(0) \ &  \overset{(a_2)}{\leq} \ \left( 1 + \frac{3n}{2} \right) \ e^{-0.5 (n-1) \frac{\beta^2}{\beta + 2} \delta } \sum_{x \in \overline{A}(\delta)} \pi_x^2 \  \overset{(c_2)}{\leq} \ \left( 1 + \frac{3n}{2} \right) \ e^{-0.5 (n-1) \frac{\beta^2}{\beta + 2} \delta } \label{eq:t1_A'} 
        \end{align}
        where we get $(a_2)$ by using \eqref{eq:P(N_x=0)_bd2}, and $(c_2)$ by using $ \sum_{x \in \overline{A}(\delta) }  \pi_x^2 \leq 1.$
        
    \item \begin{align}
        \sum_{x \in A_0} \pi_x^2 \ Q_x^n(0) \ & \overset{(a_0)}{=} \ \sum_{x \in A_0} \pi_x^2 \ (1-\pi_x)^n \ \overset{(b_0)}{\leq} \ 1/n \ \sum_{x \in A_0} \pi_x \ \overset{(c_0)}{\leq} \ 1/n \label{eq:t1_A0}
    \end{align}    
    where we get $(a_0)$ by using \eqref{eq:P(N_x=0)_bd3}, $(b_0)$ by using $\max_{p \in (0,1)} p \ (1-p)^n = \frac{1}{(n+1)} \ \left(1 - \frac{1}{n+1} \right)^n \leq \frac{1}{n},$ and $(c_0)$ by using $ \sum_{x \in  A_0}  \pi_x \leq 1.$
\end{enumerate}
Since $\cX = A(\delta) \cup \overline{A}(\delta) \cup A_0,$ 
\begin{align}
     \sum_{x \in \cX} \pi_x^2 \ Q_x^n(0) \ & = \ \sum_{x \in A_0} \pi_x^2 \ Q_x^n(0) + \sum_{x \in A(\delta)} \pi_x^2 \ Q_x^n(0) + \sum_{x \in \overline{A}(\delta)} \pi_x^2 \ Q_x^n(0) \nonumber \\
     & \overset{(a)}{\leq} \ 1/n +   2\left( 1 + \frac{9}{\beta} \right) \ \frac{(\beta  +2)}{\beta^2 n} + \left( 1 + \frac{3n}{2} \right) \  e^{-0.5 (n-1) \frac{\beta^2}{\beta + 2} \delta } \nonumber 
\end{align}
where we get $(a)$ by combining \eqref{eq:t1_A}, \eqref{eq:t1_A'} and \eqref{eq:t1_A0}.

Choosing $\delta = 8 \ \frac{(\beta + 2)}{\beta^2 (n-5)} \ln n $ and using $n-5 < n-1$ completes the proof of Lemma \ref{lem:MSE_term_bd}, part 3. 

\subsubsection{ Lemma \ref{lem:MSE_term_bd}, part 4} 

To bound $  \sum_{x \in \cX} \sum_{y \in \cX, y \neq x} T^n_{xy}, $ we divide $\{ (x,y) : x\in \cX, y \in \cX, x \neq y \}$ into two sets: 
\begin{align}
    B(\delta) \ & \triangleq \  \{ (x,y) : x,y \in \cX, \ x \neq y, \ \pi_x + \pi_y < \delta \} \nonumber \\
    \overline{B}(\delta) \ & \triangleq \ \{ (x,y) : x,y \in \cX, \  x \neq y, \ \pi_x + \pi_y \geq \delta \}  \nonumber
\end{align}
with $0 \leq \delta < \beta/5.$

In the following lemma, we bound the sum of $T^n_{xy}$ over $(x,y) \in  B(\delta)  $ and $(x,y) \in  \overline{B}(\delta).$

\begin{lemma}
For a rank-2 diagonalizable t.p.m $P$ with spectral gap $\beta,$ and $ \delta \in [0,\beta/5),$  

\begin{enumerate}

\item 
\begin{align}
    \sum_{(x,y) \in B(\delta)} T^n_{xy} \ & \leq \ O(1/n\beta^5), \label{eq:MSE_t4_2}
  \end{align}
  
  \item \begin{align}
      \sum_{(x,y) \in \overline{B}(\delta)} T^n_{xy} \ & \leq \ O(n^3) \ e^{- (n - 5) \frac{\beta^2}{2(\beta + 2)}(\delta) }  \label{eq:MSE_t4_1}
\end{align}

\end{enumerate}

\label{lem: Txy_bd }

\end{lemma}

\begin{proof}
The proof given in Section \ref{subsec:Txy_bd_Prf} uses a closed form expression for $T_{xy}^n$ obtained using the eigenvalues $\lambda_{i}^{\sim x,y},i = 1,2,$ (and their corresponding left and right eigenvectors) of the matrix $P^{\sim x,y}$ which is the matrix obtained by replacing the columns of $P$ corresponding to $x,y$ with all zero columns. \eqref{eq:MSE_t4_2} and \eqref{eq:MSE_t4_1} are obtained by using the bounds $|\lambda_i^{\sim x,y}| \leq 1 - \frac{\beta^2}{\beta+2}(\pi_x+\pi_y), i= 1,2,$ $x,y \in \cX,$ and $\Delta_{x,y} \triangleq \lambda_1^{\sim x,y} - \lambda_2^{\sim x,y} > \beta/3,$ for $(x,y) \in B(\delta),$ in the closed form expression for $T_{xy}^n.$
\end{proof}

Combining \eqref{eq:MSE_t4_1} with \eqref{eq:MSE_t4_2} and choosing $ \delta = 8 \ \frac{(\beta + 2)}{\beta^2 (n-5)} \ln n $ (so that $ \sum_{(x,y) \in \overline{B}(\delta)} T^n_{xy} \ \leq \ O(1/n) $ ) completes the proof of Lemma \ref{lem:MSE_term_bd}, part 4.

\section{Appendix B}
In this appendix, we provide the proofs for lemmas \ref{lem:x_Bias_fin}, \ref{lem: P(N_x=0)_bd} and \ref{lem: Txy_bd } used to prove theorems \ref{thm:Bias_r2} and \ref{thm:minimax_stat}.

To prove these lemmas, we utilize the diagonalized form of the t.p.m $P$ together with its rank properties. 

Consider a rank-2, $K\times K$ diagonalizable t.p.m $P$ with stationary distribution $\bpi=[\pi_1\ \cdots\ \pi_K]$ and spectral gap $\beta.$ There exist vectors $u=[u_1\ \cdots\ u_K]$ and $v=[v_1\ \cdots\ v_K]^T$ satisfying the decomposition

    $$P=RDS\text{ with }SR = \begin{bmatrix}
    1&0\\0&1
    \end{bmatrix},$$
    where $R=\begin{bmatrix}
    \one&v
    \end{bmatrix}$, $D=\begin{bmatrix}
    1&0\\0&1-\beta
    \end{bmatrix}$ and $S=\begin{bmatrix}
    \bpi \\ u
    \end{bmatrix}$. Since $P$ is a t.p.m, we have, for $1\le i,j\le K$,
    \begin{equation}
      0\le P_{ij} = \pi_j+(1-\beta)v_iu_j\le 1.  
      \label{eq:nonnegPdiag}
    \end{equation}
    The vectors $u$ and $v$ come from the standard eigenvector decomposition of $P$. Since $P$ has rank 2, the other eigenvectors correspond to eigenvalue 0. Let $\overline{a} \triangleq 1-a$ and $\psi_{wz} \triangleq \overline{\beta} v_w u_z, $ for $w,z \in \cX.$  
\begin{lemma}
\label{lem: sum_over_x} 
For a rank-2 diagonalizable t.p.m $P$ with spectral gap $\beta,$ and $x,y \in \cX,$
\begin{align}
      &  \sum_{x \in \cX} (\pi_x)^{a} \ |\psi_{xx}^{b}| \ |\psi_{yx}^{c}| \ \leq \ 3,  \text{ for }  a , b , c \in \{ 0,1,2,3, \ldots \} \text{ and }  a + b + c \geq 1. \label{eq: sum_over_x}
    \end{align}
  \label{lem:sum_over_x}
\end{lemma}
\begin{proof}
Section \ref{subsec: sum_over_x_Prf}.
\end{proof}

\subsection{Proof of Lemma \ref{lem:x_Bias_fin}}
\label{sec:x_Bias_fin_Prf}
To prove Lemma \ref{lem:x_Bias_fin}, we use the 
 following bounds on $| \Gamma_x |$ given in the lemma below.

\begin{lemma} 
For a rank-2 diagonalizable t.p.m $P$ with spectral gap $\beta,$
    \begin{enumerate}
        \item For $x \in A(\delta)$,
        \begin{align}
         \vert \Gamma_x \vert \ & \leq \ 36 \ (1/n\beta^2) \  \Big[ (4/c_{\beta}) \  ( |\psi_{xx}| + \pi_x ) \  + ( e^{-1}  + 18/\beta ) \ \pi_x \Big],  \label{eq:x_Bias_A_del_bd1} 
        \end{align}
        \label{lem_part:x_Bias_A_del}

        \item For $x \in \overline{A}(\delta)$,

        \begin{align}
             \vert \Gamma_x  \vert \ & \leq \ \pi_x \ \Big[ g(n) \ \exp{ - (n-3) \ c_{\beta} \  \delta/2 } + (1 + (3/2)  \  n ) \ \exp{ - (n-1) \ c_{\beta} \ \delta/2 } \Big],  \label{eq:x_Bias_A_til_bd1}  
        \end{align}
         where $c_{\beta} =
             \beta$ for $ \beta \in (0,1]$, $
             c_{\beta} = 1$ for $ \beta \in [1,2]$,  and $g(n) = [1 + (5/2) \  (n-1) \ (1  + (5/12)  \ ( n-2)) ]$.
             \label{lem_part:x_Bias_A_til}
             
        \item For $x \in A_0$,
           
        \begin{equation}
              \vert \Gamma_x  \vert \ \leq \ \pi_x \ O(1/n) \label{eq:x_Bias_A0_bd1}
        \end{equation}

        \label{lem_part:x_Bias_A0}
    \end{enumerate}
    \label{lem:x_Bias}
\end{lemma}
\begin{proof}
    Section \ref{sec:x_Bias_Prf}.
\end{proof}

Taking the sum of \eqref{eq:x_Bias_A_del_bd1} over $x$ in $A(\delta)$, we get 
\begin{equation}
    \sum_{x \in A(\delta)} \  \vert \Gamma_x  \vert \ \overset{(a_1)}{\leq} \  36 \ (1/n\beta^2) \ ( e^{-1} + 16/c_{\beta} +  34/\beta),  \label{eq:x_Bias_A_del_bd2} 
\end{equation}
where we get $(a_1)$ by using $\sum_{x \in A(\delta)} \pi_x \leq 1$ and lemma \ref{lem: sum_over_x} as $\sum_{x \in A(\delta)} |\psi_{xx}| \leq \sum_{x \in \cX} |\psi_{xx}| \leq 1$. 
This completes the proof of lemma \ref{lem:x_Bias_fin}, part \ref{lem_part:x_Bias_A_del_fin}. 

Taking the sum of \eqref{eq:x_Bias_A_til_bd1} over $x$ in $\overline{A}(\delta)$ and using $\sum_{x \in \overline{A}(\delta)} \pi_x \leq 1$, we get 
\begin{equation}
     \sum_{x \in \overline{A}(\delta)} \  \vert \Gamma_x  \vert \ {\leq} \ \Big[ g(n) \ \exp{ - (n-3) \ c_{\beta} \ \delta/2 } + (1 + (3/2)  \  n ) \ \exp{ - (n-1) \ c_{\beta} \  \delta/2 } \Big],  \label{eq:x_Bias_A_til_bd2}
\end{equation}
This completes the proof of lemma \ref{lem:x_Bias_fin}, part \ref{lem_part:x_Bias_A_til_fin}. 
Similarly, we get 
\begin{equation}
    \sum_{x \in A_0} \  \vert \Gamma_x  \vert \ {\leq} \ O(1/n). \label{eq:x_Bias_A0_bd2} 
\end{equation}
completing the proof of lemma \ref{lem:x_Bias_fin}.

 \subsubsection{Proof of Lemma \ref{lem:x_Bias}} 
\label{sec:x_Bias_Prf}
 
 \emph{Proof of Lemma \ref{lem:x_Bias}, part \ref{lem_part:x_Bias_A_del}}: \\ 

 For $x \in \cX \setminus A_0$,  $P_{xx} \neq \pi_x$ i.e. $v_x, u_x$ are non-zero and the matrix $P^{\sim x}$ can be diagonalized as 
    \begin{equation}
        P^{\sim x} = \sum_{i=1}^2 \ \lambda_{i,x} \ v_{i}^{\sim x} \ u_{i}^{\sim x}, \label{eq:R2_Px_diag}
    \end{equation}
    with 
    \begin{align*}
        \text{ eigenvalues } & \ \lambda_{i, x} \triangleq \ 0.5\big( \overline{\pi}_x + \overline{\beta} (1-v_x u_x) + (-1)^{i+1} \Delta_x  \big), \nonumber \\ 
        \text{ right eigenvectors } & \ v_{i}^{\sim x} =  \one + ({1}/{2 \pi_x v_x}) \  [s_x + (-1)^{i} \Delta_x] \ v, \nonumber \\ 
        \text{ left eigenvectors } & \ u_{i}^{\sim x} = ({1}/{\lambda_{i,x} \Delta_x }) \  \Big( (1/2) \ [ \Delta_x + (-1)^{i-1} s_x ] \   \bpi^{\sim x}  + (-1)^{i} \ (1-\beta) \pi_x v_x \ u^{\sim x}  \Big), \nonumber 
    \end{align*}
    for $i = 1,2$, where $\Delta_x^2 \ \triangleq \   s_x^2 + 4 \overline{\beta}\pi_x v_x u_x$  and $s_x \triangleq \beta - {\pi_x} + \overline{\beta}v_xu_x$. \\ 
   We use $u^{\sim x}$ to denote the vector $u$ with $x$-th entry set to 0. Note $\bpi v_{i}^{\sim x} = 1$ for $i = 1,2$, since $\bpi v = 0$. 
The following lemma provides a bound on $|\Gamma_x|$ using $\Delta_x$.  

\begin{lemma}
    For a rank-2 diagonalizable t.p.m $P$ with spectral gap $\beta,$ 
    \begin{equation}
        |\Gamma_x| \ \leq \ 4 \ (1/n) \ \Delta_x^{-2} \ \Big[ (4/c_{\beta}) \  ( |\psi_{xx}| + \pi_x ) \  + e^{-1} \ \pi_x + 6 \ \Delta_x^{-1} \ \pi_x \Big], \ \ x \in A(\delta), \label{eq:x_Bias_A_del_int1} 
    \end{equation}
    where $c_{\beta} =
             \beta$ for $ \beta \in (0,1]$, $
             c_{\beta} = 1$ for $ \beta \in [1,2]$. 
    \label{lem:x_Bias_A_del_int1}
\end{lemma}
\begin{proof}
    Section \ref{sec:x_Bias_A_del_int1_Prf}. 
\end{proof}
We next claim that $\Delta_x$ is bounded away from 0 for $x\in A(\delta)$. \\ 
\noindent\textbf{Claim 1:} \ $\Delta_x\ge\beta/3$, for $x\in A(\delta)$.
\begin{proof}
Using $\overline{\beta}v_xu_x\ge-\pi_x$ from \eqref{eq:nonnegPdiag}, we get
\begin{align*}
  \beta - \pi_x + \overline{\beta} v_x u_x &\geq \beta - 2\pi_x,\\
  4\pi_x \overline{\beta} v_x u_x &\geq -4\pi_x^2.
\end{align*}
Using the above in the expression for $\Delta_x^2,$
 \begin{align}
 \Delta_x^2 &= ( \beta - \pi_x + \overline{\beta} v_x u_x )^2 + 4\pi_x \overline{\beta} v_x u_x \nonumber \\ 
 & \geq (\beta - 2\pi_x)^2 - 4\pi_x^2 = \beta^2 - 4\beta\pi_x\nonumber\\
 & \overset{(a)}{>} \beta^2/5 > \beta^2/9,
 \end{align}
 where we use $\pi_x<\beta/5$ to get $(a).$
\end{proof}

Using the above lower bound on $\Delta_x$ for $x \in A(\delta)$ in \eqref{eq:x_Bias_A_del_int1} completes the proof of  Lemma \ref{lem:x_Bias}, part \ref{lem_part:x_Bias_A_del}. 

To prove the remaining parts of lemma \ref{lem:x_Bias},
we use the following lemma. 

  \begin{lemma}
       For a rank-2 diaginalizable t.p.m. with spectral gap $\beta$, 
       \begin{enumerate}
           \item For $x \in A(\delta)$,
           \begin{equation}
              \pr(N_x(X^n) =1) \  = \ n \ \pi_x \  (1-\pi_x)^{n-1}. \label{eq:Qx_1_A0}
           \end{equation}
           \item 
           For $x \in \cX \setminus A(\delta)$,
           \begin{align}
               \pr(N_x(X^n) =1) \ & \leq  \ 
n \ \pi_x \ g(n) \ \exp{ - (n-3) \ c_{\beta} \  \delta/2 }, \label{eq:Qx_1_bd} \\
\pr(N_x(X^n) =0) \ & \leq  \ 
(1 + (3/2)  \  n ) \ \exp{ - (n-1) \ c_{\beta} \ \delta/2 }, \label{eq:Qx_0_bd}
           \end{align}
           where $g(n) = [1 + (5/2) \  (n-1) \ (1  + (5/12)  \ ( n-2)) ]$.         
       \end{enumerate}
       \label{lem:exp_prob}
   \end{lemma}
\begin{proof}
    Section \ref{subsec:P(N_x=0)_bd_Prf}. 
\end{proof}

\emph{Proof of Lemma \ref{lem:x_Bias}, part \ref{lem_part:x_Bias_A_til}}:
\begin{align}
      \vert \Gamma_x \vert \ & \leq \  \frac{1}{n} \  \pr(N_x(X^n) =1) + \pi_x \ \pr(N_x(X^n) = 0) \nonumber \\ 
       & \overset{(a)}{\leq} \ \pi_x \ \Big[ g(n) \ \exp{ - (n-3) \ c_{\beta} \  \delta/2 } + (1 + (3/2)  \  n ) \ \exp{ - (n-1) \ c_{\beta} \ \delta/2 } \Big], \nonumber
   \end{align}
   where we use lemma \ref{lem:exp_prob} to get $(a)$.
This completes the proof of  Lemma \ref{lem:x_Bias}, part \ref{lem_part:x_Bias_A_til}. 

\emph{Proof of Lemma \ref{lem:x_Bias}, part \ref{lem_part:x_Bias_A0}}:  

 Using lemmas \ref{lem:exp_prob} and \ref{lem: P(N_x=0)_bd}, we get,
   \begin{equation}
     \vert \Gamma_x \vert  \  = \ \pi_x^2 \ (1-\pi_x)^{n-1} 
        \leq \ \pi_x \ O(1/n) 
     \end{equation}
completing the proof of  Lemma \ref{lem:x_Bias}. 

\subsubsection{Proof of Lemma \ref{lem:x_Bias_A_del_int1}}     \label{sec:x_Bias_A_del_int1_Prf}
    To prove \eqref{eq:x_Bias_A_del_int1}, we begin with following expression for $\Gamma_x$ obtained by using 
    \eqref{eq:R2_Px_diag} in \eqref{eq:Bias_lmda_x}. 
   \begin{align}
       \Gamma_x  \ & = \  \Bigg[ \sum_{i =1}^{2}   ( \lambda_{i,x} )^{n-1}  \  (  u^{\sim x}_{i} \one )\ ( u^{\sim x}_{i} P_x v^{\sim x}_{i} - \pi_x  \lambda_{i,x}  ) \nonumber \\
        & \qquad \qquad + (1/n) \ \Delta_x^{-1} \ [ ( \lambda_{1,x} )^{n} - ( \lambda_{2,x} )^{n}  ] \sum_{ i,j \in [2]: \  i \neq j}   \  (  u^{\sim x}_{j} \one )\ ( u^{\sim x}_{i} P_x v^{\sim x}_{j})  \Bigg] \label{eq:Gamma_x_R2}
   \end{align}
  The following lemma bounds the absolute value of the factors multiplying $\lambda_{i,x}$ powers in the above expression.
\begin{lemma}
   For a rank-2 diaginalizable t.p.m. (on $\cX$) with spectral gap $\beta$, and $x \in \cX \setminus A_0$, 
   \begin{enumerate}
       \item 
       For $i =1,2$, 
   \begin{equation}
       \Big \vert  (u_{i}^{\sim x} \one) \  (u_{i}^{\sim x} \ P_x \ v_{i}^{\sim x} - \pi_x \lambda_{i,x}) \Big \vert \ \leq \ 
       \Delta_x^{-2} \ \Big[ 4 \ ( |\psi_{xx}| + \pi_x ) \ \pi_x + 2 \ \pi_x \ \lambda_{1,x} \ (1-\lambda_{1,x}) \Big]
      \label{eq:x_Bias_A_del_int2} 
   \end{equation}
     \label{lem_part:x_Bias_A_del_int2}
  \item For $i,j$ in $\{1,2\}$ with $i \neq j$, 
  \begin{align}
      \Big \vert ( u_{i}^{\sim x} \ \one  ) \ (  u_{j}^{\sim x} \ P_x \ v_{i}^{\sim x} ) \Big \vert \ & \leq \ 6 \ \Delta_x^{-2} \ \pi_x \label{eq:x_Bias_A_del_int3}   
  \end{align}
      \label{lem_part:x_Bias_A_del_int3}
  \end{enumerate}
    \label{lem:x_Bias_A_del_int2}
   \end{lemma}
   \begin{proof}
       Section \ref{sec:x_Bias_A_del_int2_Prf}.
   \end{proof}

      Our next lemma bounds the absolute value of the eigenvalues $\lambda_1^{\sim x}, \lambda_2^{\sim x}$ using the spectral gap $\beta$ and $\pi_x.$ 

\begin{lemma}
For a rank-2 diagonalizable t.p.m $P$ with spectral gap $\beta,$ 
\begin{equation}
    |\lambda_i^{\sim x}|  \  \leq \ 1 - (c_{\beta}/2) \pi_x , \ i = 1,2,  \label{eq:e_val_bd1_x} 
    \end{equation}
    where $c_{\beta} = \beta$ for $ \beta \in (0,1]$, $
             c_{\beta} = 1$ for $ \beta \in [1,2]$.
    

\begin{equation}
    |\lambda_i^{\sim x}|  \  \leq \ 1 - \frac{\beta^2}{2(\beta + 2)} \pi_x , \ i = 1,2.  \label{eq:e_value_bd_x}
\end{equation}
     
\label{lem:e_value_bd_x}
\end{lemma}
\begin{proof}
Section \ref{subsec:e_value_bd_x_Prf}.
\end{proof}

    Using \eqref{eq:x_Bias_A_del_int2} and \eqref{eq:x_Bias_A_del_int3} in \eqref{eq:Gamma_x_R2} to bound the absolute value of $\Gamma_x$ for $x \in  A(\delta)$, we have 

    \begin{align}
        |\Gamma_x| \ & \leq \  2 \ ( \lambda_{1,x} )^{n-1} \ \Delta_x^{-2} \ \Big[ 4 \ ( |\psi_{xx}| + \pi_x ) \ \pi_x + 2 \ \pi_x \ \lambda_{1,x} \ (1-\lambda_{1,x}) \Big] \nonumber \\ 
        & \qquad \qquad \qquad  + \  (12/n) \  \Delta_x^{-3}  \ [ ( \lambda_{1,x} )^{n} -   ( \lambda_{2,x} )^{n} ] \ \pi_x  \nonumber \\
         &  \overset{(b_1)}{\leq} \  2 \ \Delta_x^{-2} \ \Big[ 4 \ ( |\psi_{xx}| + \pi_x ) \ \pi_x \ ( \lambda_{1,x} )^{n-1} + 2 \ \pi_x \ (\lambda_{1,x})^n \ (1-\lambda_{1,x}) \Big] \nonumber \\ 
        & \qquad \qquad \qquad + \ (24/n) \  \Delta_x^{-3} \ \pi_x  \nonumber \\ 
        & \overset{(b_2)}{\leq} \ 4 \ (1/n) \ \Delta_x^{-2} \ \Big[ (4/c_{\beta}) \  ( |\psi_{xx}| + \pi_x ) \  + e^{-1} \ \pi_x + 6 \ \Delta_x^{-1} \ \pi_x \Big], \nonumber
    \end{align}
   where we get $(b_1)$ by using $|\lambda_{i,x}| \leq 1$, $i = 1,2$, and $(b_2)$ by using \eqref{eq:e_val_bd1_x} along with $\max_{p \in (0,1)} p \ (1-cp)^n \leq \min \{ {e^{-1}}/{(cn)}, {1}/{c(n+1)} \}$. This completes the proof of lemma \ref{lem:x_Bias_A_del_int1}.
   \\
\subsubsection{Proof of Lemma \ref{lem:x_Bias_A_del_int2} }
 \label{sec:x_Bias_A_del_int2_Prf}
   To prove Lemma \ref{lem:x_Bias_A_del_int2},
   we use the expressions for $v_{i}^{\sim x}$ and $u_{i}^{\sim x}$ in section \ref{sec:x_Bias_Prf} to get 
   \begin{enumerate}
       \item 
       \begin{equation}
           u_{i}^{\sim x} \ \one = \frac{1}{2} \Big[ 1 + (-1)^{i-1} \frac{s_x}{\Delta_x} \Big], i=1,2. \label{eq:A_delta_int1}
       \end{equation} 
    
       \item 
       \begin{align}
           u_{i}^{\sim x} \ P_x \ v_{i}^{\sim x} - \pi_x \lambda_{i,x} 
          & \ = \ \frac{1}{\Delta_x} \ \Bigg[ \frac{1}{2} (\Delta_x + (-1)^{i} s_x ) (\overline{\beta} -  \lambda_{i,x}) \nonumber \\ 
          & \qquad \qquad  + (-1)^{i-1} \pi_x \ [ \lambda_{i,x} (1-\lambda_{i,x}) - \overline{\beta}(\pi_x + v_x u_x) ] \Bigg], i = 1,2. \label{eq:A_delta_int2}
       \end{align}
       \item \begin{align}
         u_{1}^{\sim x} \ P_x \ v_{2}^{\sim x} \ & = \  \frac{1}{\Delta_x} \ \Big[ \lambda_{1,x} - \overline{\beta} \Big] \ \Big[ \pi_x + \frac{1}{2} ( \Delta_x + s_x ) \Big] \label{eq:A_delta_int3} \\ 
          u_{2}^{\sim x} \ P_x \ v_{1}^{\sim x} \ & = \  \frac{1}{\Delta_x} \ \Big[  \overline{\beta} - \lambda_{2,x} \Big] \ \Big[ \pi_x + \frac{1}{2} ( \Delta_x - s_x ) \Big] \label{eq:A_delta_int4}
       \end{align}
   \end{enumerate}
   \emph{Proof of Lemma \ref{lem:x_Bias_A_del_int2}, part \ref{lem_part:x_Bias_A_del_int2}}: \\ 
       Using \eqref{eq:A_delta_int1} and \eqref{eq:A_delta_int2}, we have 
       \begin{align*}
           & \Big \vert  (u_{i}^{\sim x} \one) \  u_{i}^{\sim x} \ P_x \ v_{i}^{\sim x} - \pi_x \lambda_{i,x} \Big \vert \nonumber \\ 
           & \ = \ \Delta_x^{-2} \ \Big \vert (1/4) \ (\Delta_x^{2} - s_x^2) \ (\overline{\beta} -  \lambda_{i,x}) \nonumber \\ 
           & \qquad \qquad \quad +   (-1)^{i-1} \ (1/2) \ (\Delta_x +(-1)^{i-1} s_x)  \  \pi_x \ [ \lambda_{i,x} (1-\lambda_{i,x}) - \overline{\beta}(\pi_x + v_x u_x) ] \Big \vert \nonumber \\
           & \ = \ \Delta_x^{-2} \ \Big \vert \overline{\beta} \ \pi_x \   [  v_x u_x \ (\overline{\beta} -  \lambda_{i,x}) + (1/2) \  (-1)^{i} \ (\Delta_x +(-1)^{i-1} s_x)  \   (\pi_x + v_x u_x) ] \nonumber \\ 
           & \qquad \qquad \qquad +  (-1)^{i-1} \ (1/2) \ (\Delta_x +(-1)^{i-1} s_x)  \  \pi_x \ \lambda_{i,x} (1-\lambda_{i,x}) \Big \vert  \nonumber 
       \end{align*}
       Using $(1/2) \ (\Delta_x - s_x) =   \pi_x - (1 -  \lambda_{1,x} )$, $(1/2) \ (\Delta_x + s_x) =   1 - \pi_x -  \lambda_{2,x}$ along with $\psi_{xx} = \overline{\beta} v_x u_x$, $|\overline{\beta}| \leq 1$, $1-\pi_x \leq 1$ and $|\lambda_{i,x}| \leq 1$, we get
       \begin{align*}
             \Big \vert  (u_{1}^{\sim x} \one) \  u_{1}^{\sim x} \ P_x \ v_{1}^{\sim x} - \pi_x \lambda_{1,x} \Big \vert 
           \ & \leq \ \Delta_x^{-2} \ \Big[ 2( 2 |\psi_{xx}| + \pi_x) \  \pi_x + 2 \ \pi_x \  \lambda_{1,x} \ (1-\lambda_{1,x})  \Big] \\
            \Big \vert  (u_{2}^{\sim x} \one) \  u_{2}^{\sim x} \ P_x \ v_{2}^{\sim x} - \pi_x \lambda_{2,x} \Big \vert 
           \ & \leq \ \Delta_x^{-2} \ \Big[ 4(  |\psi_{xx}| +  \pi_x  ) \  \pi_x +   2 \ \pi_x \  \lambda_{1,x}  \ (1-\lambda_{1,x})  \Big]
       \end{align*}
   
   \emph{Proof of Lemma \ref{lem:x_Bias_A_del_int2}, part \ref{lem_part:x_Bias_A_del_int3}}: \\ 
   Using \eqref{eq:A_delta_int1} along with \eqref{eq:A_delta_int3} and \eqref{eq:A_delta_int4}, we have
   \begin{align*}
       \Big \vert  (u_{2}^{\sim x} \one) \ (u_{1}^{\sim x} \ P_x \ v_{2}^{\sim x}) \Big \vert & = \ \Big \vert \Delta_x^{-2} \ \pi_x \ \Big[ \lambda_{1,x} - \overline{\beta} \Big] \ \Big[ \pi_x - (1-\lambda_{1,x})   +  \psi_{xx} \Big] \Big \vert \nonumber \\ 
       & \ \overset{(d_1)}{\leq} \ 6 \  \Delta_x^{-2} \ \pi_x,  \\ 
        u_{1}^{\sim x} \one \ u_{2}^{\sim x} \ P_x \ v_{1}^{\sim x} \ & = \ \Big \vert \Delta_x^{-2} \ \pi_x \ \Big[ \overline{\beta} -  \lambda_{2,x}\Big] \ \Big[ (1 - \pi_x -\lambda_{2,x})   +  \psi_{xx}   \Big] \Big \vert  \nonumber \\ 
        & \ \overset{(d_2)}{\leq} \ 6 \  \Delta_x^{-2} \ \pi_x, 
   \end{align*}
   where we get $(d_1)$ and $(d_2)$ using $|\overline{\beta}| \leq 1$,  $|\lambda_{i,x}| \leq 1$, $1-\pi_x \leq 1$ and $ |\psi_{xx}| \leq 1$. 

  \subsubsection{Proof of Lemma \ref{lem:e_value_bd_x}}  \label{subsec:e_value_bd_x_Prf}
To bound $\lambda_{i}^{\sim x},$ we first bound $\Delta_x$ using the following lemma. 
\begin{lemma}
For a rank-2 diagonalizable t.p.m $P$ with stationary distribution $\bpi,$ and spectral gap $\beta,$ 
\begin{enumerate}
    \item  
    \begin{equation}
       \Delta_x \in  [ -( \beta(1-\pi_x) + P_{xx}) , \ ( \beta(1-\pi_x) + P_{xx})   ],  \text{ if } \beta \in [0,1] \label{eq:delta_bd_1}
    \end{equation}
    
    \item   
    \begin{equation}
        \Delta_x \in  [ -( \beta + (1-\beta) v_x u_x) , \ ( \beta + (1-\beta) v_x u_x)   ],  \text{ if } \beta \in [1,2] \label{eq:delta_bd_2}
    \end{equation}
\end{enumerate} 
\label{lem:Delta_x_bound}
\end{lemma}
\begin{proof}


To prove \eqref{eq:delta_bd_1}, we first start with the expression for $\Delta_x^2$ and bound it in the following way:
\begin{align} 
       \Delta_x^2 \ & = \ [ (1-\pi_x ) - (1-\beta) (1-v_x u_x) ]^2 + 4 (1-\beta) \pi_x v_x u_x \nonumber \\ 
       & = \ [ (1-\pi_x ) + (1-\beta) (1-v_x u_x) ]^2 - 4 (1-\beta) (1 -\pi_x - v_x u_x) \nonumber \\ 
       & \overset{(a)}{=} \ [ \beta + (1-\beta)\pi_x + (1-\beta) v_x u_x]^2 + \pi_x^2 (\beta^2 + 2\beta(1-\beta) ) + 2\beta \pi_x[  ( \beta - 2) +  (1-\beta)  v_x u_x] \nonumber \\
       & \overset{(b)}{\leq}  \ [ \beta(1-\pi_x ) + P_{xx}]^2 + \pi_x^2 (\beta^2 + 2\beta(1-\beta) ) + 2\beta \pi_x ( \beta - 2) + 2 \beta  \pi_x (1 - \pi_x) \nonumber \\ 
       & \ {=} \ [ \beta(1-\pi_x ) + P_{xx}]^2  - \beta^2 \pi_x^2 + 2\beta \pi_x ( \beta - 1) \nonumber \\
       & \overset{(c)}{\leq } \  [ \beta(1-\pi_x) + P_{xx}]^2 \nonumber 
\end{align}
where $(a)$ follows by using $[ (1-\pi_x ) + (1-\beta) (1-v_x u_x) ]^2 = [\beta + (1-\beta)\pi_x + (1-\beta) v_x u_x + \beta \pi_x - 2 ]^2 $ and simplyfying, $(b)$ follows by using $P_{xx} = \pi_x + (1-\beta) v_x u_x,  (1-\beta) v_x u_x \leq 1-\pi_x$ from \eqref{eq:nonnegPdiag}  and  $(c)$ follows from $\beta \leq 1.$ Since $ \beta(1-\pi_x) + P_{xx} \geq 0,$ we get  $ \Delta_x \  \in \ [ -( \beta(1-\pi_x) + P_{xx}) , ( \beta(1-\pi_x) + P_{xx})  ].$ This completes the proof of Part 1 of Lemma \ref{lem:Delta_x_bound}.

To prove \eqref{eq:delta_bd_2}, we again start with the expression for $\Delta_x^2$ but  bound it in a different way as shown below:
\begin{align} 
       \Delta_x^2 \ & = \ [ (1-\pi_x ) - (1-\beta) (1-v_x u_x) ]^2 + 4 (1-\beta) \pi_x v_x u_x \nonumber \\ 
       & = \ [ \beta + (1-\beta) v_x u_x -\pi_x ]^2 + 4 (1-\beta) \pi_x v_x u_x \nonumber \\  
       & = \ [ \beta + (1-\beta) v_x u_x]^2 -2\beta \pi_x +\pi_x^2 + 2 (1-\beta) \pi_x v_x u_x \nonumber \\ & \overset{(a)}{\leq} \  [ \beta + (1-\beta) v_x u_x]^2 -2\beta \pi_x +\pi_x^2 + 2 \pi_x (1-\pi_x)  \nonumber \\ 
       & = \ [ \beta + (1-\beta) v_x u_x]^2 + 2\pi_x (1-\beta) - \pi_x^2 \nonumber \\ 
        & \overset{(b)}{\leq} \ [ \beta + (1-\beta) v_x u_x]^2  \nonumber 
      \end{align}
where $(a)$ follows by using  $  (1-\beta) v_x u_x \leq 1 -\pi_x $ from \eqref{eq:nonnegPdiag} and $(b)$ follows from $\beta \geq 1.$ 
Since $ \beta + (1-\beta) v_x u_x$ is positive for $\beta \geq 1,$ we get  $ \Delta_x \  \in \ [ -( \beta + (1-\beta) v_x u_x) , ( \beta + (1-\beta) v_x u_x)   ].$
This completes the proof of Part 2 of Lemma \ref{lem:Delta_x_bound}.

\end{proof}

Using \eqref{eq:delta_bd_1} and \eqref{eq:delta_bd_2} in the expression $\lambda_{i}^{\sim x} = \frac{1}{2} \left( (1-\pi_x) + (1-\beta) (1-v_x u_x) + (-1)^{i+1} \Delta_x \right), \ i = 1,2,$ we get $ \lambda_{i}^{\sim x} \leq 1-0.5 \beta \pi_x ,  \beta \in [0,1]$ and $ \lambda_{i}^{\sim x} \leq 1-0.5  \pi_x ,  \beta \in [1,2]$ for $i = 1,2.$
Using $\frac{\beta^2}{\beta + 2} \leq \beta$ for $ \beta \in [0,1]$ and $ \beta^2 \leq \beta + 2 $  
for $ \beta \in [1,2]$
completes the proof of Lemma \ref{lem:e_value_bd_x}.

\subsection{Proof of Lemmas \ref{lem: P(N_x=0)_bd}, \ref{lem:exp_prob} } 
\label{subsec:P(N_x=0)_bd_Prf}
To prove lemmas \ref{lem: P(N_x=0)_bd}, \ref{lem:exp_prob}, we use the following expressions for $\pr(X_1 = x,N_x(X_2^m) = 0)$ and $\pr(N_x(X^n) = 0)$ in the lemma below. 

\begin{lemma}
For a rank-2 diagonalizable t.p.m $P$ with spectral gap $\beta \in (0,2]$, $m=1,\ldots,n$,  

\begin{align}
\Pr(X_1 = x,N_x(X_2^m) = 0) \ & = \  \begin{cases}
\pi_x (1-\pi_x)^{m-1}, \ \text{ if }  \pi_x = P_{xx}, \\
\frac{1}{2 } \ \pi_x \ \bigg[ \left( \lambda_{1}^{\sim x}  \right)^{m-1} + \left( \lambda_{2}^{\sim x}  \right)^{m-1} \bigg] \\ \ + \pi_x \ \Big[\frac{s_x}{2} - \overline{\beta} v_x u_x \Big] \bigg[ \sum_{l=0}^{m-2} ( \lambda_{1}^{\sim x} )^{m-2-l} \  ( \lambda_{2}^{\sim x} )^l  \bigg], \text{ otherwise } \label{eq:P(N_x = 1)_expr_no_Delta} 
\end{cases} \\ 
\pr(N_x(X^n) = 0) & = \  \begin{cases}
(1-\pi_x)^n, \ \text{ if }  \pi_x = P_{xx}, \\
\frac{1}{2 } \bigg[ \left( \lambda_{1}^{\sim x}  \right)^n + \left( \lambda_{2}^{\sim x}  \right)^n \bigg] \\ 
\ + \frac{s_x}{2} \bigg[ \sum_{l=0}^{n-1} ( \lambda_{1}^{\sim x} )^{n-1-l} \  ( \lambda_{2}^{\sim x} )^l  \bigg], \text{ otherwise } \label{eq:P(N_x = 0)_expr_no_Delta} 
\end{cases} 
\end{align}
\label{lem:P(N_x=0,1)}
\end{lemma}
\begin{proof}
Section \ref{subsec:P(N_x=0,1)_Prf }. 
\end{proof}

\noindent\textbf{Claim 2:} $|s_x| \leq 3$
\begin{proof}
Using  $| \overline{\beta} | \leq 1$, $\pi_x \leq 1,$ and $\overline{\beta}v_xu_x\le1-\pi_x$ from \eqref{eq:nonnegPdiag}  in $s_x = \overline{\pi}_x - \overline{\beta}  +  \overline{\beta} v_x u_x,$ we get $|s_x| \leq 3.$
\end{proof}

\emph{Proof of Lemma \ref{lem: P(N_x=0)_bd}}:\\ 
Since the claim 1 (in section \ref{sec:x_Bias_Prf}) implies $\lambda_1^{\sim x} - \lambda_2^{\sim x} = \Delta_x > 0$ for $x \in A(\delta),$ $\beta \in (0,2],$ we get 
\begin{equation}
  \sum_{l=0}^{n-1} ( \lambda_{1}^{\sim x} )^{n-1-l} \  ( \lambda_{2}^{\sim x} )^l \ = \   \frac{1}{ \Delta_x } \bigg[  \left( \lambda_{1}^{\sim x}  \right)^n - \left( \lambda_{2}^{\sim x}  \right)^n     \bigg], x \in A(\delta) \label{eq:P(N_x = 0)_expr}  
\end{equation}
Substituting \eqref{eq:P(N_x = 0)_expr} in \eqref{eq:P(N_x = 0)_expr_no_Delta} and applying triangle inequality on the absolute value of the R.H.S, we get 
\begin{align}
   Q_x^n(0) \ & \leq \ \frac{1}{2 } \left( 1 + \frac{|s_x|}{\Delta_x} \right) \bigg[ \left( |\lambda_{1}^{\sim x}|  \right)^n + \left( |\lambda_{2}^{\sim x}|  \right)^n \bigg] \nonumber \\
     & \overset{(a)}{\leq} \ \left( 1 + \frac{|s_x|}{\Delta_x} \right) \ \left(1 - \frac{\beta^2}{2(\beta+2)}\pi_x \right)^n \
     \overset{(b)}{\leq} \ \left( 1 + \frac{9}{\beta} \right) \ \left(1 - \frac{\beta^2}{2(\beta+2)}\pi_x \right)^n  \label{eq:P(N_x=0)_bd1_Prf}
\end{align}
where we use \eqref{eq:e_value_bd_x} to get $(a)$ and the claims 1 and 2 above to get $(b).$ \\ 
This completes the proof of Lemma \ref{lem: P(N_x=0)_bd}, part 1. 

For $x \in \overline{A}(\delta),$ using triangle inequality on the R.H.S of \eqref{eq:P(N_x = 0)_expr_no_Delta}, we get
\begin{align}
   Q_x^n(0) \ = \ \pr(N_x = 0) \ & \leq \  \frac{1}{2 } \bigg[ \left( |\lambda_{1}^{\sim x}|  \right)^n + \left( |\lambda_{2}^{\sim x}|  \right)^n \bigg] + \frac{|s_x|}{2} \bigg[ \sum_{l=0}^{n-1} ( |\lambda_{1}^{\sim x}| )^{n-1-l} \  ( |\lambda_{2}^{\sim x}| )^l  \bigg]  \nonumber \\ 
    & \overset{(a)}{\leq} \ \left(  e^{-0.5  \frac{\beta^2}{\beta + 2} \pi_x } + \frac{n|s_x|}{2} \right) \  e^{-0.5 (n-1) \frac{\beta^2}{\beta + 2} \pi_x }   \nonumber \\
    & \overset{(b)}{\leq} \ \left( 1 + \frac{3n}{2} \right) \ e^{-0.5 (n-1) \frac{\beta^2}{\beta + 2} \pi_x }   \nonumber \\
    & \overset{(c)}{\leq} \ \left( 1 + \frac{3n}{2} \right) \  e^{-0.5 (n-1) \frac{\beta^2}{\beta + 2} \delta }  \label{eq:P(N_x=0)_bd2_Prf} 
\end{align}
where we use \eqref{eq:e_value_bd_x} along with $1-z \leq e^{-z}$ to get $(a)$, $e^{-z} \leq 1$ and the claim 2 above, to get $(b),$ and $\pi_x \geq \delta$ for $x \in \overline{A}(\delta)$ to get $(c).$ This completes the proof of Lemma \ref{lem: P(N_x=0)_bd}, part 2. 

\emph{Proof of Lemma \ref{lem:exp_prob}}:\\ 
To prove \eqref{eq:Qx_1_A0} and \eqref{eq:Qx_1_bd} in lemma \ref{lem:exp_prob}, we use the following lemma.

\begin{lemma}\cite[Lemma 8]{Chandra20_ISIT}
\label{lem:swap}
For a stationary Markov chain $X^n$, $m = 2,\ldots, n$,
\begin{equation}
 \pr(X_m{=}x, N_x(X_1^{m-1})= 0) = \pr(X_1{=}x, N_x(X_2^{m})= 0 ) \label{eq:m_1_swap}
\end{equation}
\label{lem:m_1_swap}
\end{lemma}

Our next lemma provides an expression for $\pr(N_x(X^n) = 1)$ in terms of $\pr(X_1{=}x, N_x(X_2^{m})= 0)$ using \eqref{eq:m_1_swap}.

\begin{lemma}
   For a stationary Markov chain $X^n$,
   \begin{equation}
     \pr(N_x(X^n) = 1) \ = \ \pi_x^{-1} \  \sum_{m=1}^n \    \pr(X_1 = x,N_x(X_2^{m}) = 0 ) \
    \pr(X_1 = x, N_x(X_{2}^{n-m+1}) = 0 )  \label{eq:Qx_1_factored}
   \end{equation}
\end{lemma}

\begin{proof}

\begin{align}
    \pr(N_x(X^n) = 1) \ & = \ \sum_{m=1}^n \ \Pr(X_m = x,N_x(X_1^{m-1}) = N_x(X_{m+1}^{n}) = 0) \nonumber \\ 
    & \overset{(a)}{=} \ \sum_{m=1}^n \ \pr(X_m = x,N_x(X_1^{m-1}) = 0 ) \ \pr(N_x(X_{m+1}^{n}) = 0 | X_m = x) \nonumber \\
    & \overset{(b)}{=} \ \pi_x^{-1} \ \sum_{m=1}^n \  \pr(X_1 = x,N_x(X_2^{m}) = 0 ) \
    \pr(X_1 = x, N_x(X_{2}^{n-m+1}) = 0 )  \nonumber \\ 
\end{align}
where we get $(a)$ by using the Markov property, $(b)$ using lemma \ref{lem:m_1_swap} and noting that $\pr(N_x(X_{m+1}^{n}) = 0 | X_m = x) = \pi_x^{-1} \ \pr(X_1 = x, N_x(X_{2}^{n-m+1}) = 0)$. 
\end{proof}

Using \eqref{eq:P(N_x = 1)_expr_no_Delta} with $\pi_x = P_{xx}$ (i.e. for $x$ in $A_0$) in \eqref{eq:Qx_1_factored}, we get \eqref{eq:Qx_1_A0}. 

To bound $\pr(N_x(X^n) = 1)$ for $x$ in $\overline{A}(\delta)$, we use the following claim. 

\noindent\textbf{Claim 3:} For $x$ in $\overline{A}$, \ $m = 1,\ldots,n$, \\ 
$\Pr(X_1 = x,N_x(X_2^m) = 0) \leq \pi_x \ (1+(5/2) \ (m-1)) \ \exp{-(m-2) \ c_{\beta} \ \delta/2}$. 
\begin{proof}
 To get the above bound, we bound \eqref{eq:P(N_x = 1)_expr_no_Delta} using $|s_x| \leq 3$ (claim 2 above), $|\overline{\beta}v_x u_x| \leq 1$, and $|\lambda_{i,x}| \leq 1-c_{\beta} \delta/2$ (from \eqref{eq:e_val_bd1_x} along with $1-t \leq e^{-t}$. 
\end{proof}

Using the above claim in \eqref{eq:Qx_1_factored}, we get 
\begin{align}
  \pr(N_x(X^n) = 1)  & {\leq} \ \sum_{m=1}^n \ \pi_x \  (1+(5/2) \ (m-1)) \ \exp{-(m-2) \ c_{\beta} \ \delta/2} \nonumber \\ 
    & \qquad \qquad (1+(5/2) \ (n-m)) \ \exp{-(n-m-1) \ c_{\beta} \ \delta/2} \nonumber \\ 
    & = \ n \ \pi_x \ g(n) \  \exp{-(n-3) \ c_{\beta} \ \delta/2} \nonumber 
 \end{align}
 where $g(n) = [1 + (5/2) \  (n-1) \ (1  + (5/12)  \ ( n-2)) ]$. This completes the proof of \eqref{eq:Qx_1_bd} in lemma \ref{lem:exp_prob}. 

The proof of \eqref{eq:Qx_0_bd} in lemma \ref{lem:exp_prob} is direct by bounding \eqref{eq:P(N_x = 0)_expr_no_Delta}  using $|s_x| \leq 3$ (claim 2 above) and $|\lambda_{i,x}| \leq 1-c_{\beta} \delta/2$ (from \eqref{eq:e_val_bd1_x} along with $1-t \leq e^{-t}$. This completes the proof of lemma \ref{lem:exp_prob}.

\subsubsection{Proof of Lemma \ref{lem:P(N_x=0,1)} } 
\label{subsec:P(N_x=0,1)_Prf }
To prove Lemma \ref{lem:P(N_x=0,1)}, we consider $P^{\sim x},$ the matrix obtained by setting the $x$-th column in the rank-2 diagonalizable t.p.m $P$ to zeros. Since $P = RDS,$ we have $P^{\sim x} = RDS^{\sim x}$ where $S^{\sim x}$ is obtained by setting the $x$-th column in $S$ to zeros. $\bpi^{\sim x}$ is the vector obtained by setting the $x$-entry in $\bpi$ to 0. 
$Q_x^n(0)$ can be written as 
\begin{align}
    Q_x^n(0) \ & \overset{(a)}{=} \ \bpi^{\sim x} \ (P^{\sim x})^{n-1} \ \one \
     = \ (\bpi^{\sim x}) \ R \ D \ (S^{\sim x} R D)^{n-2} \ S^{\sim x} \ \one \nonumber \\ 
     & \qquad \qquad \qquad \qquad \overset{(b)}{=} \ \begin{bmatrix}
    1-\pi_x & -(1-\beta)\pi_x v_x
    \end{bmatrix} \ (S^{\sim x} R D)^{n-2} \ \begin{bmatrix}
    1 - \pi_x \\ -u_x
    \end{bmatrix},  \label{eq:P(Fx=0)_int}
\end{align}
where we get $(a)$ by noting that the entry of the ($K \times 1$) vector $(P^{\sim x})^{n-1} \ \one $ corresponding to any state $z \in \cX$ is the probability of not passing through the state $x$ in the next $n-1$ steps, given the present state is $z,$   $(b)$ by using $\bpi \one = 1, \bpi^ v = 0, $ and $u \one = 0.$
Similarly, 
\begin{align}
  \Pr(X_1 = x,N_x(X_2^m) = 0) \  = \ \pi_x \ e_x \ (P^{\sim x})^{m-1} \ \one \ & = \ \pi_x \ e_x \ R \ D \ (S^{\sim x} R D)^{m-2} \ S^{\sim x} \ \one \nonumber \\
  & = \ \pi_x \ \begin{bmatrix}
    1 & \overline{\beta}v_x
    \end{bmatrix}  \ (S^{\sim x} R D)^{m-2} \ \begin{bmatrix}
    1 - \pi_x \\ -u_x
    \end{bmatrix},  \label{eq:P(Fx=1)_int}
\end{align}
where $e_x$ is a $1 \times K$ vector with $x$-th entry as 1 and all other entries as 0. 

The following lemma gives a closed form expression for $(S^{\sim x} R D)^{l} \ ( l \geq 1 ) $
for various values of $\pi_x, v_x$ and $u_x.$

 \begin{lemma} $ (S^{\sim x}RD)^l:$
  
  \begin{enumerate}
      \item For $x \in \cX,$ with $v_x = u_x = 0,$ $ (S^{\sim x} R   D)^l 
      = \begin{bmatrix} \overline{\pi_x}^l & 0 \\ 0 &  \overline{\beta}^l \end{bmatrix}.$ \\
      
      \item For $x \in \cX,$ with $v_x = 0, u_x \neq 0,$ 
      \begin{enumerate}
          \item $\pi_x = \beta:$ 
           \begin{align}
       & (S^{\sim x} R   D)^l  = \begin{bmatrix} \overline{\pi_x}^l & 0 \\ -l u_x \overline{\pi_x}^{l-1} & \overline{\pi_x}^l \end{bmatrix}  \nonumber 
       \end{align}
          \item $\pi_x \neq \beta : $
      \begin{align}
        (S^{\sim x} R   D)^l \ & = \ \begin{bmatrix} 1 & 0 \\ \frac{u_x}{\pi_x - \beta} & 1 \end{bmatrix} \begin{bmatrix} \overline{\pi_x}^l & 0 \\ 0 & \overline{\beta}^l \end{bmatrix}  \begin{bmatrix} 1 & 0 \\ -\frac{u_x}{\pi_x - \beta} & 1 \end{bmatrix} \label{eq:diag_vx_0}
       \end{align}
       \end{enumerate}

      \item For $x \in \cX,$ with  $v_x \neq 0, u_x = 0,$
      \begin{enumerate}
          \item $\pi_x = \beta:$ 
           \begin{align}
       & (S^{\sim x} R   D)^l \ = \ \begin{bmatrix} \overline{\pi_x}^l & -l  \pi_x v_x \overline{\pi_x}^l \\ 0 & \overline{\pi_x}^l \end{bmatrix}  \nonumber 
       \end{align}
      \item  $\pi_x \neq \beta: $
      \begin{align}
        & (S^{\sim x} R   D)^l  \ = \ \begin{bmatrix} 1 & -\frac{ \overline{\beta}\pi_x v_x  }{\pi_x - \beta} \\ 0 & 1 \end{bmatrix} 
       \begin{bmatrix} \overline{\pi_x}^l & 0 \\ 0 & \overline{\beta}^l \end{bmatrix}
       \begin{bmatrix} 1 & \frac{ \overline{\beta} \pi_x v_x }{\pi_x - \beta} \\ 0 & 1 \end{bmatrix} \label{eq:diag_ux_0}
       \end{align}
        \end{enumerate}
       
       \item For $x \in \cX,$ with $v_x \neq 0, u_x \neq 0,$
       \begin{align}
        (S^{\sim x} R   D)^l \ & = \ V^{\sim x} \begin{bmatrix} (\lambda_1^{\sim x})^l & 0 \\ 0 & (\lambda_2^{\sim x})^l  \end{bmatrix} (V^{\sim x})^{-1} \label{eq:diag_vx_ux_nz}
      \\
       \text{ where } 
       V^{\sim x} \ & = \ \begin{bmatrix}
       1 & 1 \\ -\frac{\Delta_x - s_x}{ 2(1-\beta) \pi_x v_x  } & \frac{\Delta_x +s_x}{ 2 (1-\beta) \pi_x v_x  } \end{bmatrix}, \nonumber
       \end{align} $s_x = 1-\pi_x - (1-\beta)(1-v_x u_x),$ 
       $\lambda_1^{\sim x},\lambda_2^{\sim x}$ and $\Delta_x$ are as defined in Lemma \ref{lem:P(N_x=0,1)}.
       \label{lem_part: v_x_u_x_nz}
       \end{enumerate}
       \label{lem:Diagonalization_x}
        \end{lemma} 
        \begin{proof}
        Since $SR = I,$ we have $S^{\sim x} R = \begin{bmatrix}
\overline{\pi_x} & -\pi_x v_x \\ -u_x & 1-v_x u_x
\end{bmatrix}$ and 
\begin{equation}
    S^{\sim x} RD = \begin{bmatrix}
\overline{\pi_x} & -\overline{\beta}\pi_x v_x \\ -u_x & \overline{\beta} (1-v_x u_x)  
\end{bmatrix} \label{eq:SRD_x}
\end{equation}

\begin{enumerate}
    \item We substitute $v_x = u_x = 0$ in \eqref{eq:SRD_x} and raise the power on both sides to $l$.
    
    \item 
    \begin{enumerate}
        \item Substituting $v_x =0$, $\beta = \pi_x$ in \eqref{eq:SRD_x}, we get $ S^{\sim x} RD = \begin{bmatrix}
\overline{\pi_x} & 0 \\ -u_x & \overline{\pi_x}
\end{bmatrix}.$ Using induction on the exponent $l,$ we get  
 $ (S^{\sim x} R   D)^l  = \begin{bmatrix} \overline{\pi_x}^l & 0 \\ -l u_x \overline{\pi_x}^{l-1} & \overline{\pi_x}^l \end{bmatrix}.$
 
 \item Substituting $v_x =0$ in \eqref{eq:SRD_x}, we get $ S^{\sim x} RD = \begin{bmatrix}
\overline{\pi_x} & 0 \\ -u_x & \overline{\beta}
\end{bmatrix}.$ We observe that $\overline{\pi_x}, \overline{\beta} $ are the eigenvalues of $ S^{\sim x} RD$ with  $ \begin{bmatrix}
1  \\ \frac{u_x}{\pi_x - \beta} \end{bmatrix} $ and   $ \begin{bmatrix}
0  \\ 1 \end{bmatrix} $ as their respective right eigenvectors resulting in the diagonalised form in \eqref{eq:diag_vx_0}.
 \end{enumerate}
 
 \item 
 \begin{enumerate}
        \item Substituting $u_x =0$, $\beta = \pi_x$ in \eqref{eq:SRD_x}, we get $ S^{\sim x} RD = \begin{bmatrix}
\overline{\pi_x} & -\overline{\pi_x} \pi_x v_x \\ 0 & \overline{\pi_x}
\end{bmatrix}.$ Using induction on the exponent $l,$ we get  
 $ (S^{\sim x} R   D)^l  = \begin{bmatrix} \overline{\pi_x}^l & -l\pi_x v_x \overline{\pi_x}^{l}  \\ 0 & \overline{\pi_x}^l \end{bmatrix}.$
 
 \item Substituting $u_x =0$ in \eqref{eq:SRD_x}, we get $ S^{\sim x} RD = \begin{bmatrix}
\overline{\pi_x} & -\overline{\beta}\pi_x v_x \\ 0 & \overline{\beta}
\end{bmatrix}.$ We observe that $\overline{\pi_x}, \overline{\beta} $ are the eigenvalues of $ S^{\sim x} RD$ with  $ \begin{bmatrix}
1  \\ 0 \end{bmatrix} $ and   $ \begin{bmatrix}
-\frac{ \overline{\beta} \pi_x v_x }{\pi_x - \beta}  \\ 1 \end{bmatrix} $ as their respective right eigenvectors resulting in the diagonalised form in \eqref{eq:diag_ux_0}.
 \end{enumerate}
 
 \item Solving det$(S^{\sim x} RD - \lambda I) = 0,$ we get $\lambda_{i}^{\sim x} = \ 0.5\big( \overline{\pi}_x + \overline{\beta} (1-v_x u_x) + (-1)^{i+1} \Delta_x  \big)$, $i = 1,2$, as the eigenvalues of $S^{\sim x} RD$ with $  \ \begin{bmatrix}
       1 \\ -\frac{\Delta_x - s_x}{ 2(1-\beta) \pi_x v_x  } \end{bmatrix} , \begin{bmatrix} 1 \\  \frac{\Delta_x +s_x}{ 2 (1-\beta) \pi_x v_x  } \end{bmatrix} $ as their respective right eigenvectors, where  $\Delta_x^2  = s_x^2 + 4 \overline{\beta}\pi_x v_x u_x, s_x = 1-\pi_x - (1-\beta)(1-v_x u_x),$  resulting in the diagonalised form in \eqref{eq:diag_vx_ux_nz}.
 
\end{enumerate}
        \end{proof}
For the states $x \in \cX$ that fall in any of the first three categories in Lemma \ref{lem:Diagonalization_x} (which makes $P_{xx} = \pi_x$), substituting the corresponding form of $(S^{\sim x} R D)^{n-2}$ in \eqref{eq:P(Fx=0)_int} and \eqref{eq:P(Fx=1)_int} gives $Q_x^n(0) = (1-\pi_x)^n$, $\pr(X_1 =x, N_x(X_2^m) = 0) = \pi_x \ (1-\pi_x)^{m-1}$. For states $x$, with $\pi_x \neq P_{xx},$ substituting the diagonalized form of $(S^{\sim x} R D)^{n-2}$ from \eqref{eq:diag_vx_ux_nz} into \eqref{eq:P(Fx=0)_int}, \eqref{eq:P(Fx=1)_int} and simplifying, we get \eqref{eq:P(N_x = 0)_expr_no_Delta} and \eqref{eq:P(N_x = 1)_expr_no_Delta} respectively. This completes the proof of Lemma \ref{lem:P(N_x=0,1)}. 

\subsection{Proof of Lemma \ref{lem: Txy_bd }} 
\label{subsec:Txy_bd_Prf}
To prove Lemma \ref{lem: Txy_bd }, we begin with the following expression for $T^n_{xy}.$

\begin{lemma}
 For a rank-2 diagonalizable t.p.m $P$ with spectral gap $\beta,$ 
\begin{align}
    T^n_{xy} \ & = \ \frac{1}{n(n-1)}  \sum_{m_1 = 1}^{n-1} \sum_{m_2 = m_1 + 1}^{n}  \bpi^{\sim x,y}  R \  D \ M_{ m_1 - 2 } \left[ L_x \ M_{m_2 - m_1 -1} \ L_y \right. \nonumber \\ 
    & \qquad \qquad \qquad \qquad \qquad \qquad \qquad \qquad \qquad \qquad \qquad \left. + L_y \ M_{m_2 - m_1 -1} \ L_x  \right]   M_{1,n-m_2}  
    \label{eq:T_xy_expr}  
\end{align}
with $\bpi^{\sim x,y}  =  \bpi - \pi_x e_x - \pi_y e_y,$ ($e_x, e_y$ respectively being the $1 \times K$ vectors with 1 only in the entries corresponding to $x,y$ and zeros elsewhere), and 
\begin{align}
  L_x \ & = \ \pi_x H_{x,y} -  \begin{bmatrix}
  0 & \overline{\beta} \pi_x v_x \\ u_x & \overline{\beta} v_x u_x 
  \end{bmatrix}, \
  L_y \  = \  \pi_y H_{x,y} -  \begin{bmatrix}
  0 & \overline{\beta} \pi_y v_y \\ u_y & \overline{\beta} v_y u_y 
  \end{bmatrix}, \nonumber \\ 
 H_{x,y} \ & = \  \begin{bmatrix} -( \pi_x + \pi_y) &  -\overline{\beta}  (\pi_x v_x + \pi_y v_y ) \\ - (u_x+u_y)  & \overline{\beta}  ( 1- v_x u_x - v_y u_y)  \end{bmatrix},   \nonumber \\
  M_{l} \ & = \ [ M_{1,l} \ M_{2,l} ], \  l \in \{0,1,2, \ldots\}, \nonumber \\ 
   M_{1,l} \ & = \ 
  \begin{bmatrix}
  \frac{1}{2} & \frac{s_{x,y}}{2}  \\ 
  0 &  -(u_x + u_y)
  \end{bmatrix} \ 
  W_l, \ 
  M_{2,l} \  = \ \begin{bmatrix}
  0 & -\overline{\beta}(\pi_x v_x + \pi_y v_y)   \\ 
  \frac{1}{2} &  - \frac{s_{x,y}}{2 } 
  \end{bmatrix} \ W_l, \text{ and } 
  \nonumber \\
  W_l \ & = \ \begin{bmatrix}
   \left( \lambda_{1}^{\sim x,y}  \right)^l + \left( \lambda_{2}^{\sim x,y}  \right)^l  \\ 
   \sum_{m=0}^{l-1} \ ( \lambda_{1}^{\sim x,y} )^{l-1-m} \  ( \lambda_{2}^{\sim x,y} )^m  
  \end{bmatrix}, \nonumber \end{align}
  where 
   $\lambda_{i}^{\sim x,y } \  = \ 0.5\big( (1-\pi_x-\pi_y) + \overline{\beta} \ (1-v_x u_x -v_y u_y) \pm \Delta_{x,y}  \big), i = 1,2,$ are the eigenvalues of the matrix $ \left( I_{2 \times 2} - \begin{bmatrix} \pi_x & \pi_y \\ u_x & u_y \end{bmatrix} \begin{bmatrix} 1 & v_x \\ 1 & v_y  \end{bmatrix} \right) D ,$
    $\Delta_{x,y}^2  \triangleq \  s_{x,y}^2 + 4 \ \overline{\beta} \ (\pi_x v_x + \pi_y v_y) \ (u_x + u_y),$ and $ s_{x,y}  \triangleq  (\beta -\pi_x -\pi_y) + \overline{\beta}  (v_x u_x + v_y u_y).$
    \label{lem:Txy_expr}
    \end{lemma}
\begin{proof}
Section \ref{subsec:Txy_expr_Prf}.
\end{proof}

Our next lemma bounds the eigenvalues $\lambda^{\sim x,y}_{1}, \lambda^{\sim x,y}_{2}$ in absolute value using the spectral gap $\beta$ and $\pi_x + \pi_y.$
\begin{lemma}
 For a rank-2 diagonalizable t.p.m $P$ with spectral gap $\beta,$ 
\begin{align}
    |\lambda_i^{\sim x,y}|  \ & \leq \ 1 -   \frac{\beta^2}{2(\beta + 2)} (\pi_x + \pi_y), \ i = 1,2.  \label{eq:e_value_bd_xy}
\end{align}
\label{lem:e_value_bd_xy}
\end{lemma}
\begin{proof}
Section \ref{subsec:e_value_bd_xy_Prf}.
\end{proof}

As described in section \ref{subsec:MSE_term_bd_Prf}, we divide $\{ (x,y) : x\in \cX, y \in \cX, x \neq y \}$ into two sets: 
\begin{align}
    B(\delta) \ & \triangleq \  \{ (x,y) : x,y \in \cX, \ x \neq y, \ \pi_x + \pi_y < \delta \} \nonumber \\
    \overline{B}(\delta) \ & \triangleq \ \{ (x,y) : x,y \in \cX, \  x \neq y, \ \pi_x + \pi_y \geq \delta \}  \nonumber
\end{align}
with $0 \leq \delta < \beta/5.$ \\
Note that $ \lambda_1^{\sim x,y} - \lambda_2^{\sim x,y} = \Delta_{x,y}.$ We next claim that $\Delta_{x,y}$ is bounded away from 0 for $(x,y) \in B(\delta).$ \\
\textbf{Claim 4:} $\Delta_{x,y} \ \geq \ \beta/3,$ for $(x,y) \in B(\delta).$ 
\begin{proof}
Similar to Claim 1.
\end{proof}

\textbf{Claim 5:} $|s_{x,y}| \leq 4.$ 
\begin{proof}
Similar to Claim 2.
\end{proof}

\textbf{Proof of Lemma \ref{lem: Txy_bd }, Part 1}: Recall that $\psi_{wz} \triangleq (1-\beta)v_w u_z, $ for $w,z \in \cX.$ The matrix product $\bpi^{\sim x,y}  R \  D \ M_{ m_1 - 2 } \left[ L_x \ M_{m_2 - m_1 -1} \ L_y + L_y \ M_{m_2 - m_1 -1} \ L_x  \right]   M_{1,n-m_2},$ in the expression for $T_{xy}^n$ in \eqref{eq:T_xy_expr}, can be written as a summation of (a constant number, not growing with $n,$ of) terms of the form:
 \begin{enumerate}
     \item 
     \begin{equation}
         \zeta_{xy} \triangleq \pi_x^{a_1} \ \pi_y^{a_2} \ \psi_{xx}^{b_1} \  \psi_{yx}^{c_1} \  \psi_{xy}^{b_2} \ \psi_{yy}^{c_2} \ \frac{(\Delta_{x,y} \pm s_{x,y})^d}{ \Delta_{x,y}^3} \ \left( \lambda_j^{\sim x,y} \right)^{n-2}, \ j = 1,2, \label{eq: Txy_general_term_1}
     \end{equation}
    where (i) $a_i , b_i , c_i \in \{0, 1,2,3, \ldots \},$ and $a_i + b_i +c_i \geq 1,$ for $i =1,2,$ (ii) $a_i + b_i + c_i > 1$ and $a_i \geq 1,$ for atleast one of $i = 1,2,$ and (iii) $d \in \{ 0,1,2,3 \},$ and

    \item  
    \begin{equation}
        \Upsilon_{xy,m_1,m_2} \ \triangleq \ \pi_x^{a_1} \ \pi_y^{a_2} \ \psi_{xx}^{b_1} \  \psi_{yx}^{c_1} \  \psi_{xy}^{b_2} \ \psi_{yy}^{c_2} \ \frac{(\Delta_{x,y} \pm s_{x,y})^d}{ \Delta_{x,y}^3} \ \left( \lambda_1^{\sim x,y} \right)^{d_1} \ \left( \lambda_2^{\sim x,y} \right)^{d_2},   \label{eq: Txy_general_term_2}
    \end{equation}
    with (i) $a_i , b_i , c_i \in \{0, 1,2,3, \ldots \},$ and  $a_i + b_i +c_i \geq 1$ for $i =1,2,$ (ii) $d \in \{ 0,1,2,3 \},$ and (iii) $(d_1, d_2) \in \{ (m_2 - 2, n-m_2), (n-m_2, m_2 - 2), (n-m_1-1, m_1-1), (m_1-1, n-m_1-1), (n-m_2+m_1-1, m_2 - m_1 -1), ( m_2 - m_1 -1, n-m_2+m_1-1 )   \}.$
    
 \end{enumerate}
Therefore to prove $ \sum_{(x,y) \in {B}(\delta)} T^n_{xy} \ \leq \ O(1/n\beta^5),$ it is enough to show that \\ $ \sum_{(x,y) \in {B}(\delta)} \sum_{m_1 = 1}^{n-1} \sum_{m_2 = m_1 + 1}^{n} \zeta_{xy}$,  $ \sum_{(x,y) \in {B}(\delta)} \sum_{m_1 = 1}^{n-1} \sum_{m_2 = m_1 + 1}^{n} \Upsilon_{xy,m_1, m_2}$ are less than \\ $ n ( n-1) \  O(1/n\beta^5).$ \\ 
We first consider terms of the form $\zeta_{xy},$ in \eqref{eq: Txy_general_term_1}, with $a_1 + b_1 + c_1 > 1$ and $a_1 \geq 1.$ Summing up the absolute value of such a  $\zeta_{xy}$  over all $(x,y) \in {B}(\delta),$ we get 
\begin{align}
     & \sum_{(x,y) \in {B}(\delta)}   \  |\zeta_{xy}| \ \overset{(a)}{\leq} \ \sum_{(x,y) \in {B}(\delta)}  \pi_x^{a_1-1} \ \pi_y^{a_2} \ |\psi_{xx}^{b_1} \  \psi_{yx}^{c_1} \  \psi_{yx}^{b_2} \ \psi_{yy}^{c_2}| \ \frac{|(\Delta_{x,y} \pm s_{x,y})|^d}{ \Delta_{x,y}^3} \ \pi_x \ \left( |\lambda_j^{\sim x,y}| \right)^{n-2} \nonumber \\ 
     & \ \overset{(b)}{\leq} \  \sum_{(x,y) \in {B}(\delta)}  \pi_x^{a_1-1} \ \pi_y^{a_2} \ |\psi_{xx}^{b_1} \  \psi_{yx}^{c_1} \  \psi_{yx}^{b_2} \ \psi_{yy}^{c_2}| \ \frac{|(\Delta_{x,y} \pm s_{x,y})|^d}{ \Delta_{x,y}^3} \ \pi_x \ \left( 1- \frac{\beta^2}{2(\beta+2)}(\pi_x + \pi_y) \right)^{n-2} \nonumber \\ 
     & \ \overset{(c)}{\leq} \  6^d \left(\frac{3}{\beta}\right)^3 \sum_{(x,y) \in {B}(\delta)}  \pi_x^{a_1-1} \ \pi_y^{a_2} \ |\psi_{xx}^{b_1} \  \psi_{yx}^{c_1} \  \psi_{yx}^{b_2} \ \psi_{yy}^{c_2}| \  (\pi_x + \pi_y) \ \left( 1- \frac{\beta^2}{2(\beta+2)}(\pi_x + \pi_y) \right)^{n-2} \nonumber \\
     & \ \overset{(d)}{\leq} \  2^{d+1} \ 3^{d+3}\frac{(\beta+2)}{(n-1)\beta^5}  \sum_{(x,y) \in {B}(\delta)}  \pi_x^{a_1-1} \ \pi_y^{a_2} \ |\psi_{xx}^{b_1} \  \psi_{yx}^{c_1} \  \psi_{xy}^{b_2} \ \psi_{yy}^{c_2}| \nonumber \\ 
     & \ \overset{(e)}{\leq} \  2^{d+1} \ 3^{d+3} \frac{ (\beta+2)}{(n-1)\beta^5} \sum_{x \in \cX} \pi_x^{a_1-1} \  |\psi_{xx}^{b_1} \  \psi_{yx}^{c_1} | \sum_{y \in \cX} \pi_y^{a_2} \ |\psi_{xy}^{b_2} \ \psi_{yy}^{c_2}| \ \overset{(f)}{\leq} \ 2^{d+1} \ 3^{d+5} \frac{ (\beta+2)}{(n-1)\beta^5}, \label{eq: zeta_xy_bd} 
     \end{align}
where we get $(a)$ by using the assumption  $a_1 \geq 1,$ $(b)$ by using \eqref{eq:e_value_bd_xy}, $(c)$ by using $\Delta_{x,y} = \lambda_{1}^{\sim x,y} - \lambda_{2}^{\sim x,y} \leq 2,$ claims 4 and 5, and $\pi_x \leq \pi_x + \pi_y,$ $(d)$ by using $\max_{p \in [0,1]} p (1-cp)^{n-2} <  \frac{1}{c(n-1)},$ $(e)$ by using  $\sum_{(x,y) \in {B}(\delta)} |f(x)| \ |g(y)| \leq \sum_{(x,y) \in \cX^2} |f(x)| \ |g(y)| =  \sum_{x \in \cX} |f(x)| \sum_{y \in \cX} |g(y)|,$ and $(f)$ by using Lemma \ref{lem: sum_over_x} together with the assumptions $a_1 + b_1 + c_1 > 1$ and $a_2 + b_2 + c_2 \geq 1.$  \\

Using \eqref{eq: zeta_xy_bd}, we get  $ \sum_{(x,y) \in {B}(\delta)} \sum_{m_1 = 1}^{n-1} \sum_{m_2 = m_1 + 1}^{n} \zeta_{xy}=$  $  \sum_{m_1 = 1}^{n-1} \sum_{m_2 = m_1 + 1}^{n} \sum_{(x,y) \in {B}(\delta)} \zeta_{xy}$  $\leq 2^d \ 3^{d+5} n \frac{ (\beta+2)}{\beta^5},$ for $\zeta_{xy}$ with  $a_1 + b_1 + c_1 > 1$ and $a_1 \geq 1.$ \\  

Following a similar method, we can show that \\ $ \sum_{(x,y) \in {B}(\delta)} \sum_{m_1 = 1}^{n-1} \sum_{m_2 = m_1 + 1}^{n} |\zeta_{xy}| \leq 2^d \ 3^{d+5} n \frac{ (\beta+2)}{\beta^5},$ for $\zeta_{xy}$ with  $a_2 + b_2 + c_2 > 1$ and $a_2 \geq 1.$ \\

To bound $ \sum_{(x,y) \in {B}(\delta)} \sum_{m_1 = 1}^{n-1} \sum_{m_2 = m_1 + 1}^{n} \Upsilon_{xy,m_1,m_2},$ we use the expressions for \\ $ \sum_{m_1 = 1}^{n-1} \sum_{m_2 = m_1 + 1}^{n} \left( \lambda_1^{\sim x,y} \right)^{d_1} \ \left( \lambda_2^{\sim x,y} \right)^{d_2}  $ from the lemma below. 

\begin{lemma}
For a rank-2 diagonalizable t.p.m $P$ with spectral gap $\beta,$ 

\begin{enumerate}
    \item For  $(d_1, d_2) \in \{ (m_2 - 2, n-m_2), (n-m_1-1, m_1-1),  (n-m_2+m_1-1, m_2 - m_1 -1) \},$
    \begin{align}
         & \sum_{m_1 = 1}^{n-1} \sum_{m_2 = m_1 + 1}^{n} \left( \lambda_1^{\sim x,y} \right)^{d_1} \ \left( \lambda_2^{\sim x,y} \right)^{d_2}  \nonumber \\ 
          & = \ \frac{1}{\Delta_{x,y}} \left[ (n-1) \left( \lambda_1^{\sim x,y} \right)^{n-1} - \frac{\lambda_2^{\sim x,y}}{ \Delta_{x,y}} \left( \left( \lambda_1^{\sim x,y} \right)^{n-1} - \left(\lambda_2^{\sim x,y} \right)^{n-1} \right) \right], \label{eq: lambda_GP_sum_1} 
    \end{align}
    \item For  $(d_1, d_2) \in  \{ ( n-m_2, m_2 - 2 ), (m_1-1, n-m_1-1 ),  ( m_2 - m_1 -1, n-m_2+m_1-1 ) \},$
    \begin{align}
        & \sum_{m_1 = 1}^{n-1} \sum_{m_2 = m_1 + 1}^{n} \left( \lambda_1^{\sim x,y} \right)^{d_1} \ \left( \lambda_2^{\sim x,y} \right)^{d_2}  \nonumber \\ 
         & = \ \frac{1}{\Delta_{x,y}} \left[ -(n-1) \left( \lambda_2^{\sim x,y} \right)^{n-1} + \frac{\lambda_1^{\sim x,y}}{ \Delta_{x,y}} \left( \left( \lambda_1^{\sim x,y} \right)^{n-1} - \left(\lambda_2^{\sim x,y} \right)^{n-1} \right) \right]. \label{eq: lambda_GP_sum_2} 
    \end{align}
\end{enumerate}
\end{lemma}
\begin{proof}
The proof is direct using the formulae for the sums of a finite geometric series, and a finite arithmetico-geometric series.
\end{proof}
We first consider terms of the form $\Upsilon_{xy,m_1,m_2},$ in \eqref{eq: Txy_general_term_2}, with $(d_1, d_2) \in \{ (m_2 - 2, n-m_2), (n-m_1-1, m_1-1),  (n-m_2+m_1-1, m_2 - m_1 -1) \}.$ 
Summing up such a $\Upsilon_{xy,m_1,m_2}$ over $m_1, m_2,$ and all $(x,y) \in B(\delta),$ we get
\begin{align}
    & \sum_{(x,y) \in {B}(\delta)} \sum_{m_1 = 1}^{n-1} \sum_{m_2 = m_1 + 1}^{n} \Upsilon_{xy,m_1,m_2} \nonumber \\
    & \ \overset{(a)}{=} \ \sum_{(x,y) \in {B}(\delta)}  \pi_x^{a_1} \ \pi_y^{a_2} \ \psi_{xx}^{b_1} \  \psi_{yx}^{c_1} \  \psi_{xy}^{b_2} \ \psi_{yy}^{c_2} \ \frac{(\Delta_{x,y} \pm s_{x,y})^d}{ \Delta_{x,y}^3} \nonumber \\ 
    & \qquad \qquad \qquad \qquad \qquad  \frac{1}{\Delta_{x,y}} \left[ (n-1) \left( \lambda_1^{\sim x,y} \right)^{n-1} - \frac{\lambda_2^{\sim x,y}}{ \Delta_{x,y}} \left( \left( \lambda_1^{\sim x,y} \right)^{n-1} - \left(\lambda_2^{\sim x,y} \right)^{n-1} \right) \right] \nonumber \\
    & \ \overset{(b)}{\leq} \ \sum_{(x,y) \in {B}(\delta)}  \pi_x^{a_1} \ \pi_y^{a_2} \ |\psi_{xx}^{b_1} \  \psi_{yx}^{c_1}| \  |\psi_{xy}^{b_2} \ \psi_{yy}^{c_2}| \ \frac{(|\Delta_{x,y} \pm s_{x,y}|)^d}{ \Delta_{x,y}^4} \ \left[ (n-1) + \frac{2}{ \Delta_{x,y}} \right] \nonumber \\
    & \ \overset{(c)}{\leq} \ 6^d \left(\frac{3}{\beta}\right)^4  \ \left[ (n-1) + \frac{6}{ \beta} \right] \ \sum_{(x,y) \in {B}(\delta)}  \pi_x^{a_1} \ \pi_y^{a_2} \ |\psi_{xx}^{b_1} \  \psi_{yx}^{c_1}| \  |\psi_{xy}^{b_2} \ \psi_{yy}^{c_2}| \nonumber \\ 
    & \  \overset{(d)}{\leq} \ 6^d \left(\frac{3}{\beta}\right)^4  \ \left[ (n-1) + \frac{6}{ \beta} \right] \ \sum_{x \in \cX}  \pi_x^{a_1}  \ |\psi_{xx}^{b_1} \  \psi_{yx}^{c_1}| \sum_{y \in \cX} \pi_y^{a_2}  \  |\psi_{xy}^{b_2} \ \psi_{yy}^{c_2}| \nonumber \\ 
    & \  \overset{(e)}{\leq} \  2^d \ 3^{d+2} \left(\frac{3}{\beta}\right)^4  \ \left[ (n-1) + \frac{6}{ \beta} \right], \label{eq: Upsilon_xy_bd} 
\end{align}
where we get $(a)$ by using \eqref{eq: lambda_GP_sum_1}, $(b)$ by taking absolute value and using 
$|\lambda_j^{\sim x,y}| \leq 1, $ for $j = 1,2,$  $(c)$ by using $\Delta_{x,y} = \lambda_{1}^{\sim x,y} - \lambda_{2}^{\sim x,y} \leq 2,$ claims 4 and 5, $(d)$ by using  $\sum_{(x,y) \in {B}(\delta)} |f(x)| \ |g(y)| \leq \sum_{(x,y) \in \cX^2} |f(x)| \ |g(y)| =  \sum_{x \in \cX} |f(x)| \sum_{y \in \cX} |g(y)|,$ and $(e)$ by using Lemma \ref{lem: sum_over_x}.\\

Following a similar method, we can show that \\   $  \sum_{(x,y) \in {B}(\delta)} \sum_{m_1 = 1}^{n-1} \sum_{m_2 = m_1 + 1}^{n} \Upsilon_{xy,m_1,m_2}  \leq 2^d \ 3^{d+2} \left(\frac{3}{\beta}\right)^4 \  \left[ (n-1) +  \frac{6}{  \beta} \right],$ for $ \Upsilon_{xy,m_1,m_2} $ with \\ $(d_1, d_2)  \in  \{ ( n-m_2, m_2 - 2 ), (m_1-1, n-m_1-1 ),  ( m_2 - m_1 -1, n-m_2+m_1-1 ) \}.$ 
This completes the proof of Lemma \ref{lem: Txy_bd }, part 1. \\

\textbf{Proof of Lemma \ref{lem: Txy_bd }, Part 2}: The matrix product \\ $\bpi^{\sim x,y}  R \  D \ M_{ m_1 - 2 } \left[ L_x \ M_{m_2 - m_1 -1} \ L_y + L_y \ M_{m_2 - m_1 -1} \ L_x  \right]   M_{1,n-m_2},$ in the definition of $T_{xy}$ in \eqref{eq:T_xy_expr}, can also be written as a summation of (a constant number, not growing with $n,$ of) terms of the form, 
\begin{align}
     t_{xy,m_1,m_2}  & \triangleq  (\pi_x)^{a_1} \ (\pi_y)^{a_2} \ \psi_{xx}^{b_1}  \psi_{yx}^{c_1} \    \psi_{xy}^{b_2}  \psi_{yy}^{c_2} \ (s_{x,y})^{d_0} \  W_{m_1-1,d_1}   W_{m_2-m_1-1,d_2}  W_{n-m_2,d_3}, \label{eq: Txy_general_term_3}  
\end{align}
 where (i) $a_i , b_i , c_i \in \{0, 1,2,3, \ldots \},$ and  $a_i + b_i + c_i \geq 1,$ for $i = 1,2,$ (ii) $ d_0 \in \{0,1,2,3 \},$ $d_j \in \{1,2\}$ for $j = 1,2,$ and 3, and (iii)  $W_{t,1}, W_{t,2}$ are the first and second elements of the column vector $W_t$ defined in Lemma \ref{lem:Txy_expr}. Therefore to prove $ \sum_{(x,y) \in \overline{B}(\delta)} T_{xy}^n \ \leq \ O(n^3) e^{-\frac{\beta^2}{2(\beta+2)}(n-5)\delta},$ it is enough to show that $ \frac{1}{n(n-1)} \sum_{(x,y) \in \overline{B}(\delta)} \sum_{m_1 = 1}^{n-1} \sum_{m_2 = m_1 + 1}^{n} t_{xy,m_1,m_2} \ \leq \ O(n^3) e^{-\frac{\beta^2}{2(\beta+2)}(n-5)\delta}.$

Our next lemma bounds the absolute value of the product $W_{m_1-1,d_1} \  W_{m_2-m_1-1,d_2} \ W_{n-m_2,d_3}.$  

\begin{lemma}
\label{lem: W_bd} 
For a rank-2 diagonalizable t.p.m $P$ with spectral gap $\beta,$ 
\begin{enumerate}
    \item 
    \begin{equation}
       \sum_{m_1 = 1}^{n-1} \sum_{m_2 = m_1 + 1}^{n}  |W_{m_1-1,1} \  W_{m_2-m_1-1,1} \ W_{n-m_2,1}| \ \leq \ O(n^2) \ e^{ -(n-2) \frac{\beta^2}{2(\beta + 2)}(\pi_x + \pi_y)}, \label{eq: E_bd_n-2}  
    \end{equation} 
    
    \item 
    \begin{align}
       \begin{rcases} \sum_{m_1 = 1}^{n-1} \sum_{m_2 = m_1 + 1}^{n} |W_{m_1-1,1} \  W_{m_2-m_1-1,1} \ W_{n-m_2,2}|,  \\   \sum_{m_1 = 1}^{n-1} \sum_{m_2 = m_1 + 1}^{n} |W_{m_1-1,1} \  W_{m_2-m_1-1,2} \ W_{n-m_2,1}|,  \text{ and }  \\ 
        \sum_{m_1 = 1}^{n-1} \sum_{m_2 = m_1 + 1}^{n} |W_{m_1-1,2} \  W_{m_2-m_1-1,1} \ W_{n-m_2,1}| \end{rcases}
         &  \leq  O(n^3)  e^{ -(n-3) \frac{\beta^2}{2(\beta + 2)}(\pi_x + \pi_y)}, \label{eq: E_bd_n-3}  \end{align}
        
    \item 
    \begin{align}
       \begin{rcases} \sum_{m_1 = 1}^{n-1} \sum_{m_2 = m_1 + 1}^{n} |W_{m_1-1,1} \  W_{m_2-m_1-1,2} \ W_{n-m_2,2}|,  \\   \sum_{m_1 = 1}^{n-1} \sum_{m_2 = m_1 + 1}^{n} |W_{m_1-1,2} \  W_{m_2-m_1-1,1} \ W_{n-m_2,2}|,  \text{ and }  \\ 
        \sum_{m_1 = 1}^{n-1} \sum_{m_2 = m_1 + 1}^{n} |W_{m_1-1,2} \  W_{m_2-m_1-1,2} \ W_{n-m_2,1}| \end{rcases}
        & \leq  O(n^4) e^{ -(n-4) \frac{\beta^2}{2(\beta + 2)}(\pi_x + \pi_y)}, \label{eq: E_bd_n-4}  \end{align}

    \item \begin{align}
       \sum_{m_1 = 1}^{n-1} \sum_{m_2 = m_1 + 1}^{n} |W_{m_1-1,2} \  W_{m_2-m_1-1,2} \ W_{n-m_2,2}| \ &\leq \ O(n^5) \ e^{ -(n-5) \frac{\beta^2}{2(\beta + 2)}(\pi_x + \pi_y)}, \label{eq: E_bd_n-5}  
    \end{align}
\end{enumerate}
\end{lemma}
\begin{proof}
We first prove the bound on $\sum_{m_1 = 1}^{n-1} \sum_{m_2 = m_1 + 1}^{n} |W_{m_1-1,1} \  W_{m_2-m_1-1,2} \ W_{n-m_2,2}|.$
\begin{align}
    & \sum_{m_1 = 1}^{n-1} \sum_{m_2 = m_1 + 1}^{n} |W_{m_1-1,1} \  W_{m_2-m_1-1,2} \ W_{n-m_2,2}| \nonumber \\ 
    & \overset{(a)}{=} \sum_{m_1 = 1}^{n-1} \sum_{m_2 = m_1 + 1}^{n} \left| \left( \lambda_{1}^{\sim x,y}  \right)^{m_1 - 1} + \left( \lambda_{2}^{\sim x,y}  \right)^{m_1 - 1}  \right| \ \left|  \sum_{l=0}^{m_2-m_1-2} \ ( \lambda_{1}^{\sim x,y} )^{m_2-m_1-2-l} \  ( \lambda_{2}^{\sim x,y} )^l   \right| \nonumber \\ 
    & \qquad \qquad \qquad \qquad \qquad  \left|  \sum_{l=0}^{n-m_2-1} \ ( \lambda_{1}^{\sim x,y} )^{n-m_2-1-l} \  ( \lambda_{2}^{\sim x,y} )^l  \right| \nonumber \\ 
    & \overset{(b)}{\leq}  \sum_{m_1 = 1}^{n-1} \sum_{m_2 = m_1 + 1}^{n} 2 e^{ -(m_1-1) \frac{\beta^2}{2(\beta + 2)}(\pi_x + \pi_y)} \ e^{ -(m_2-m_1-2) \frac{\beta^2}{2(\beta + 2)}(\pi_x + \pi_y)} \ ( m_2-m_1-1 ) \nonumber \\ 
    & \qquad \qquad \qquad \qquad \qquad e^{ -(n-m_2-1) \frac{\beta^2}{2(\beta + 2)}(\pi_x + \pi_y)} \ (n-m_2) \nonumber \nonumber \\ 
    & {\leq} 2 \ e^{ -(n-4) \frac{\beta^2}{2(\beta + 2)}(\pi_x + \pi_y)} \ \sum_{m_1 = 1}^{n-1} \sum_{m_2 = m_1 + 1}^{n} ( m_2-m_1-1 ) \ (n-m_2) \nonumber \\
    & \overset{(c)}{\leq} O(n^4) \ e^{ -(n-4) \frac{\beta^2}{2(\beta + 2)}(\pi_x + \pi_y)}, \nonumber 
\end{align}
where we get $(a)$ by substituting the expressions for $W_{t,j},$ $j = 1,2$, $t = m_1-1, m_2-m_1-1$ and $ n-m_2,$  from Lemma \ref{lem:Txy_expr}, $(b)$ by using \eqref{eq:e_value_bd_xy}, and $(c)$ by using the standard formulae for the sums of integral powers of the first $m$ natural numbers, for appropriately chosen $m,$ and simplifying.
\end{proof}

Summing up the absolute value of the term $t_{xy,m_1,m_2}$ over $m_1,m_2,$ and over all $(x,y) \in \overline{B}(\delta),$ we get 
\begin{align}
 & \sum_{(x,y) \in \overline{B}(\delta)}  \sum_{m_1 = 1}^{n-1} \sum_{m_2 = m_1 + 1}^{n} \  |t_{xy,m_1,m_2}| \nonumber \\  
 & \  \overset{(a)}{\leq} \ \sum_{(x,y) \in \overline{B}(\delta)}   (\pi_x)^{a_1} (\pi_y)^{a_2} \ |\psi_{xx}^{b_1}| \ | \psi_{yx}^{c_1}|  \   |\psi_{xy}^{b_2}| \ |\psi_{yy}^{c_2}| \  |(s_{x,y})|^{d_0} \  O(n^5) \ e^{ -(n-5) \frac{\beta^2}{2(\beta + 2)}(\pi_x + \pi_y)}
\nonumber \\
& \  \overset{(b)}{\leq} \ 4^{d_0} \ O(n^5) \ e^{ -(n-5) \frac{\beta^2}{2(\beta + 2)}\delta} \sum_{(x,y) \in \overline{B}(\delta)}   \ (\pi_x)^{a_1} \ (\pi_y)^{a_2} \ |\psi_{xx}^{b_1}| \  |\psi_{yx}^{c_1}|  \   |\psi_{xy}^{b_2}| \  |\psi_{yy}^{c_2}| \nonumber \\
 & \  \overset{(c)}{=} O(n^5) \ e^{ -(n-5) \frac{\beta^2}{2(\beta + 2)}\delta} \ \left( \sum_{x \in \cX}   \ (\pi_x)^{a_1} \ |\psi_{xx}^{b_1}| \  |\psi_{yx}^{c_1}| \right) \ \left(\sum_{y \in \cX}  (\pi_y)^{a_2} \   |\psi_{xy}^{b_2}| \  |\psi_{yy}^{c_2}| \right) \nonumber \\
 &  \  \overset{(d)}{\leq} O(n^5) \ e^{ -(n-5) \frac{\beta^2}{2(\beta + 2)}\delta}, \label{eq: txy_m1_m2_bd}
\end{align}
where we get $(a)$ by using Lemma \ref{lem: W_bd}, $(b)$ by using $\pi_x + \pi_y \geq \delta$ for $(x,y) \in \overline{B}(\delta)$ and claim 5, $(c)$ by using $\sum_{(x,y) \in \overline{B}(\delta)} |f(x)| \ |g(y)| \leq \sum_{(x,y) \in \cX^2} |f(x)| \ |g(y)| =  \sum_{x \in \cX} |f(x)| \sum_{y \in \cX} |g(y)|,$ and $(d)$ by using Lemma \ref{lem: sum_over_x}.  \\
This implies $ \frac{1}{n(n-1)} \sum_{(x,y) \in \overline{B}(\delta)}  \sum_{m_1 = 1}^{n-1} \sum_{m_2 = m_1 + 1}^{n} \  |t_{xy,m_1,m_2}| \leq O(n^3) \ e^{ -(n-5) \frac{\beta^2}{2(\beta + 2)}\delta}.$ 
This completes the proof of Lemma \ref{lem: Txy_bd }, part 2.

\section{Appendix C} 

\subsection{Proof of Lemma \ref{lem: sum_over_x} } 
\label{subsec: sum_over_x_Prf}
\begin{enumerate}
    \item We first consider the case with $a \geq 1.$
    \begin{align}
        \sum_{x \in \cX} (\pi_x)^{a} \ |\psi_{xx}^{b}| \ |\psi_{yx}^{c}| \ & \overset{(f)}{\leq} \  \sum_{x \in \cX} (\pi_x)^{a} 
        \  \overset{(g)}{\leq} \  \sum_{x \in \cX} \pi_x \ = \ 1, \nonumber
    \end{align}
    where we get $(f)$ by using $|\psi_{xx}| , |\psi_{yx}| \leq 1$ (from \eqref{eq:nonnegPdiag}) and $(g)$ by using $\pi_x^{a} \leq \pi_x.$ 
    \item We now consider the case with $a = 0.$ Since $a+b+c \geq 1,$ atleast one of $b,c$ must be $\geq 1.$
    \begin{enumerate}
        \item Say $c \geq 1.$ 
    \begin{align}
        \sum_{x \in \cX}  \ |\psi_{xx}^{b}| \ |\psi_{yx}^{c}| \  \overset{(a)}{\leq} \  \sum_{x \in \cX}  |\psi_{yx}| \ & = \   \sum_{x \in \cX: \psi_{yx} < 0}  |\psi_{yx}|  +  \sum_{x \in \cX: \psi_{yx} \geq 0}  \psi_{yx} \nonumber \\ 
        & \overset{(b)}{=} \ 2 \sum_{x \in \cX: \psi_{yx} < 0}  |\psi_{yx}| 
        \ \overset{(c)}{\leq} \ 2 \sum_{x \in \cX: \psi_{yx} < 0}  \pi_x \ \leq \ 2, \nonumber 
    \end{align}
    where we get $(a)$ by using $|\psi_{xx}| , |\psi_{yx}| \leq 1$ (from \eqref{eq:nonnegPdiag}), $(b)$ by using $\sum_{x \in \cX} \psi_{yx} = 0,$  and $(c)$ by using $-\psi_{yx} \leq \pi_x$ (from \eqref{eq:nonnegPdiag}).
    
    \item Say $b \geq 1.$ 
    \begin{align}
        \sum_{x \in \cX}  \ |\psi_{xx}^{b}| \ |\psi_{yx}^{c}| \  \overset{(a)}{\leq} \  \sum_{x \in \cX}  |\psi_{xx}| \ & = \   \sum_{x \in \cX: \psi_{xx} < 0}  |\psi_{xx}|  +  \sum_{x \in \cX: \psi_{xx} \geq 0}  \psi_{xx} \nonumber \\ 
        & \overset{(b)}{=} \ (1-\beta)  + 2 \sum_{x \in \cX: \psi_{xx} < 0}  |\psi_{xx}| 
        \nonumber \\ 
        & \overset{(c)}{\leq} \ 1 + 2 \sum_{x \in \cX: \psi_{xx} < 0}  \pi_x \ \leq \ 3, \nonumber 
    \end{align}
    where we get $(a)$ by using $|\psi_{xx}| , |\psi_{yx}| \leq 1$ (from \eqref{eq:nonnegPdiag}), $(b)$ by using $\sum_{x \in \cX} \psi_{xx} = (1-\beta),$  and $(c)$ by using $-\psi_{xx} \leq \pi_x$ (from \eqref{eq:nonnegPdiag}).
    \end{enumerate}
\end{enumerate}
This completes the proof of Lemma \ref{lem: sum_over_x}.

\subsection{Proof of Lemma \ref{lem:Txy_expr} } \label{subsec:Txy_expr_Prf}
To prove Lemma \ref{lem:Txy_expr}, we consider $P^{\sim x,y},$ the matrix obtained by replacing the columns corresponding to the states $x,y$ in the t.p.m $P$ with zeros. Let $P_z, z \in \cX$ be the matrix obtained by replacing all the columns of $P$ other than the column corresponding to the state $z$ with zeros. Let $\bpi^{\sim x,y}$ be the vector obtained by setting the entries corresponding to the states $x,y$ in the vector $\bpi$ to zeros. Let $e_z$ be the length$-K$ vector with $1$ in the entry corresponding to $z$ and zeros elsewhere. \\

Recall that $T^n_{xy} \  = \ \pi_x \pi_y Q_{xy}(0,0) - \frac{1}{n} (\pi_y Q_{xy}(1,0) + \pi_x Q_{xy}(0,1) )  + \frac{1}{n (n-1) }  Q_{xy}(1,1),$ where $Q_x^n(a) \triangleq \pr(N_x(X^n) = a),$ and $ Q_{x,y}^n(a,b) \triangleq \pr(N_x(X^n) = a, N_y(X^n) = b) $ for $x, y \in \cX$ and $a,b \in \{ 0 ,1 , \ldots \}.$ The following lemma expresses $\pi_x \pi_y Q_{xy}(0,0) - \frac{1}{n} \pi_y Q_{xy}(1,0)  ,$ and $ \frac{1}{n} \pi_x Q_{xy}(0,1)  - \frac{1}{n (n-1) }  Q_{xy}(1,1)   $  in terms of $P^{\sim x,y}, P_x, P_y$ and $\bpi^{\sim x,y}.$

  \begin{lemma}
  For a stationary Markov chain $X^n$ with state distribution $\bpi$ and t.p.m $P,$ 
  \begin{enumerate}
      \item 
      \begin{align}
          & \pi_y \left( \pi_x Q_{xy}(0,0) - \frac{1}{n} Q_{xy}(1,0) \right) \nonumber \\
          & \quad = \frac{\pi_y}{n} \bigg( \pi_x \ ( \bpi^{\sim x,y} - e_x ) \ \left( P^{\sim x,y} \right)^{n-1}  \nonumber \\ & \qquad \qquad \qquad  + \ \sum_{m_1=2}^n \bpi^{\sim x,y} \left( P^{\sim x,y} \right)^{m_{1}-2} ( \pi_x P^{\sim x,y} - P_x) \left( P^{\sim x,y} \right)^{n - m_{1}}   \bigg)  \one \label{eq:T_xy_int_1}
      \end{align}
      
      \item 
      \begin{align}
            &  \frac{1}{n} \left( \pi_x Q_{xy}(0,1)  - \frac{1}{ (n-1) } Q_{xy}(1,1)  \right) \nonumber \\ 
             & \quad = \ \frac{1}{n(n-1)} \ \bigg(  \sum_{m_2=2}^n \pi_x \ ( \bpi^{\sim x,y} -  e_x) \ \left( P^{\sim x,y} \right)^{m_{2}-2}  P_y \left( P^{\sim x,y} \right)^{n - m_{2}} 
  \nonumber \\ 
  & \qquad +  \sum_{m_1= 2}^{n - 1} \sum_{m_2=m_1 + 1}^{n} \bpi^{\sim x,y}  \left( P^{\sim x,y} \right)^{m_1 - 2} ( \pi_x P^{\sim x,y} - P_x) \nonumber \\
  & \qquad \qquad \qquad \qquad \qquad \qquad \qquad \qquad \quad  \left( P^{\sim x,y} \right)^{m_{2} - m_{1} - 1} P_y \left( P^{\sim x,y} \right)^{n - m_{2} }  \nonumber \\
  & \qquad + \sum_{m_1=2}^n \pi_y e_y  \left( P^{\sim x,y} \right)^{m_1 - 2} ( \pi_x P^{\sim x,y} - P_x)  \left( P^{\sim x,y} \right)^{n - m_1} \nonumber \\ 
  & \qquad + \sum_{m_2=2}^{n-1} \sum_{m_1= m_2+1}^n \bpi^{\sim x,y} \left( P^{\sim x,y} \right)^{m_{2}-2} P_y
   \left( P^{\sim x,y} \right)^{m_{1} - m_{2} - 1} \nonumber \\ 
  & \qquad \qquad \qquad \qquad \qquad \qquad \qquad \qquad ( \pi_x P^{\sim x,y} - P_x )  \left( P^{\sim x,y} \right)^{n - m_{1} }  \bigg) \one 
  \label{eq:T_xy_int_2}
      \end{align}
  \end{enumerate}
  \label{lem:T_xy_int}
  \end{lemma}
  \begin{proof}
  Section \ref{subsec:T_xy_int_Prf}.
  \end{proof}
  Plugging the expressions from the R.H.S of \eqref{eq:T_xy_int_1} and \eqref{eq:T_xy_int_2} into the expression for $T^n_{xy}$ and simplyfying, we get  
\begin{align}
    T^n_{xy} \ & = \ \frac{1}{n(n-1)} \bigg( \sum_{m_2 = 2}^n \pi_x \ ( \bpi^{\sim x,y} - e_x ) \ \left( P^{\sim x,y} \right)^{m_2-2} ( \pi_y P^{\sim x,y} - P_y  ) \left( P^{\sim x,y} \right)^{n-m_2}   \nonumber \\ 
    & \ +  \sum_{m_1= 2}^{n - 1} \sum_{m_2=m_1 + 1}^{n}  \bpi^{\sim x,y}  \left( P^{\sim x,y} \right)^{m_1 - 2} ( \pi_x P^{\sim x,y} - P_x) \left( P^{\sim x,y} \right)^{m_{2} - m_{1} - 1} \nonumber \\ & \qquad \qquad \qquad \qquad   ( \pi_y P^{\sim x,y}  - P_y) \left( P^{\sim x,y} \right)^{n - m_{2} }  \nonumber \\
    & \ + \sum_{m_1 = 2}^n  \pi_y \ ( \bpi^{\sim x,y} - e_y ) \ \left( P^{\sim x,y} \right)^{m_1-2} ( \pi_x P^{\sim x,y} - P_x  ) \left( P^{\sim x,y} \right)^{n-m_1}  \nonumber \\ 
    & \ + \sum_{m_2=2}^{n-1} \sum_{m_1= m_2+1}^n \bpi^{\sim x,y} \left( P^{\sim x,y} \right)^{m_{2}-2} ( \pi_y P^{\sim x,y}  - P_y)
   \left( P^{\sim x,y} \right)^{m_{1} - m_{2} - 1} \nonumber \\ 
   & \qquad \qquad \qquad \qquad  ( \pi_x P^{\sim x,y} - P_x )  \left( P^{\sim x,y} \right)^{n - m_{1} } 
    \bigg) \ \one \label{eq:T_xy_int} 
\end{align}
Let $L_x$ and $L_y$ be the respective matrix representations of the transforms $ \pi_x P^{\sim x,y} - P_x$ and $ \pi_y P^{\sim x,y} - P_y$ restricted to the set of all linear combinations of the vectors $\one$ and $(1-\beta)v.$ Using $\bpi^T \one = u^Tv = 1$ and $ \bpi^T v = u^T\one = 0,$ we get 
\begin{align}
  L_x \ & = \ \pi_x H_{x,y} -  \begin{bmatrix}
  0 & \overline{\beta} \pi_x v_x \\ u_x & \overline{\beta} v_x u_x 
  \end{bmatrix}, \
  L_y \  = \  \pi_y H_{x,y} -  \begin{bmatrix}
  0 & \overline{\beta} \pi_y v_y \\ u_y & \overline{\beta} v_y u_y 
  \end{bmatrix}, \nonumber \\ 
\text{ where } H_{x,y} \ & \triangleq \  \begin{bmatrix} -( \pi_x + \pi_y) &  -\overline{\beta}  (\pi_x v_x + \pi_y v_y ) \\ - (u_x+u_y)  & \overline{\beta}  ( 1- v_x u_x - v_y u_y)  \end{bmatrix}
\end{align}
Since $P = RDS,$ we have $P^{\sim x,y} = RDS^{\sim x,y},$ where $S^{\sim x,y}$ is the matrix obtained by replacing the columns corresponding to the states $x$ and $y$ in the t.p.m $P$ with zeros.
Let $M_{1,t} \triangleq (S^{\sim x,y} R D)^{t-1} \ S^{\sim x,y} \ \one,$ $M_{2,t} \triangleq (S^{\sim x,y} R D)^{t-1} \ S^{\sim x,y} \ (1-\beta) \ v$ and $M_{t} \triangleq \begin{bmatrix} M_{1,t} & M_{2,t} \end{bmatrix},$ $t \in \mathbb{Z}.$
Our next lemma  gives closed form expressions for $M_{t} = \begin{bmatrix} M_{1,t} & M_{2,t} \end{bmatrix}$ in terms of  $\lambda_{1}^{\sim x,y}, \lambda_{2}^{\sim x,y},$ the eigenvalues of the matrix $S^{\sim x,y} RD =  \left( I_{2 \times 2} - \begin{bmatrix} \pi_x & \pi_y \\ u_x & u_y \end{bmatrix} \begin{bmatrix} 1 & v_x \\ 1 & v_y  \end{bmatrix} \right) D.$ 

\begin{lemma}
For $t \in \mathbb{Z},$
\begin{align}
      M_{1,t} \ & = \ 
  \begin{bmatrix}
  0.5 & {0.5s_{x,y}}  \\ 
  0 &  -(u_x + u_y)
  \end{bmatrix} \ 
  W_t, \ 
  M_{2,t} \  = \ \begin{bmatrix}
  0 & -(1-\beta)(\pi_x v_x + \pi_y v_y)   \\ 
  0.5 &  - {0.5s_{x,y}} 
  \end{bmatrix} \ W_t, \text{ where } 
  \nonumber \\
  W_t \ & \triangleq \ \begin{bmatrix}
   \left( \lambda_{1}^{\sim x,y}  \right)^t + \left( \lambda_{2}^{\sim x,y}  \right)^t  \\ 
   \sum_{m=0}^{t-1} \ ( \lambda_{1}^{\sim x,y} )^{t-1-m} \  ( \lambda_{2}^{\sim x,y} )^m  
  \end{bmatrix}  \nonumber \end{align}
\begin{proof}
Section \ref{subsec:M_lambda_expr_Prf}. 
\end{proof}
\label{lem:M_lambda_expr}
\end{lemma} 

The following lemma expresses each of the summands in \eqref{eq:T_xy_int} in terms of $M_t, L_x$ and $L_y.$

\begin{lemma}
For a rank-2 diagonalizable t.p.m $P$ with stationary distribution $\pi,$ 

\begin{enumerate}
    \item For $m_1 = 1, m_2 \geq 2,$
 \begin{align}
     & \pi_x \ ( \bpi^{\sim x,y} - e_x ) \ \left( P^{\sim x,y} \right)^{m_2-2} ( \pi_y P^{\sim x,y} - P_y  ) \left( P^{\sim x,y} \right)^{n-m_2} \one \nonumber \\ 
      & \ {=} \ \bpi^{\sim x,y} \ R \ D \   M_{m_1 -2} \  L_x \   M_{m_2 - m_1 - 1} \  L_y \   M_{1,n-m_2}  \label{eq:T_xy_mtr_3}
 \end{align}
   \label{lem_part:m2>m1=1} 
    \item For $m_2 = 1, m_1 \geq 2,$
    \begin{align}
   & \pi_y \ ( \bpi^{\sim x,y} - e_y ) \ \left( P^{\sim x,y} \right)^{m_1-2} ( \pi_x P^{\sim x,y} - P_x  ) \left( P^{\sim x,y} \right)^{n-m_1} \one \nonumber \\  
   & \ = \ \bpi^{\sim x,y} \ R \ D \   M_{m_2 -2} \  L_y \   M_{m_1 - m_2 - 1} \  L_x \   M_{1,n-m_1}  \label{eq:T_xy_mtr_4}
\end{align}
    \label{lem_part:m1>m2=1} 
    \item For $ m_1 \geq 2, m_2 > m_1,$
    \begin{align}
       & \bpi^{\sim x,y}  \left( P^{\sim x,y} \right)^{m_1 - 2} ( \pi_x P^{\sim x,y} - P_x) \left( P^{\sim x,y} \right)^{m_{2} - m_{1} - 1}  ( \pi_y P^{\sim x,y}  - P_y) \left( P^{\sim x,y} \right)^{n - m_{2} } \one \nonumber \\ 
       & \ = \ \bpi^{\sim x,y} \ R \ D \   M_{m_1 -2} \  L_x \   M_{m_2 - m_1 - 1} \  L_y \   M_{1,n-m_2}  \label{eq:T_xy_mtr_1}
    \end{align}
    \label{lem_part:m2>m1>1} 
    \item For $ m_2 \geq 2, m_1 > m_2,$
    \begin{align}
     & \bpi^{\sim x,y}  \left( P^{\sim x,y} \right)^{m_2 - 2} ( \pi_y P^{\sim x,y} - P_y) \left( P^{\sim x,y} \right)^{m_{1} - m_{2} - 1}  ( \pi_x P^{\sim x,y}  - P_x) \left( P^{\sim x,y} \right)^{n - m_{1} } \one \nonumber \\
     & \ {=} \ \bpi^{\sim x,y} \ R \ D \   M_{m_2 -2} \  L_y \   M_{m_1 - m_2 - 1} \  L_x \   M_{1,n-m_1}   \label{eq:T_xy_mtr_2}
 \end{align}
  \label{lem_part:m1>m2>1} 
\end{enumerate}

\begin{proof}
Section \ref{subsec:T_xy_term_expr_Prf}
\end{proof}
\label{lem: T_xy_term_expr}
\end{lemma}

 Using \eqref{eq:T_xy_mtr_3}, \eqref{eq:T_xy_mtr_4}, \eqref{eq:T_xy_mtr_1}, and \eqref{eq:T_xy_mtr_2} in \eqref{eq:T_xy_int}, we get 
 \begin{align}
   T_{xy} \ & =  \ \frac{1}{n(n-1)} \ \bigg( \sum_{m_1 = 1}^{n-1} \sum_{m_2 = m_1 + 1}^{n} \bpi^{\sim x,y} \ R \ D \   M_{m_1 -2} \  L_x \   M_{m_2 - m_1 - 1} \  L_y \   M_{1,n-m_2} \nonumber \\ 
   & \quad + \sum_{m_2 = 1}^{n-1} \sum_{m_1 = m_2 + 1}^{n} \bpi^{\sim x,y} \ R \ D \   M_{m_2 -2} \  L_y \   M_{m_1 - m_2 - 1} \  L_x \   M_{1,n-m_1} \bigg)  \label{eq:T_xy_mtr_5}
 \end{align}
 Interchanging the variables $m_1$ and $m_2$ in the second term of \eqref{eq:T_xy_mtr_5} completes the proof of Lemma \ref{lem:Txy_expr}. 
 
\subsection{Proof of Lemma \ref{lem:e_value_bd_xy}}
\label{subsec:e_value_bd_xy_Prf}
Recall that 
$\lambda_{i}^{\sim x,y} = \frac{1}{2} \left( (1-\pi_x - \pi_y) + \overline{\beta} (1-v_x u_x -v_y u_y) + (-1)^{i+1} \Delta_{x,y} \right), \ i = 1,2,$  with
$ \Delta_{x,y}^2 = [ (1 - \pi_x - \pi_y) - \overline{\beta} (1-v_x u_x - v_y u_y) ]^2 + 4 \ \overline{\beta} \ (\pi_x v_x + \pi_y v_y) \ (u_x + u_y),$ are the eigenvalues of the matrix $\left(I_{2 \times } - \begin{bmatrix}
\pi_x & \pi_y \\ u_x & u_y \end{bmatrix} \begin{bmatrix}
1 & v_x \\ 1 & v_y
\end{bmatrix} \right)D.$   \\ 
To bound $\lambda_{i}^{\sim x,y},$ we first bound $\Delta_{x,y}$ using the following lemma. 
\begin{lemma}
For a rank-2 diagonalizable t.p.m $P$ with stationary distribution $\bpi,$ and spectral gap $\beta,$ 
\begin{align}
    \Delta_{x,y} \ & \in \  \left[ -\left( \beta -\frac{\beta^2}{\beta + 2}(\pi_x + \pi_y) + P_{xx} + P_{yy}\right) , \  \beta-\frac{\beta^2}{\beta + 2}(\pi_x + \pi_y) + P_{xx} +P_{yy}   \right], \ \beta \in [0,2] \label{eq:Delta_xy_bound}
\end{align}
    \label{lem:Delta_xy_bound}
\end{lemma}
\begin{proof}
Section \ref{subsec:Delta_xy_bound_Prf}.
\end{proof}
Using the above bound in the expression for $\lambda_{i}^{\sim x}, (i = 1,2),$ we get $ \lambda_{i}^{\sim x,y} \leq 1-0.5 \frac{\beta^2}{\beta + 2} (\pi_x+\pi_y),  \beta \in [0,2],$  for $i = 1,2.$ This completes the proof of Lemma \ref{lem:e_value_bd_xy}. 

\section{Appendix D} 

\subsection{Proof of Lemma \ref{lem:T_xy_int}  }  
\label{subsec:T_xy_int_Prf}
We use the following lemma to prove Lemma \ref{lem:T_xy_int}. Let $X^{n}_{\sim m} \triangleq (X_1, \ldots,X_{m-1}, X_{m+1},\ldots, X_n),$ denote the samples $X_1^n$ except $X_m,$ $ 1 \leq m \leq n.$

\begin{lemma}
For a stationary Markov chain $X^n$ with state distribution $\bpi$ and t.p.m $P,$ 
\begin{enumerate}
    \item \begin{equation}
        \pr(N_x = N_y = 0) \ = \ \pi_x \bpi^{\sim x,y} \left( P^{\sim x,y} \right)^{n-1} \one \label{eq:P(N_xy = 0)}
    \end{equation}
         
    \item \begin{align}
        \pr( X_{1} = x, N_x(X^n_{\sim 1}) =  N_y = 0 ) \ & = \ \pi_x e_x  \left( P^{\sim x,y} \right)^{n - 1} \one \label{eq:P(X_1 = x, N_y = 0)} \\
        \pr( X_{m_1} = x, N_x(X^n_{\sim m_1}) =  N_y = 0 ) \ & = \ \bpi^{\sim x,y} \left( P^{\sim x,y} \right)^{m_{1}-2} P_x \left( P^{\sim x,y} \right)^{n - m_{1}} \one, \nonumber \\ 
        & \qquad \qquad \qquad \qquad \qquad \qquad \qquad \qquad  2 \leq m_1 \leq n \label{eq:P(X_m1 = x, N_y = 0)} 
    \end{align}
    
    \item \begin{align}
       & \pr( X_{1} = x, X_{m_2} = y, N_x(X^n_{\sim 1}) =  N_y( X^n_{\sim m_2} ) = 0 ) \nonumber \\
       & \qquad \qquad \qquad \qquad \qquad = \ \pi_x e_x \left( P^{\sim x,y} \right)^{m_{2}-2}  P_y \left( P^{\sim x,y} \right)^{n - m_{2}} \one, \ 2 \leq m_2 \leq n \label{eq:P(X_1 = x, X_m2 = y)} \\
        & \pr( X_{m_1} = x, X_{m_2} = y, N_x(X^n_{\sim m_1}) =  N_y( X^n_{\sim m_2} ) = 0 ) \nonumber \\  & \qquad \qquad \qquad =  \ \bpi^{\sim x,y} \left( P^{\sim x,y} \right)^{m_{1}-2} P_x \left( P^{\sim x,y} \right)^{m_{2} - m_{1} - 1} P_y \left( P^{\sim x,y} \right)^{n - m_{2} } \one, \nonumber \\ 
       & \qquad \qquad \qquad \qquad \qquad \qquad \qquad \qquad \qquad \qquad \qquad \qquad \qquad \qquad 2 \leq m_1 < m_2 \leq n \label{eq:P(X_m1 = x, X_m2 = y)}
    \end{align}
\end{enumerate}
\label{lem: prob_expr}
\end{lemma}
\begin{proof}
Section \ref{subsec:prob_expr_Prf}.
\end{proof}

We first prove part 1 of Lemma \ref{lem:T_xy_int}.
 \begin{align}
     & \pi_y \left( \pi_x \pr(N_x = N_y = 0) - \frac{1}{n}  \pr(N_x = 1, N_y = 0) \right) \nonumber \\ 
   & \quad \overset{(a_0)}{=} {\pi_y} \left( \pi_x \pr( N_x = N_y = 0 ) - \frac{1}{n} \sum_{m_1=1}^n \pr( X_{m_1} = x, N_x(X^n_{\sim m_1}) =  N_y = 0 ) \right) \nonumber \\ 
   & \quad \overset{(b_0)}{=} {\pi_y} \left(  \pi_x \bpi^{\sim x,y} \left( P^{\sim x,y} \right)^{n-1} \one  - \frac{1}{n} \pi_x e_x  \left( P^{\sim x,y} \right)^{n - 1} \one \right. \nonumber \\
   & \quad \qquad \qquad \qquad \qquad \qquad \quad   \left.  - \frac{1}{n} \sum_{m_1=2}^n \bpi^{\sim x,y} \left( P^{\sim x,y} \right)^{m_{1}-2} P_x \left( P^{\sim x,y} \right)^{n - m_{1}} \one \right) \nonumber \\ 
   & \quad \overset{(c_0)}{=} \frac{\pi_y}{n} \bigg( \pi_x \ ( \bpi^{\sim x,y} - e_x ) \ \left( P^{\sim x,y} \right)^{n-1} \one   \nonumber \\ & \qquad \qquad \qquad  + \ \sum_{m_1=2}^n \bpi^{\sim x,y} \left( P^{\sim x,y} \right)^{m_{1}-2} ( \pi_x P^{\sim x,y} - P_x) \left( P^{\sim x,y} \right)^{n - m_{1}} \one  \bigg), \nonumber 
 \end{align}
 where we get $(a_0)$ by expressing $\pr(N_x = 1, N_y = 0)$ as the sum over the probabilities of the sample paths $X^n$ with $x$ in one of the $X_i$s, $(b_0)$ by using Lemma \ref{lem: prob_expr}, and $(c_0)$ by using \\  $ \pi_x \bpi^{\sim x,y} \left( P^{\sim x,y} \right)^{n-1} \one = \frac{1}{n} \ \sum_{m_1 = 1}^n \pi_x \bpi^{\sim x,y} \left( P^{\sim x,y} \right)^{n-1} \one$ in $(b_0)$ and simplyfying the difference. \\
 
 We next prove part 2 of Lemma \ref{lem:T_xy_int}.
 \begin{align}
      &  \frac{1}{n} \left( \pi_x \pr(N_x = 0, N_y = 1) - \frac{1}{ (n-1) }  \pr(N_x = N_y = 1) \right) \nonumber \\ 
  & \quad \overset{(a_1)}{=} \ \frac{1}{n}  \sum_{m_2=1}^n \bigg(   \pi_x \pr(X_{m_2} = y, N_x(X^n)= 0, N_y( X^n_{\sim m_2} ) = 0 ) \nonumber \\ 
  & \qquad \qquad \qquad - \frac{1}{ (n-1) } \sum_{m_1=1, m_1 \neq m_2}^n  \pr( X_{m_1} = x, X_{m_2} = y, N_x(X^n_{\sim m_1}) =  N_y( X^n_{\sim m_2} ) = 0 ) \bigg) \nonumber \\ 
  & \quad \overset{(b_1)}{=} \ \frac{1}{n} \bigg( \pi_x \pi_y e_y  \left( P^{\sim x,y} \right)^{n - 1} \one - \frac{1}{ (n-1) } \sum_{m_1=2}^n \pi_y e_y  \left( P^{\sim x,y} \right)^{m_{1} - 2} P_x \left( P^{\sim x,y} \right)^{n - m_{1} } \one  \bigg)
  \nonumber \\ 
  & \qquad  + \frac{1}{n  }  \sum_{m_2=2}^n \bigg( \pi_x \bpi^{\sim x,y} \left( P^{\sim x,y} \right)^{m_{2}-2} P_y \left( P^{\sim x,y} \right)^{n - m_{2}} \one \nonumber \\ 
  &\qquad \qquad \ \  - \frac{1}{n-1} \pi_x e_x \left( P^{\sim x,y} \right)^{m_{2}-2}  P_y \left( P^{\sim x,y} \right)^{n - m_{2}} \one \nonumber \\ 
  & \qquad \qquad \ \  - \frac{1}{n-1} \sum_{m_1=2}^{m_2-1}  \bpi^{\sim x,y} \left( P^{\sim x,y} \right)^{m_{1}-2} P_x \left( P^{\sim x,y} \right)^{m_{2} - m_{1} - 1} P_y \left( P^{\sim x,y} \right)^{n - m_{2} } \one \nonumber \\ 
  & \qquad \qquad   \ \  - \frac{1}{n-1} \sum_{m_1=m_2+1}^{n}  \bpi^{\sim x,y} \left( P^{\sim x,y} \right)^{m_{2}-2} P_y \left( P^{\sim x,y} \right)^{m_{1} - m_{2} - 1} P_x \left( P^{\sim x,y} \right)^{n - m_{1} } \one
  \bigg) \nonumber 
  \end{align}
 where we get $(a_1)$ by expressing $\pr(N_x = 0, N_y = 1)$ as the sum over the probabilities of the sample paths $X^n$ with $y$ in one of the $X_i$s and $\pr(N_x = 1, N_y = 1)$ as the sum over the probabilities of the sample paths $X^n$ with $x, y$ occurring once, $(b_1)$ by separating out the term corresponding to $m_2 = 1$ in $(a_1)$ and using Lemma \ref{lem: prob_expr}.  
 
 Using $ \pi_x \pi_y e_y \left( P^{\sim x,y} \right)^{n-1} \one = \frac{1}{n-1} \ \sum_{m_1 = 2}^n \pi_x \pi_y e_y \left( P^{\sim x,y} \right)^{n-1} \one$, $ \pi_x \bpi^{\sim x,y} \left( P^{\sim x,y} \right)^{m_2-2} P_y \\ \left( P^{\sim x,y} \right)^{n - m_2}  \one  = \frac{1}{n-1} \ \sum_{m_1 = 2}^n \pi_x \bpi^{\sim x,y} \left( P^{\sim x,y} \right)^{m_2-2} P_y \left( P^{\sim x,y} \right)^{n - m_2}  \one$ in $(b_1)$ and simplyfying the difference, we get 
 \begin{align}
      &  \frac{1}{n} \left( \pi_x \pr(N_x = 0, N_y = 1) - \frac{1}{ (n-1) }  \pr(N_x = N_y = 1) \right) \nonumber \\ 
      & \quad {=} \ \frac{1}{n(n-1)} \bigg( \sum_{m_1=2}^n \pi_y e_y  \left( P^{\sim x,y} \right)^{m_1 - 2} ( \pi_x P^{\sim x,y} - P_x)  \left( P^{\sim x,y} \right)^{n - m_1} \one \nonumber \\
  & \qquad +  \sum_{m_2=2}^n \pi_x \ ( \bpi^{\sim x,y} -  e_x) \ \left( P^{\sim x,y} \right)^{m_{2}-2}  P_y \left( P^{\sim x,y} \right)^{n - m_{2}} \one 
  \nonumber \\ 
  & \qquad + \sum_{m_2=2}^{n} \sum_{m_1= 2}^{m_2 - 1} \bpi^{\sim x,y}  \left( P^{\sim x,y} \right)^{m_1 - 2} ( \pi_x P^{\sim x,y} - P_x) \left( P^{\sim x,y} \right)^{m_{2} - m_{1} - 1} P_y \left( P^{\sim x,y} \right)^{n - m_{2} } \one \nonumber \\
  & \qquad + \sum_{m_2=2}^{n-1} \sum_{m_1= m_2+1}^n \bpi^{\sim x,y} \left( P^{\sim x,y} \right)^{m_{2}-2} P_y
   \left( P^{\sim x,y} \right)^{m_{1} - m_{2} - 1} ( \pi_x P^{\sim x,y} - P_x )  \left( P^{\sim x,y} \right)^{n - m_{1} } \one \bigg), 
  \nonumber 
\end{align}
 
 \subsection{Proof of Lemma \ref{lem:M_lambda_expr}} \label{subsec:M_lambda_expr_Prf}
 To prove Lemma \ref{lem:M_lambda_expr}, we begin with our next lemma which gives a closed form expression for $(S^{\sim x,y} R D)^l,$ $(l \geq 1),$ for various values of $\pi_x+\pi_y, \pi_x v_x+\pi_y v_y,$ and $u_x + u_y.$ 

\begin{lemma}
$(S^{\sim x,y}RD)^l:$ 
\begin{enumerate}
    \item For $x,y \in \cX$ with $\pi_x v_x + \pi_y v_y = u_x + u_y = 0,$
    \item For $x,y \in \cX$ with $\pi_x v_x + \pi_y v_y = 0, u_x + u_y \neq 0,$
    
    \begin{enumerate}
    
        \item $1-\pi_x-\pi_y = \overline{\beta} (1-v_x u_x - v_y u_y):$ 
        \begin{align}
            (S^{\sim x,y} R D)^l \ & = \ \begin{bmatrix}
    (1-\pi_x-\pi_y)^l & 0 \\
    -l(u_x+u_y)(1-\pi_x-\pi_y)^{l-1} & (1-\pi_x-\pi_y)^l
    \end{bmatrix} \label{eq:non_diag_vxy_0_rep_ev}
        \end{align}
        
     \item $1-\pi_x-\pi_y \neq \overline{\beta} (1-v_x u_x - v_y u_y):$ 
        \begin{align}
    (S^{\sim x,y} R D)^l \ & = \ V_1^{\sim x,y} \ 
    \begin{bmatrix}
    (\overline{\pi_x + \pi_y})^l & 0 \\
    0 & \overline{\beta}^l(1-v_x u_x - v_y u_y)^l
    \end{bmatrix} \ (V_1^{\sim x,y})^{-1}, \label{eq:diag_vxy_0_dist_ev}
        \end{align}
        where $V_1^{\sim x,y} = \begin{bmatrix}
    1 & 0 \\
    \frac{(u_x + u_y)}{\overline{\beta}(1-v_x u_x-v_y u_y) - (\overline{\pi_x + \pi_y}) } & 1
    \end{bmatrix}.$
    \end{enumerate}
    
    \item For $x,y \in \cX$ with $\pi_x v_x + \pi_y v_y \neq 0, u_x + u_y = 0,$
    
     \begin{enumerate}
        \item $1-\pi_x-\pi_y = \overline{\beta} (1-v_x u_x - v_y u_y):$ 
         \begin{align}
         (S^{\sim x,y} R D)^l \ & = \ \begin{bmatrix}
    (1-\pi_x-\pi_y)^l &  -l(1-\beta)( \pi_x u_x + \pi_y u_y)(1-\pi_x-\pi_y)^{l-1} \\
    0 & (1-\pi_x-\pi_y)^l
    \end{bmatrix}  \label{eq:non_diag_uxy_0_rep_ev}
    \end{align}
    
        \item $1-\pi_x-\pi_y \neq \overline{\beta} (1-v_x u_x - v_y u_y):$ 
        \begin{align}
    (S^{\sim x,y} R D)^l \ & = \ V_2^{\sim x,y} \ 
    \begin{bmatrix}
     (\overline{\pi_x + \pi_y})^l & 0 \\
    0 & \overline{\beta}^l (1-v_x u_x - v_y u_y)^l
    \end{bmatrix} \ (V_2^{\sim x,y})^{-1}, \label{eq:diag_uxy_0_dist_ev}
    \end{align}
    where $V_2^{\sim x,y} = \begin{bmatrix}
    1 &   \frac{\overline{\beta}(\pi_x v_x + \pi_y v_y)}{(\overline{\pi_x + \pi_y}) - \overline{\beta}(1-v_x u_x-v_y u_y) } \\
    0 & 1
    \end{bmatrix}.$
      \end{enumerate}
      
    \item For $x,y \in \cX,$ with $\pi_x v_x + \pi_y v_y \neq 0, u_x + u_y \neq 0,$
       \begin{align}
        (S^{\sim x,y} R   D)^l \ & = \ V^{\sim x,y} \begin{bmatrix} (\lambda_1^{\sim x,y})^l & 0 \\ 0 & (\lambda_2^{\sim x,y})^l  \end{bmatrix} (V^{\sim x,y})^{-1}, \label{eq:diag_vxy_uxy_nz}
      \\
       \text{ where } 
       V^{\sim x,y} \ & = \ \begin{bmatrix}
       1 & 1 \\ -\frac{\Delta_{x,y} - s_{x,y}}{ 2(1-\beta) (\pi_x v_x + \pi_y v_y)  } & \frac{\Delta_{x,y} +s_{x,y}}{ 2 (1-\beta)  (\pi_x v_x + \pi_y v_y)   } \end{bmatrix}, \nonumber 
       \end{align}
       $\lambda_1^{\sim x,y}, \lambda_2^{\sim x,y}, \Delta_{x,y}$ and $s_{x,y}$ are as defined in Lemma \ref{lem:Txy_expr}.
     
\end{enumerate}
\label{lem:Diagonalization_xy}
\end{lemma}
\begin{proof}
Section \ref{subsec:lem_Diag_xy_proof}
\end{proof}
  For every $(x,y) \in \cX^2$ with $x\neq y,$ substituting the corresponding closed form of $ (S^{\sim x,y} R   D)^{t-1} $ from the above lemma in the definition of $M_{1,t}$ and $M_{2,t}$ gives 
  \begin{align}
      M_{1,t} \ & = \ 
  \begin{bmatrix}
  0.5 & {0.5s_{x,y}}  \\ 
  0 &  -(u_x + u_y)
  \end{bmatrix} \ 
  W_t, \ 
  M_{2,t} \  = \ \begin{bmatrix}
  0 & -(1-\beta)(\pi_x v_x + \pi_y v_y)   \\ 
  0.5 &  - {0.5s_{x,y}} 
  \end{bmatrix} \ W_t, \text{ where } 
  \nonumber \\
  W_t \ & \triangleq \ \begin{bmatrix}
   \left( \lambda_{1}^{\sim x,y}  \right)^t + \left( \lambda_{2}^{\sim x,y}  \right)^t  \\ 
   \sum_{m=0}^{t-1} \ ( \lambda_{1}^{\sim x,y} )^{t-1-m} \  ( \lambda_{2}^{\sim x,y} )^m  
  \end{bmatrix}, \ \lambda_{1}^{\sim x,y}, \lambda_{2}^{\sim x,y}, \text{ are the eigenvalues of the}  \nonumber \end{align}  matrix $S^{\sim x,y} RD =  \left( I_{2 \times 2} - \begin{bmatrix} \pi_x & \pi_y \\ u_x & u_y \end{bmatrix} \begin{bmatrix} 1 & v_x \\ 1 & v_y  \end{bmatrix} \right) D$ as described in Lemma \ref{lem:Txy_expr}. 
 This completes the proof of Lemma \ref{lem:M_lambda_expr}.
 
 \subsection{ Proof of Lemma \ref{lem: T_xy_term_expr} } \label{subsec:T_xy_term_expr_Prf}
 We first begin with the proof of lemma \ref{lem: T_xy_term_expr}, part \ref{lem_part:m2>m1>1}. 
 For $m_1 \geq 2$ and $m_2 > m_1,$
 \begin{align}
     & \bpi^{\sim x,y}  \left( P^{\sim x,y} \right)^{m_1 - 2} ( \pi_x P^{\sim x,y} - P_x) \left( P^{\sim x,y} \right)^{m_{2} - m_{1} - 1}  ( \pi_y P^{\sim x,y}  - P_y) \left( P^{\sim x,y} \right)^{n - m_{2} } \one \nonumber \\ 
     & \overset{(a)}{=} \  \bpi^{\sim x,y} R D \left( S^{\sim x,y} R D\right)^{m_1 - 3} S^{\sim x,y} ( \pi_x P^{\sim x,y} - P_x) R D \left( S^{\sim x,y} R D\right)^{m_2 - m_1 - 2} \nonumber \\ 
     & \qquad \qquad S^{\sim x,y} ( \pi_y P^{\sim x,y}  - P_y) R D \left( S^{\sim x,y} R D\right)^{n - m_2 - 1} S^{\sim x,y} \one \nonumber  \\
     & \overset{(b)}{=} \ \bpi^{\sim x,y} R D \left( S^{\sim x,y} R D\right)^{m_1 - 3} S^{\sim x,y} ( \pi_x P^{\sim x,y} - P_x) \ \begin{bmatrix} \one & (1-\beta)v \end{bmatrix}  \nonumber \\ 
     &  \qquad \qquad  \left( S^{\sim x,y} R D\right)^{m_2 - m_1 - 2} \ S^{\sim x,y} ( \pi_y P^{\sim x,y}  - P_y) \ \begin{bmatrix} \one & (1-\beta)v \end{bmatrix} \  M_{1,n-m_2} \nonumber \\
     & \overset{(c)}{=} \ \bpi^{\sim x,y} R D \left( S^{\sim x,y} R D\right)^{m_1 - 3} S^{\sim x,y} ( \pi_x P^{\sim x,y} - P_x) \  \begin{bmatrix} \one & (1-\beta)v \end{bmatrix} \nonumber \\  
   &  \qquad \qquad \left( S^{\sim x,y} R D\right)^{m_2 - m_1 - 2} \ S^{\sim x,y}  \begin{bmatrix} \one & (1-\beta)v \end{bmatrix} L_y \   M_{1,n-m_2} \nonumber \\
   & \overset{(d)}{=} \ \bpi^{\sim x,y} R D \left( S^{\sim x,y} R D\right)^{m_1 - 3} S^{\sim x,y} ( \pi_x P^{\sim x,y} - P_x) \  \begin{bmatrix} \one & (1-\beta)v \end{bmatrix}  \  M_{m_2 - m_1 - 1} \nonumber \\  
   &  \qquad \qquad  L_y \  M_{1,n-m_2} \nonumber \\ 
   & \overset{(e)}{=} \ \bpi^{\sim x,y} \ R \ D \   M_{m_1 -2} \  L_x \   M_{m_2 - m_1 - 1} \  L_y \   M_{1,n-m_2}  \tag{\ref{eq:T_xy_mtr_1}}
 \end{align}
 where we get $(a)$ by using $P^{\sim x,y} = RDS^{\sim x,y},$ $(b)$ by using the definition of $M_{1,t}$ with $t = n-m_2$ and $RD = \begin{bmatrix} \one & (1-\beta)v \end{bmatrix},$ $(c)$ by using the definition of $L_y,$ $(d)$ by using the definition of $M_{t}$ with $t = m_2 - m_1 - 1$ and $(e)$ by following steps similar to $(c)$ and $(d).$  This completes the proof of lemma \ref{lem: T_xy_term_expr} part \ref{lem_part:m2>m1>1}. The proof for lemma \ref{lem: T_xy_term_expr}, part \ref{lem_part:m1>m2>1} is similar. \\
 
 We next prove lemma \ref{lem: T_xy_term_expr}, part \ref{lem_part:m2>m1=1}.
 For $m_2 \geq 2,$
 \begin{align}
     & \pi_x \ ( \bpi^{\sim x,y} - e_x ) \ \left( P^{\sim x,y} \right)^{m_2-2} ( \pi_y P^{\sim x,y} - P_y  ) \left( P^{\sim x,y} \right)^{n-m_2} \one \nonumber \\ 
     & \overset{(a_0)}{=} \ \pi_x \ ( \bpi^{\sim x,y} - e_x ) \ R D \  M_{m_2 -  2} \  L_y \   M_{1,n-m_2} \nonumber \\
     & \overset{(a_1)}{=} \bpi^{\sim x,y} \ R \ D \   M_{-1} \  L_x \   M_{m_2 - 2} \  L_y \   M_{1,n-m_2} \nonumber \\
     & \overset{(a_2)}{=} \bpi^{\sim x,y} \ R \ D \   M_{m_1 -2} \  L_x \   M_{m_2 - m_1 - 1} \  L_y \   M_{1,n-m_2}  \tag{\ref{eq:T_xy_mtr_3}}
 \end{align}
 where we get $(a_0)$ by following the same steps as $(a), (b), (c) $ and $(d)$ above, $(a_1)$ by observing that $ \bpi^{\sim x,y} \ R \ D \   M_{-1} \  L_x = \pi_x \ ( \bpi^{\sim x,y} - e_x ) \ R D,$ and $(a_2)$ by using $m_1 = 1.$ This completes the proof of lemma \ref{lem: T_xy_term_expr} part \ref{lem_part:m2>m1=1}. The proof for lemma \ref{lem: T_xy_term_expr}, part \ref{lem_part:m1>m2=1} is similar.

\subsection{Proof of Lemma \ref{lem:Delta_xy_bound} }
\label{subsec:Delta_xy_bound_Prf}
To get the bound in \eqref{eq:Delta_xy_bound} on $\Delta_{x,y}$, we first bound $\Delta_{x,y}^2$ using  $ \tau_{x,y} \triangleq 1 -  \frac{ (\pi_y u_x - \pi_x u_y) }{ (\pi_x + \pi_y) } \ (v_x - v_y)$ in the following lemma.

\begin{lemma}
For a rank-2 diagonalizable t.p.m $P$ with stationary distribution $\bpi,$ and spectral gap $\beta,$
\begin{enumerate}
    \item 
\begin{align}
    \Delta_{x,y}^2 \ & \leq \ \left[ \beta - \frac{\beta^2}{ \beta + 2} (\pi_x + \pi_y) + P_{xx} +P_{yy} \right]^2 -  \frac{\beta^4}{ (\beta + 2)^2}  \ (\pi_x + \pi_y)^2 \nonumber \\ 
    & \qquad + 2 \  (\pi_x + \pi_y) \ \left( (1-\beta) \ \tau_{x,y} \ \left(2 -  \frac{\beta^2}{ \beta + 2}\right)  - 2 \ \left( 1 - \frac{\beta^2}{ \beta + 2 } \right) \right)   \label{eq:Delta_tau_bd} 
\end{align}

\item \begin{align}
    (1-\beta) \ \tau_{x,y} \ & \leq \ 1 - \frac{\beta^2}{ \beta + 2}, \ \beta \in [0,2] \label{eq:tau_xy_bd}
\end{align}

\end{enumerate}
\label{lem:Delta_tau_xy_bd}
\end{lemma}

\begin{proof}
Section \ref{subsec:Delta_tau_xy_bd_Prf}.
\end{proof}

Using \eqref{eq:tau_xy_bd} in \eqref{eq:Delta_tau_bd}, we get 
$ \Delta_{x,y}^2  \ \leq \ [ \beta - \frac{\beta^2}{ \beta + 2} (\pi_x + \pi_y) + P_{xx} +P_{yy} ]^2 .$ This completes the proof of Lemma \ref{lem:Delta_xy_bound}.

\section{Appendix E}

\subsection{Proof of Lemma \ref{lem: prob_expr}} \label{subsec:prob_expr_Prf}

$\left( P^{\sim x,y} \right)^{l - 1 } \one, l > 1$ is a $K \times 1$ vector with its entry corresponding to any state $z \in  X$ being the probability of not passing through the states $x$ and $y$ in the next $l-1$ steps, given the present state is $z.$
\begin{enumerate}
    \item Taking the dot product of $\left( P^{\sim x,y} \right)^{n - 1 } \one$ with $\bpi^{\sim x,y}$ gives the probability of not passing through through the states $x$ and $y$ in $n$ steps of the stationary Markov chain $X^n.$ \newline 
    
    \item 
    \begin{enumerate}
        \item Similarly the dot product of $\left( P^{\sim x,y} \right)^{n - 1 } \one$ with $\pi_x e_x$ gives the probability that the stationary Markov chain $X^n$ starts in state $x$ and never visits the states $x$ and $y$ in the next $n-1$ steps.  \newline
        \item Recall that $P_x$ is the matrix obtained by replacing all the columns of the t.p.m $P,$ except the column corresponding to the state $x \in \cX,$ with zeros. \\ 
        
        $ P_x \left( P^{\sim x,y} \right)^{n - m_1 } \one$ is a $ K \times 1 $ vector with its entry corresponding to any state $z \in  X$ being the probability of going to the state $x$ in the next step and not passing through the states $x$ and $y$ in the subsequent $n-m_1$ steps, given the present state is $z.$ \\ 
        
        $  \left( P^{\sim x,y} \right)^{m_1 -2 } P_x \left( P^{\sim x,y} \right)^{n - m_1 } \one$ is a $ K \times 1 $ vector with its entry corresponding to any state $z \in  X$ being the probability of not passing through the states $x$ and $y$ in the next $m_1 - 2$ steps and going to the state $x$ in the $(m_1 -1)-$th step and not passing through the states $x$ and $y$ in the subsequent $n-m_1$ steps, given the present state is $z.$ Taking dot product with $\bpi^{\sim x,y}$ gives $\pr( X_{m_1} = x, N_x(X^n_{\sim m_1}) =  N_y = 0 ).$
    \end{enumerate}
   
   \item Similar to above.  
\end{enumerate}

\subsection{Proof of Lemma \ref{lem:Diagonalization_xy}} \label{subsec:lem_Diag_xy_proof}
Since $SR = I,$ we have $S^{\sim x,y} R = \begin{bmatrix}
1-{\pi_x}-\pi_y & -(\pi_x v_x + \pi_y v_y) \\ -(u_x + u_y) & 1-v_x u_x - v_y u_y
\end{bmatrix}$ and 
\begin{equation}
    S^{\sim x} RD = \begin{bmatrix}
1-{\pi_x}-\pi_y & -\overline{\beta}(\pi_x v_x + \pi_y v_y) \\ -(u_x + u_y) & \overline{\beta} (1-v_x u_x - v_y u_y)  
\end{bmatrix} \label{eq:SRD_xy}
\end{equation}

\begin{enumerate}
    \item We substitute $ \pi_x v_x + \pi_y v_y = u_x + u_y = 0$ in \eqref{eq:SRD_xy} and raise the power on both sides to $l$.
    
    \item 
    \begin{enumerate}
        \item Substituting $ \pi_x v_x + \pi_y v_y =0$, $\overline{\beta} (1-v_x u_x - v_y u_y)   = 1-{\pi_x}-\pi_y  $ in \eqref{eq:SRD_xy}, we get $ S^{\sim x,y} RD = \begin{bmatrix}
1-{\pi_x}-\pi_y & 0 \\ -(u_x+u_y) & 1-{\pi_x}-\pi_y
\end{bmatrix}.$ Using induction on the exponent $l,$ we get \eqref{eq:non_diag_vxy_0_rep_ev}.  
 
 \item Substituting $  \pi_x v_x + \pi_y v_y =0$ in \eqref{eq:SRD_xy}, we get $ S^{\sim x,y} RD = \\ \begin{bmatrix}
1- \pi_x - \pi_y & 0 \\ -(u_x+u_y) & \overline{\beta} (1-v_x u_x - v_y u_y)  
\end{bmatrix}.$ We observe that $1-{\pi_x}-\pi_y, \overline{\beta}(1-v_x u_x - v_y u_y)   $ are the eigenvalues of $ S^{\sim x,y} RD$ with  $ \begin{bmatrix}
1  \\ \frac{u_x + u_y}{\overline{\beta}(1-v_x u_x - v_y u_y) - \overline{ (\pi_x + \pi_y) } } \end{bmatrix} $ and   $ \begin{bmatrix}
0  \\ 1 \end{bmatrix} $ as their respective right eigenvectors resulting in the diagonalised form in \eqref{eq:diag_vxy_0_dist_ev}.
 \end{enumerate}
 
 \item 
 \begin{enumerate}
        \item Substituting $u_x + u_y =0$, $\overline{\beta} (1-v_x u_x - v_y u_y)   = 1-{\pi_x}-\pi_y  $ in \eqref{eq:SRD_xy}, we get $ S^{\sim x,y} RD = \begin{bmatrix}
1-{\pi_x}-\pi_y & -\overline{\beta} (\pi_x v_x + \pi_y v_y) \\ 0 & 1-{\pi_x}-\pi_y
\end{bmatrix}.$ Using induction on the exponent $l,$ we get \eqref{eq:non_diag_uxy_0_rep_ev}.  
 
 \item Substituting $u_x + u_y =0$ in \eqref{eq:SRD_xy}, we get $ S^{\sim x,y} RD = \begin{bmatrix}
\overline{(\pi_x+\pi_y)} & -\overline{\beta}(\pi_x v_x + \pi_y v_y) \\ 0 & \overline{\beta}
\end{bmatrix}.$ We observe that $1-{\pi_x}-\pi_y, \overline{\beta}(1-v_x u_x - v_y u_y)  $ are the eigenvalues of $ S^{\sim x,y} RD$ with  $ \begin{bmatrix}
1  \\ 0 \end{bmatrix} $ and   $ \begin{bmatrix}
\frac{ \overline{\beta} (\pi_x v_x + \pi_y v_y) }{ \overline{ (\pi_x + \pi_y) } -  \overline{\beta}(1-v_x u_x - v_y u_y) }  \\ 1 \end{bmatrix} $ as their respective right eigenvectors resulting in the diagonalised form in \eqref{eq:diag_uxy_0_dist_ev}.
 \end{enumerate}
 
 \item Solving det$(S^{\sim x,y} RD - \lambda I) = 0,$ we get $\lambda_{i}^{\sim x,y} = \ 0.5\big( 1-{\pi}_x-\pi_y + \overline{\beta} (1-v_x u_x -v_y u_y) + (-1)^{i+1} \Delta_{x,y}  \big)$, $i = 1,2$, as the eigenvalues of $S^{\sim x,y} RD$ with $  \ \begin{bmatrix}
       1 \\ -\frac{\Delta_{x,y} - s_{x,y}}{ 2(1-\beta) (\pi_x v_x + \pi_y v_y)  } \end{bmatrix} , \\  \begin{bmatrix} 1 \\  \frac{\Delta_{x,y} +s_{x,y}}{ 2 (1-\beta) (\pi_x v_x + \pi_y v_y) } \end{bmatrix} $ as their respective right eigenvectors, where  $\Delta_{x,y}^2  = s_{x,y}^2 + 4 \overline{\beta} ( \pi_x v_x + \pi_y v_y )  (u_x+u_y), s_{x,y} = 1-\pi_x - \pi_y - (1-\beta)(1-v_x u_x -v_y u_y),$  resulting in the diagonalised form in \eqref{eq:diag_vxy_uxy_nz}.
 
\end{enumerate}
This completes the proof of Lemma \ref{lem:Diagonalization_xy}.

\subsection{Proof of Lemma \ref{lem:Delta_tau_xy_bd}}
\label{subsec:Delta_tau_xy_bd_Prf}
\begin{enumerate}
    \item \textbf{Lemma  \ref{lem:Delta_tau_xy_bd}, Part 1}:\\ 
    To show \eqref{eq:Delta_tau_bd}, we first start with the expression for $\Delta^2_{x,y}.$ 
    \begin{align}
        \Delta_{x,y}^2 \ & = \  [ (1 - \pi_x - \pi_y) - \overline{\beta} (1-v_x u_x - v_y u_y) ]^2 + 4 \ \overline{\beta} \ (\pi_x v_x + \pi_y v_y) \ (u_x + u_y) \nonumber \\ 
        & \overset{(a)}{=} \ [ \beta + P_{xx} + P_{yy} -2    ]^2 - 4 \ \overline{\beta} \ (1 -v_x u_x - v_y u_y) + 4 \ \overline{\beta} \  (\pi_x + \pi_y) \ \tau_{x,y} \nonumber \\ 
         & = \ \left[ \beta + P_{xx} + P_{yy} - c(\pi_x + \pi_y) +c(\pi_x + \pi_y)   -2    \right]^2 - 4 \ \overline{\beta} \ (1 -v_x u_x - v_y u_y) \nonumber \\ 
         & \qquad + 4 \ \overline{\beta} \  (\pi_x + \pi_y) \ \tau_{x,y} \nonumber \\ 
         & {=} \ \left[ \beta + P_{xx} + P_{yy} - c(\pi_x + \pi_y) \right]^2 + (2c - c^2) \ (\pi_x + \pi_y)^2 - 2 \ ( 2 - c ) \ (\pi_x + \pi_y) \nonumber \\ 
         & \qquad + 2 \ c \  \overline{\beta} \ (\pi_x v_x + \pi_y v_y) \ (u_x + u_y) + 2 \  \overline{\beta} \ (\pi_x + \pi_y) \ \tau \ (2- c) \nonumber \\
         & \overset{(b)}{\leq} \left[ \beta + P_{xx} + P_{yy} - c(\pi_x + \pi_y) \right]^2 + (2c - c^2) \ (\pi_x + \pi_y)^2 - 2 \ ( 2 - c ) \ (\pi_x + \pi_y) \nonumber \\  
         & \qquad + 2 \ c \ (\pi_x + \pi_y) \ (1 - \pi_x - \pi_y )  + 2 \  \overline{\beta} \ (\pi_x + \pi_y) \ \tau \ (2- c) \nonumber \\
         & = \  \left[ \beta + P_{xx} + P_{yy} - c(\pi_x + \pi_y) \right]^2 - c^2 \ (\pi_x + \pi_y)^2 - 4 \ ( 1 - c ) \ (\pi_x + \pi_y) \nonumber \\   
         & \qquad  + 2 \  \overline{\beta} \ (\pi_x + \pi_y) \ \tau \ (2- c) \nonumber 
    \end{align}
    where we get $(a)$ by using the definition of $\tau_{x,y}$$ P_{xx} = \pi_x + \overline{\beta}v_x u_x, P_{yy} = \pi_y + \overline{\beta}v_y u_y,$ and $(b)$ by using $P(X_2 = x \text{ or } y |X_1 = x) = P_{xx} + P_{xy} = \pi_x + \pi_y + \overline{\beta} \ v_x \ (u_x + u_y) \leq 1 \Rightarrow \overline{\beta} \ v_x \ (u_x + u_y) \leq 1 - ( \pi_x + \pi_y ).$
    Choosing $c = \beta^2/(\beta + 2)$ completes the proof. 
    
    \item  \textbf{Lemma  \ref{lem:Delta_tau_xy_bd}, Part 2}:\\ 
    We begin with the following expression for $ (1-\beta) \ \tau_{x,y}:$ 
    \begin{align}
        (1-\beta) \ \tau_{x,y} \ & = \ (1 - \beta) \ \left( 1 -  \frac{ (\pi_y u_x - \pi_x u_y) }{ (\pi_x + \pi_y) } \ (v_x - v_y) \right) \nonumber \\ 
        & = \ (1-\beta) \ + \ \frac{(1-\beta)}{ (\pi_x + \pi_y) } \ (  \pi_x( v_x u_y - v_y u_y ) + \pi_y (v_y u_x - v_x u_x   )  ) \nonumber \\
        & \overset{(a)}{=} \ (1-\beta) \ + \ \frac{ \pi_{x} ( P_{xy} - P_{yy} ) + \pi_y (P_{yx} - P_{xx} ) }{ (\pi_x + \pi_y) } \label{eq:tau_xy_P_xy_expr}
    \end{align}
    where we get $(a)$ by using $ P_{kl} = \pi_l + \overline{\beta}v_k u_l,$ for $k,l \in \{ x,y\}.$
\end{enumerate}
The following lemma bounds the sum of $ P_{xy} - P_{yy} $ and $ P_{yx} - P_{xx}.$ 
\begin{lemma}
For a rank-2 diagonalizable t.p.m $P$ with spectral gap $\beta,$ and $x,y \in \cX,$
\begin{equation}
    P_{xy} - P_{yy} + P_{yx} - P_{xx} \ \leq \ \beta, \  \beta \in [0,2]. \label{eq:P_xy_bd}
\end{equation}
\label{lem:P_xy_bd}
\end{lemma}
\begin{proof}
Section \ref{subsec:P_xy_bd_Prf}.
\end{proof}
To prove \eqref{eq:tau_xy_bd}, we bound 
\begin{align}
    H_0 \ & \triangleq \  \max_{P_{xx},P_{xy},P_{yx},P_{yy}} \frac{ \pi_x(P_{xy} - P_{yy}) + \pi_y (P_{yx} - P_{xx} ) }{\pi_x + \pi_y } \nonumber 
 \end{align}
 subject to the constraint $P_{xy} - P_{yy} + P_{yx} -P_{xx} \  \leq \ \beta.$
 Since $\bpi$ is the stationary distribution of the t.p.m $P,$ we have $\bpi P = \bpi $ implying
 \begin{align}
     \sum_{u \in \cX} \pi_u P_{uk} \ & = \ \pi_k \nonumber \\ 
     \Rightarrow \pi_k P_{kk} + \pi_l P_{lk} \ & \leq \ \pi_k, \ (l \neq k) \nonumber \\ 
     \Rightarrow  (\pi_k + \pi_l) P_{kk} + \pi_l (P_{lk} -P_{kk}) \ &   \leq \ \pi_k,  \nonumber \\
     \Rightarrow \pi_l (P_{lk} -P_{kk}) \ & \overset{(a)}{\leq} \ \pi_k,   \nonumber \\
     \Rightarrow {\pi_l}/{ (\pi_l + \pi_k) } \ & \overset{(b)}{\leq} \ 1/{(P_{lk} + 1 - P_{kk}) }, \ (l \neq k), \label{eq: const_int }
 \end{align}
 where we get $(a)$ by using $P_{kk} \geq 0$, $(b)$ by adding $\pi_l$ on both the sides of $(a).$

Setting $\Psi_1 \triangleq P_{xy} - P_{yy} $, $\Psi_2 \triangleq P_{yx} - P_{xx},$ and using \eqref{eq: const_int } with $k,l \in \{x, y\},$ we get 
  $H_0 \leq \max_{\Psi_1,\Psi_2} \frac{\Psi_1}{\Psi_1 + 1} + \frac{\Psi_2}{\Psi_2 + 1},$ subject to the constraints $\Psi_1 , \Psi_2 \in [-1,1], \ \Psi_1 + \Psi_2 \leq \beta.$ Since $f(z) = \frac{z}{z+1}$ is an increasing function of $z,$ we use $\Psi_1 \leq \beta - \Psi_2$ to get
  \begin{align*}
      H_0 \ & \leq \ \max_{\Psi_2 \in [-1,1]} f(\beta - \Psi_2) + f(\Psi_2) = \max_{\Psi_2 \in [-1,1] } \left( \frac{\beta - \Psi_2}{\beta - \Psi_2 + 1} + \frac{\Psi_2}{\Psi_2 + 1} \right).
  \end{align*}
  $f'(\beta - \Psi_2) + f'(\Psi_2) =  \frac{ (\beta - 2\Psi_2) ( \beta + 2) }{ (\beta - \Psi_2 + 1)^2 (\Psi_2 + 1)^2 }$ implies $f(\beta -\Psi_2) + f(\Psi_2)$ achieves its maximum at $\Psi_2 = \beta/2. $ Therefore $ H_0 \leq 2\beta/(\beta+2).$ Plugging this into \eqref{eq:tau_xy_P_xy_expr}, we get $(1-\beta) \ \tau_{x,y} \leq 1 - \beta^2/(\beta+2).$ This completes the proof of  Lemma  \ref{lem:Delta_tau_xy_bd}, Part 2. 

\subsection{Proof of Lemma \ref{lem:P_xy_bd} }
 \label{subsec:P_xy_bd_Prf}
 For any rank-2 t.p.m $P$ on the alphabet $\cX = [K]$ states, there exist length-$K$ probability vectors $q \triangleq [q_1, q_2, \ldots, q_K]$ , $r \triangleq [r_1, r_2, \ldots, r_K]$ such that any row of $P$ is a convex linear combination of $q$ and $r.$ \\ 
 To prove \eqref{eq:P_xy_bd}, we begin with the trace of $P,$ which is the sum of the diagonal elements of $P$ and is also equal to the sum of the eigenvalues of $P$. 
 \begin{align}
     \text{ trace of } P \ \overset{(a)}{=} 2-\beta \ & {=} \ P_{11} +  \ldots + P_{ii} + \ldots + P_{jj} + \ldots + P_{KK} \nonumber \\ 
     & \overset{(b)}{\leq} \ q_1 + r_1  + \ldots + q_{i-1} + r_{i-1} + P_{ii} + \ldots \nonumber \\ 
     & \qquad + q_{j-1} + r_{j-1} + P_{jj} + \ldots + q_{K} + r_{K} \nonumber \\
     & = \ 1 - q_{i} - q_{j} + 1 - r_{i} - r_{j} + P_{ii} + P_{jj} \nonumber \\
   \Rightarrow  q_{i} + r_{i}  + q_{j} + r_{j} - P_{ii} - P_{jj} \ & \leq \ \beta \nonumber \\
     \Rightarrow P_{ji} - P_{ii} + P_{ij} - P_{jj} \ & \overset{(c)}{\leq} \ \beta \nonumber
 \end{align}
where we get $(a)$ by noting that $1,1-\beta$ are the only non-zero eigenvalues of a rank-2 t.p.m $P$ with spectral gap $\beta,$ $(b), (c)$ by using $P_{ij} \leq q_j + r_j$ for any $i,j \in \cX.$

\end{document}